\documentclass[final,5p,times,twocolumn]{elsarticle}

\usepackage{anyfontsize}
\usepackage{graphicx}
\usepackage{t1enc}
\usepackage[utf8]{inputenc}
\DeclareUnicodeCharacter{2010}{-}
\DeclareUnicodeCharacter{221E}{0xE2}
\usepackage{amsfonts}
\usepackage{flushend}
\usepackage{xcolor}
\usepackage{epstopdf}
\usepackage{amsbsy}
\usepackage{amssymb}
\usepackage{amsmath}
\usepackage{arydshln}
\usepackage{wrapfig}
\usepackage{booktabs}
\usepackage{nomencl} 
\usepackage{accents}
\usepackage{amsthm}
\usepackage{framed} 
\usepackage{nomencl} 

\immediate\write18{%
makeindex -s nomencl.ist -o \jobname.nls -t \jobname.nlg \jobname.nlo%
}

\makenomenclature

\usepackage{bm}

\DeclareBoldMathCommand{\bfa}{\bm a}
\DeclareBoldMathCommand{\bfb}{\bm b}
\DeclareBoldMathCommand{\bfc}{\bm c}
\DeclareBoldMathCommand{\bfd}{\bm d}
\DeclareBoldMathCommand{\bfe}{\bm e}
\DeclareBoldMathCommand{\bff}{\bm f}
\DeclareBoldMathCommand{\bfg}{\bm g}
\DeclareBoldMathCommand{\bfh}{\bm h}
\DeclareBoldMathCommand{\bfi}{\bm i}
\DeclareBoldMathCommand{\bfj}{\bm j}
\DeclareBoldMathCommand{\bfk}{\bm k}
\DeclareBoldMathCommand{\bfl}{\bm l}
\DeclareBoldMathCommand{\bfm}{\bm m}
\DeclareBoldMathCommand{\bfn}{\bm n}
\DeclareBoldMathCommand{\bfo}{\bm o}
\DeclareBoldMathCommand{\bfp}{\bm p}
\DeclareBoldMathCommand{\bfq}{\bm q}
\DeclareBoldMathCommand{\bfr}{\bm r}
\DeclareBoldMathCommand{\bfs}{\bm s}
\DeclareBoldMathCommand{\bft}{\bm t}
\DeclareBoldMathCommand{\bfu}{\bm u}
\DeclareBoldMathCommand{\bfv}{\bm v}
\DeclareBoldMathCommand{\bfw}{\bm w}
\DeclareBoldMathCommand{\bfx}{\bm x}
\DeclareBoldMathCommand{\bfy}{\bm y}
\DeclareBoldMathCommand{\bfz}{\bm z}

\DeclareBoldMathCommand{\bfA}{\bm A}
\DeclareBoldMathCommand{\bfB}{\bm B}
\DeclareBoldMathCommand{\bfC}{\bm C}
\DeclareBoldMathCommand{\bfD}{\bm D}
\DeclareBoldMathCommand{\bfE}{\bm E}
\DeclareBoldMathCommand{\bfF}{\bm F}
\DeclareBoldMathCommand{\bfG}{\bm G}
\DeclareBoldMathCommand{\bfH}{\bm H}
\DeclareBoldMathCommand{\bfI}{\bm I}
\DeclareBoldMathCommand{\bfJ}{\bm J}
\DeclareBoldMathCommand{\bfK}{\bm K}
\DeclareBoldMathCommand{\bfL}{\bm L}
\DeclareBoldMathCommand{\bfM}{\bm M}
\DeclareBoldMathCommand{\bfN}{\bm N}
\DeclareBoldMathCommand{\bfO}{\bm O}
\DeclareBoldMathCommand{\bfP}{\bm P}
\DeclareBoldMathCommand{\bfQ}{\bm Q}
\DeclareBoldMathCommand{\bfR}{\bm R}
\DeclareBoldMathCommand{\bfS}{\bm S}
\DeclareBoldMathCommand{\bfT}{\bm T}
\DeclareBoldMathCommand{\bfU}{\bm U}
\DeclareBoldMathCommand{\bfV}{\bm V}
\DeclareBoldMathCommand{\bfW}{\bm W}
\DeclareBoldMathCommand{\bfX}{\bm X}
\DeclareBoldMathCommand{\bfY}{\bm Y}
\DeclareBoldMathCommand{\bfZ}{\bm Z}

\DeclareBoldMathCommand{\bfalpha}{\bm\alpha}
\DeclareBoldMathCommand{\bfbeta}{\bm\beta}
\DeclareBoldMathCommand{\bfgamma}{\bm\gamma}
\DeclareBoldMathCommand{\bfdelta}{\bm\delta}
\DeclareBoldMathCommand{\bfepsilon}{\bm\epsilon}
\DeclareBoldMathCommand{\bfvarepsilon}{\bm\varepsilon}
\DeclareBoldMathCommand{\bfzeta}{\bm\zeta}
\DeclareBoldMathCommand{\bfeta}{\bm\eta}
\DeclareBoldMathCommand{\bftheta}{\bm\theta}
\DeclareBoldMathCommand{\bfvartheta}{\bm\vartheta}
\DeclareBoldMathCommand{\bfiota}{\bm\iota}
\DeclareBoldMathCommand{\bfkappa}{\bm\kappa}
\DeclareBoldMathCommand{\bflambda}{\bm\lambda}
\DeclareBoldMathCommand{\bfmu}{\bm\mu}
\DeclareBoldMathCommand{\bfnu}{\bm\nu}
\DeclareBoldMathCommand{\bfxi}{\bm\xi}
\DeclareBoldMathCommand{\bfpi}{\bm\pi}
\DeclareBoldMathCommand{\bfvarpi}{\bm\varpi}
\DeclareBoldMathCommand{\bfrho}{\bm\rho}
\DeclareBoldMathCommand{\bfvarrho}{\bm\varrho}
\DeclareBoldMathCommand{\bfsigma}{\bm\sigma}
\DeclareBoldMathCommand{\bfvarsigma}{\bm\varsigma}
\DeclareBoldMathCommand{\bftau}{\bm\tau}
\DeclareBoldMathCommand{\bfupsilon}{\bm\upsilon}
\DeclareBoldMathCommand{\bfphi}{\bm\phi}
\DeclareBoldMathCommand{\bfvarphi}{\bm\varphi}
\DeclareBoldMathCommand{\bfchi}{\bm\chi}
\DeclareBoldMathCommand{\bfpsi}{\bm\psi}
\DeclareBoldMathCommand{\bfomega}{\bm\omega}

\DeclareBoldMathCommand{\bfGamma}{\bm\Gamma}
\DeclareBoldMathCommand{\bfDelta}{\bm\Delta}
\DeclareBoldMathCommand{\bfTheta}{\bm\Theta}
\DeclareBoldMathCommand{\bfLambda}{\bm\Lambda}
\DeclareBoldMathCommand{\bfXi}{\bm\Xi}
\DeclareBoldMathCommand{\bfPi}{\bm\Pi}
\DeclareBoldMathCommand{\bfSigma}{\bm\Sigma}
\DeclareBoldMathCommand{\bfUpsilon}{\bm\Upsilon}
\DeclareBoldMathCommand{\bfPhi}{\bm\Phi}
\DeclareBoldMathCommand{\bfPsi}{\bm\Psi}
\DeclareBoldMathCommand{\bfOmega}{\bm\Omega}

\DeclareBoldMathCommand{\zero}{\bm 0}
\DeclareBoldMathCommand{\Zero}{\bm O}
\DeclareBoldMathCommand{\bfimath}{\bm\imath}
\DeclareBoldMathCommand{\bfjmath}{\bm\jmath}

\DeclareBoldMathCommand{\bma}{\mathbf{a}}
\DeclareBoldMathCommand{\bmb}{\mathbf{b}}
\DeclareBoldMathCommand{\bmc}{\mathbf{c}}
\DeclareBoldMathCommand{\bmd}{\mathbf{d}}
\DeclareBoldMathCommand{\bme}{\mathbf{e}}
\DeclareBoldMathCommand{\bmf}{\mathbf{f}}
\DeclareBoldMathCommand{\bmg}{\mathbf{g}}
\DeclareBoldMathCommand{\bmh}{\mathbf{h}}
\DeclareBoldMathCommand{\bmi}{\mathbf{i}}
\DeclareBoldMathCommand{\bmj}{\mathbf{j}}
\DeclareBoldMathCommand{\bmk}{\mathbf{k}}
\DeclareBoldMathCommand{\bml}{\mathbf{l}}
\DeclareBoldMathCommand{\bmm}{\mathbf{m}}
\DeclareBoldMathCommand{\bmn}{\mathbf{n}}
\DeclareBoldMathCommand{\bmo}{\mathbf{o}}
\DeclareBoldMathCommand{\bmp}{\mathbf{p}}
\DeclareBoldMathCommand{\bmq}{\mathbf{q}}
\DeclareBoldMathCommand{\bmr}{\mathbf{r}}
\DeclareBoldMathCommand{\bms}{\mathbf{s}}
\DeclareBoldMathCommand{\bmt}{\mathbf{t}}
\DeclareBoldMathCommand{\bmu}{\mathbf{u}}
\DeclareBoldMathCommand{\bmv}{\mathbf{v}}
\DeclareBoldMathCommand{\bmw}{\mathbf{w}}
\DeclareBoldMathCommand{\bmx}{\mathbf{x}}
\DeclareBoldMathCommand{\bmy}{\mathbf{y}}
\DeclareBoldMathCommand{\bmz}{\mathbf{z}}

\DeclareBoldMathCommand{\bmA}{\mathbf{A}}
\DeclareBoldMathCommand{\bmB}{\mathbf{B}}
\DeclareBoldMathCommand{\bmC}{\mathbf{C}}
\DeclareBoldMathCommand{\bmD}{\mathbf{D}}
\DeclareBoldMathCommand{\bmE}{\mathbf{E}}
\DeclareBoldMathCommand{\bmF}{\mathbf{F}}
\DeclareBoldMathCommand{\bmG}{\mathbf{G}}
\DeclareBoldMathCommand{\bmH}{\mathbf{H}}
\DeclareBoldMathCommand{\bmI}{\mathbf{I}}
\DeclareBoldMathCommand{\bmJ}{\mathbf{J}}
\DeclareBoldMathCommand{\bmK}{\mathbf{K}}
\DeclareBoldMathCommand{\bmL}{\mathbf{L}}
\DeclareBoldMathCommand{\bmM}{\mathbf{M}}
\DeclareBoldMathCommand{\bmN}{\mathbf{N}}
\DeclareBoldMathCommand{\bmO}{\mathbf{O}}
\DeclareBoldMathCommand{\bmP}{\mathbf{P}}
\DeclareBoldMathCommand{\bmQ}{\mathbf{Q}}
\DeclareBoldMathCommand{\bmR}{\mathbf{R}}
\DeclareBoldMathCommand{\bmS}{\mathbf{S}}
\DeclareBoldMathCommand{\bmT}{\mathbf{T}}
\DeclareBoldMathCommand{\bmU}{\mathbf{U}}
\DeclareBoldMathCommand{\bmV}{\mathbf{V}}
\DeclareBoldMathCommand{\bmW}{\mathbf{W}}
\DeclareBoldMathCommand{\bmX}{\mathbf{X}}
\DeclareBoldMathCommand{\bmY}{\mathbf{Y}}
\DeclareBoldMathCommand{\bmZ}{\mathbf{Z}}

\renewcommand{\labelenumi}{A\theenumi.}
\newtheorem{cremark}{Remark}
\newtheorem{proposition}{Proposition}
\newtheorem{assumption}{Assumption}

\newtheorem{lemma}{Lemma}

\newcommand\blfootnote[1]{%
  \begingroup
  \renewcommand\thefootnote{}\footnote{#1}%
  \addtocounter{footnote}{-1}%
  \endgroup
}

\journal{European Journal of Control}

\begin{document}

\begin{frontmatter}




\title{Robust output-feedback VFO-ADR control of  underactuated spatial vehicles \newline in the task of following non-parametrized paths}


\author{Krzysztof {\L}akomy$^*$}
\ead{krzysztof.pi.lakomy@doctorate.put.poznan.pl}
\cortext[cor1]{Corresponding author}

\author{Maciej Marcin Micha{\l}ek}
\ead{maciej.michalek@put.poznan.pl}

\address{Institute of Automatic Control and Robotics,
              Poznan University of Technology (PUT),
              Piotrowo 3A, 60-965 Pozna\'n, Poland}

\begin{abstract}
   This article concerns the development of the Vector Field Orientation - Active Disturbance Rejection (VFO-ADR) cascaded path-following controller for underactuated vehicles moving in a 3-dimensional space. The proposed concept of a cascaded control structure decouples system kinematics from system dynamics, resembling the approach utilized for nonholonomic systems. Thanks to the use of an ADR control approach in the dynamic-level controller, the proposed control structure is robust to even significant model uncertainties and external disturbances. Application of an error-based form of the Extended State Observer (ESO), implemented within the ADR inner-loop controller, implies the output-feedback characteristic of the control structure, i.e., only position and attitude of the vehicle body are expected to be measured. The kinematic-level controller is designed according to the VFO method utilizing the non-parametrized path representation to calculate the commanded velocities. The description of the proposed control structure is followed by the theoretical analysis utilizing the Input-to-State Stability (ISS) theorem and the simulation verification of the proposed solution.
\end{abstract}




\end{frontmatter}


\newcommand{\todo}{\color{magenta}TODO\color{black}: }

\newtheorem{uwaga}{Uwaga}

\newtheorem{zalozenie}{Założenie}


\newcommand{\xLocalAxis}{x^B}
\newcommand{\yLocalAxis}{y^B}
\newcommand{\zLocalAxis}{z^B}

\newcommand{\xGlobalAxis}{x^G}
\newcommand{\yGlobalAxis}{y^G}
\newcommand{\zGlobalAxis}{z^G}

\newcommand{\origin}{\mathcal{O}}

\newcommand{\globalCoordinateSystem}{\{G\}}
\newcommand{\localCoordinateSystem}{\{B\}}





\newcommand{\massCenter}{C}


\newcommand{\configurationVector}{\bfeta}
\newcommand{\reducedConfigurationVector}{\bar{\bfeta}}
\newcommand{\positionVector}{\bfeta_p}
\newcommand{\positionVectorDerivative}{\dot{\bfeta}_p}
\newcommand{\orientationVector}{\bfeta_o}

\newcommand{\configurationVectorDerivative}{\dot{\bfeta}}
\newcommand{\reducedConfigurationVectorDerivative}{\dot{\bar{\bfeta}}}
\newcommand{\configurationVectorSecondDerivative}{\ddot{\bfeta}}
\newcommand{\orientationVectorDerivative}{\dot{\bfeta}_o}

\newcommand{\xPosition}{x}
\newcommand{\yPosition}{y}
\newcommand{\zPosition}{z}
\newcommand{\roll}{\phi}
\newcommand{\pitch}{\theta}
\newcommand{\yaw}{\psi}

\newcommand{\xPositionDerivative}{\dot{x}}
\newcommand{\yPositionDerivative}{\dot{y}}
\newcommand{\zPositionDerivative}{\dot{z}}
\newcommand{\rollDerivative}{\dot{\phi}}
\newcommand{\pitchDerivative}{\dot{\theta}}
\newcommand{\yawDerivative}{\dot{\psi}}



\newcommand{\localVelocityVector}{\bfnu}
\newcommand{\localVelocityVectorDerivative}{\dot{\bfnu}}
\newcommand{\globalVelocityVector}{\dot{\bfeta}}

\newcommand{\localLongitudinalVelocityVector}{\bfnu_p}
\newcommand{\localAngularVelocityVector}{\bfnu_o}
\newcommand{\globalLongitudinalVelocityVector}{\dot{\bfeta}_p}
\newcommand{\globalAngularVelocityVector}{\dot{\bfeta}_o}

\newcommand{\xLocalVelocity}{u}
\newcommand{\yLocalVelocity}{v}
\newcommand{\zLocalVelocity}{w}
\newcommand{\rollLocalVelocity}{p}
\newcommand{\pitchLocalVelocity}{q}
\newcommand{\yawLocalVelocity}{r}

\newcommand{\commandedVelocityVectorLocal}{\bfnu_c}
\newcommand{\commandedNominalVelocityVectorLocal}{\bfnu_{cn}}
\newcommand{\reducedCommandedVelocityVectorLocal}{\bar{\bfnu}_c}
\newcommand{\reducedCommandedNominalVelocityVectorLocal}{\bar{\bfnu}_{cn}}
\newcommand{\reducedCommandedVelocityVectorLocalEstimate}{\hat{\bar{\bfnu}}_c}
\newcommand{\longitudinalCommandedVelocityVectorLocal}{\bfnu_{p c}}
\newcommand{\angularCommandedVelocityVectorLocal}{\bfnu_{o c}}
\newcommand{\commandedXLocalVelocity}{u_c}
\newcommand{\commandedYLocalVelocity}{v_c}
\newcommand{\commandedZLocalVelocity}{w_c}
\newcommand{\commandedRollLocalVelocity}{p_c}
\newcommand{\commandedPitchLocalVelocity}{q_c}
\newcommand{\commandedYawLocalVelocity}{r_c}

\newcommand{\commandedXNominalLocalVelocity}{u_{cn}}
\newcommand{\commandedPitchNominalLocalVelocity}{q_{cn}}
\newcommand{\commandedYawNominalLocalVelocity}{r_{cn}}

\newcommand{\rollControllerFunction}{f_\phi}
\newcommand{\rollErrorUltimateBound}{r^\infty_\roll}

\newcommand{\desiredXLocalVelocity}{u_d}
\newcommand{\desiredPitchLocalVelocity}{q_d}
\newcommand{\desiredYawLocalVelocity}{r_d}


\newcommand{\rotationMatrix}{\bfR}
\newcommand{\rotationMatrixResidue}{\bfGamma_R}
\newcommand{\auxiliaryRotationResidueMatrix}{\bfW}
\newcommand{\angularVelocityTransformationMatrix}{\bfT}
\newcommand{\velocityTransformationMatrix}{\bfJ}
\newcommand{\simplifiedVelocityTransformationMatrix}{\bfG}
\newcommand{\velocityTransformationMatrixDerivative}{\dot{\bfJ}}


\newcommand{\massMatrixGlobal}{\bfM_\eta}
\newcommand{\dynamicPhenomenaVectorGlobal}{\bfmu_\eta}
\newcommand{\controlSignalGlobal}{\bftau_\eta}
\newcommand{\perturbationsVectorGlobal}{\bftau_\eta^*}
\newcommand{\perturbationsVectorGlobalDerivative}{\dot{\bftau}_\eta^*}
\newcommand{\actuationMatrixGlobal}{\bfGamma_\eta}

\newcommand{\massMatrixLocal}{\bfM}
\newcommand{\mass}{m}
\newcommand{\dynamicPhenomenaVectorLocal}{\bfmu}
\newcommand{\controlSignalLocal}{\bftau}
\newcommand{\perturbationsVectorLocal}{\bftau^*}
\newcommand{\perturbationsVectorLocalDerivative}{\dot{\bftau}^*}
\newcommand{\actuationMatrixLocal}{\bfGamma}
\newcommand{\actuationMatrixLocalElement}[1]{\gamma_#1}

\newcommand{\controlSignalLocalAlongX}{\tau_u}
\newcommand{\controlSignalLocalAlongY}{\tau_v}
\newcommand{\controlSignalLocalAlongZ}{\tau_w}
\newcommand{\controlSignalLocalAroundX}{\tau_p}
\newcommand{\controlSignalLocalAroundY}{\tau_q}
\newcommand{\controlSignalLocalAroundZ}{\tau_r}
\newcommand{\controlSignalLocalUnknown}{\tau_i}

\newcommand{\actuationAlongYLocal}{\gamma_v}
\newcommand{\actuationAlongZLocal}{\gamma_w}
\newcommand{\actuationAroundXLocal}{\gamma_p}
\newcommand{\actuationUndefinedLocal}{\gamma_i}

\newcommand{\perturbationsVectorLocalLimit}{\bar{m}_\tau}
\newcommand{\perturbationsVectorGlobalLimit}{\bar{m}_{\tau\eta}}
\newcommand{\perturbationsVectorLocalDerivativeLimit}{\bar{M}_\tau}
\newcommand{\perturbationsVectorGlobalDerivativeLimit}{\bar{M}_{\tau\eta}}

\newcommand{\controlSignalUpperBoundlLocal}{\bar{\bftau}}
\newcommand{\controlSignalUpperBoundLocalAlongX}{\bar{\tau}_u}
\newcommand{\controlSignalUpperBoundLocalAlongY}{\bar{\tau}_v}
\newcommand{\controlSignalUpperBoundLocalAlongZ}{\bar{\tau}_w}
\newcommand{\controlSignalUpperBoundLocalAroundX}{\bar{\tau}_p}
\newcommand{\controlSignalUpperBoundLocalAroundY}{\bar{\tau}_q}
\newcommand{\controlSignalUpperBoundLocalAroundZ}{\bar{\tau}_r}
\newcommand{\controlSignalUpperBoundLocalUnknown}{\bar{\tau}_i}

\newcommand{\controlAuthorityMatrix}{\bar{\bfA}}
\newcommand{\controlAuthorityAlongX}{a_u}
\newcommand{\controlAuthorityAlongY}{a_v}
\newcommand{\controlAuthorityAlongZ}{a_w}
\newcommand{\controlAuthorityAroundX}{a_p}
\newcommand{\controlAuthorityAroundY}{a_q}
\newcommand{\controlAuthorityAroundZ}{a_r}
\newcommand{\controlAuthorityUnknown}{a_i}



\newcommand{\pathPointSet}{\mathcal{S}_d}
\newcommand{\pathPositionVector}{\bfeta_{pd}}
\newcommand{\pathXPosition}{x_d}
\newcommand{\pathYPosition}{y_d}
\newcommand{\pathZPosition}{z_d}
\newcommand{\admissiblePositionSet}{\mathcal{D}}
\newcommand{\levelSurface}[1]{s_#1}
\newcommand{\levelSurfaceDerivative}[1]{\dot{s}_#1}
\newcommand{\levelSurfaceDerivativeEstimate}[1]{\hat{\dot{s}}_#1}
\newcommand{\levelSurfaceDerivativeError}[1]{\tilde{\dot{s}}_#1}
\newcommand{\pathVelocityProfile}{u_d}
\newcommand{\pathVelocityDirection}{\xi_d}
\newcommand{\pathVelocityMagnitude}{U}

\newcommand{\levelSurfaceGradient}[1]{\nabla s_#1}
\newcommand{\levelSurfaceGradientTranspose}[1]{\nabla^\top s_#1}
\newcommand{\levelSurfaceGradientElement}[2]{s_{#1 #2}}
\newcommand{\levelSurfaceHessian}[1]{\nabla(\levelSurfaceGradientTranspose{j})}
\newcommand{\levelSurfaceHessianWithArgument}[1]{\nabla(\levelSurfaceGradientTranspose{j}(\positionVector))}
\newcommand{\levelSurfaceHessianUpperBound}[1]{\bar{M}_#1}
\newcommand{\levelSurfaceGradientUpperBound}[1]{\bar{m}_#1}
\newcommand{\levelSurfaceGradientLowerBound}[1]{\underaccent{\bar}{m}_#1}
\newcommand{\levelSurfaceHessianElement}[3]{s_{#1 #2 #3}}
\newcommand{\levelSurfaceGradientCrossProduct}{\bfw_\bot}
\newcommand{\levelSurfaceGradientCrossProductDerivative}{\dot{\bfw}_\bot}
\newcommand{\levelSurfaceGradientCrossProductDerivativeEstimate}{\hat{\dot{\bfw}}_\bot}
\newcommand{\levelSurfaceGradientCrossProductDerivativeError}{\tilde{\dot{\bfw}}_\bot}

\newcommand{\angleBetweenlevelSurfaceGradients}{\kappa}

\newcommand{\pathNormalUnitVector}[1]{\bfvartheta_#1}
\newcommand{\pathReducedNormalUnitVector}[1]{\bar{\bfvartheta}_#1}
\newcommand{\pathNormalUnitVectorDerivative}[1]{\dot{\bfvartheta}_#1}
\newcommand{\pathNormalUnitVectorDerivativeEstimate}[1]{\hat{\dot{\bfvartheta}}_#1}
\newcommand{\pathNormalUnitVectorDerivativeError}[1]{\tilde{\dot{\bfvartheta}}_#1}
\newcommand{\pathTangentialUnitVector}{\bfvartheta_\bot}
\newcommand{\pathTangentialUnitVectorDerivative}{\dot{\bfvartheta}_\bot}
\newcommand{\pathTangentialUnitVectorDerivativeEstimate}{\hat{\dot{\bfvartheta}}_\bot}
\newcommand{\pathTangentialUnitVectorDerivativeError}{\tilde{\dot{\bfvartheta}}_\bot}
\newcommand{\pathNormalUnitVectorElement}[2]{\vartheta_{#1 #2}}
\newcommand{\pathTangentialUnitVectorElement}[1]{\vartheta_{\bot #1}}
\newcommand{\pathMotionDirection}{\sigma}
\newcommand{\pathReducedTangentialUnitVector}{\bar{\bfvartheta}_\bot}
\newcommand{\pathReducedTangentialUnitVectorLowerBound}{\underaccent{\bar}{m}_{\bot}}


\newcommand{\timePoint}{t}

\newcommand{\realNumbers}{\mathbb{R}}
\newcommand{\nonNegativeRealNumbers}{\mathbb{R}_{\geq0}}

\newcommand{\maxEigenvalue}[1]{\lambda_\textrm{max}(#1)}
\newcommand{\minEigenvalue}[1]{\lambda_\textrm{min}(#1)}

\newcommand{\integerNumbers}{\mathbb{Z}}

\newcommand{\positiveIntegerNumbers}{\mathbb{Z}_+}

\newcommand{\kappaFunction}{\mathcal{K}}
\newcommand{\kappaLambdaFunction}{\mathcal{KL}}

\newcommand{\zeroMatrix}{\pmb{0}}

\newcommand{\identityMatrix}{\bfI}

\newcommand{\timeDerivative}{\frac{d}{dt}}

\let\cos\undefined
\let\sin\undefined
\let\tan\undefined
\newcommand{\cos}{\textrm{c}}
\newcommand{\sin}{\textrm{s}}
\newcommand{\tan}{\textrm{t}}

\newcommand{\diag}{\textrm{diag}}

\newcommand{\atan}{\textrm{arctan}}
\newcommand{\atantwo}{\textrm{Atan2}}
\newcommand{\atantwoc}{\textrm{Atan2c}}

\newcommand{\modulo}{\textrm{mod}}
\newcommand{\module}[1]{\|#1\|}
\newcommand{\oneNorm}[1]{\left\|#1\right\|_1}
\newcommand{\abs}[1]{|#1|}

\newcommand{\ls}{\limsup_{t\rightarrow\infty}}
\newcommand{\lsShort}{\textrm{ls}_\infty}

\newcommand\eqtext[1]{\stackrel{{{#1}}}{=}}

\newcommand{\blkdiag}{\textrm{blkdiag}}

\newcommand{\simulationTime}{T_s}

\newcommand{\eqdef}{\triangleq}
\newcommand{\eq}[1]{\stackrel{(\ref{#1})}{=}}


\newcommand{\observerInputRowVector}{\bfb}
\newcommand{\localInputMatrixEstimate}{\hat{\bfB}}
\newcommand{\localInputMatrixEstimateDiagonalElement}[1]{\hat{b}_{#1}}

\newcommand{\globalInputMatrixEstimate}{\hat{\bfB}_\eta}

\newcommand{\extendedStateVector}[1]{\bfx_#1}
\newcommand{\extendedStateVectorDerivative}[1]{\dot{\bfx}_#1}
\newcommand{\extendedStateVectorElement}[2]{x_{#1#2}}
\newcommand{\extendedStateVectorElementObservationError}[2]{\tilde{x}_{#1#2}}

\newcommand{\configurationVectorElement}[1]{\eta_#1}

\newcommand{\commandedConfigurationVector}{\bfeta_{c}}
\newcommand{\commandedPositionVectorDerivative}{\dot{\bfeta}_{pc}}
\newcommand{\commandedOrientationVectorDerivative}{\dot{\bfeta}_{oc}}
\newcommand{\commandedConfigurationVectorDerivative}{\dot{\bfeta}_{c}}
\newcommand{\commandedConfigurationVectorSecondDerivative}{\ddot{\bfeta}_{c}}
\newcommand{\commandedConfigurationVectorElement}[1]{\eta_{c#1}}

\newcommand{\velocityErrorVector}{\bfepsilon}
\newcommand{\reducedVelocityErrorVector}{\bar{\bfepsilon}}
\newcommand{\longitudinalVelocityErrorVector}{\bfepsilon_p}
\newcommand{\longitudinalVelocityErrorVectorEstimate}{\hat{\bfepsilon}_p}
\newcommand{\longitudinalVelocityErrorVectorEstimateDerivative}{\dot{\hat{\bfepsilon}}_p}
\newcommand{\longitudinalVelocityErrorVectorEstimateDerivativeEstimate}{\hat{\dot{\hat{\bfepsilon}}}_p}
\newcommand{\longitudinalVelocityErrorVectorObservationError}{\tilde{\bfepsilon}_p}
\newcommand{\angularVelocityErrorVector}{\bfepsilon_o}
\newcommand{\reducedAngularVelocityErrorVector}{\bar{\bfepsilon}_o}
\newcommand{\reducedLongitudinalVelocityErrorVector}{\bar{\bfepsilon}_p}
\newcommand{\reducedLongitudinalVelocityErrorVectorEstimate}{\hat{\bar{\bfepsilon}}_p}
\newcommand{\reducedAngularVelocityErrorVectorEstimate}{\hat{\bar{\bfepsilon}}_o}
\newcommand{\angularVelocityErrorVectorEstimate}{\hat{{\bfepsilon}}_o}
\newcommand{\reducedAngularVelocityErrorVectorObservationError}{\tilde{\bar{\bfepsilon}}_o}
\newcommand{\velocityErrorVectorEstimate}{\hat{\bfepsilon}}
\newcommand{\velocityErrorVectorObservationError}{\tilde{\bfepsilon}}
\newcommand{\velocityErrorVectorObservationErrorElement}[1]{\tilde{\epsilon}_#1}
\newcommand{\velocityErrorVectorDerivative}{\dot{\bfepsilon}}

\newcommand{\velocityErrorVectorLocal}{\bfepsilon_\nu}
\newcommand{\longitudinalVelocityErrorVectorLocal}{\bfepsilon_{\nu p}}
\newcommand{\angularVelocityErrorVectorLocal}{\bfepsilon_{\nu o}}

\newcommand{\velocityErrorVectorElement}[1]{\epsilon_#1}
\newcommand{\velocityErrorVectorElementEstimate}[1]{\hat{\epsilon}_#1}
\newcommand{\velocityErrorVectorElementObservationError}[1]{\tilde{\epsilon}_#1}

\newcommand{\totalDisturbanceVector}{\bfd}
\newcommand{\totalDisturbanceVectorEstimate}{\hat{\bfd}}
\newcommand{\totalDisturbanceVectorObservationError}{\tilde{\bfd}}
\newcommand{\totalDisturbanceVectorDerivative}{\dot{\bfd}}
\newcommand{\totalDisturbanceVectorElement}[1]{d_#1}
\newcommand{\totalDisturbanceVectorElementDerivative}[1]{\dot{d}_#1}

\newcommand{\modifiedTotalDisturbanceVector}{\bfd^*}
\newcommand{\modifiedTotalDisturbanceVectorElement}[1]{d^*_#1}
\newcommand{\modifiedTotalDisturbanceVectorElementDerivative}[1]{\dot{d}_#1}

\newcommand{\observerStateMatrix}{\bfA}

\newcommand{\observerOutputMatrix}{\bfC}

\newcommand{\cutoffMatrix}[1]{\bfzeta_#1^\top}
\newcommand{\cutoffMatrixElement}[2]{\zeta_{#1#2}}

\newcommand{\observerOutput}[1]{y_#1}
\newcommand{\observerAggregatedOutput}{\bfy}
\newcommand{\observerAggregatedOutputEstimate}{\hat{\bfy}}

\newcommand{\observerEstimatedStateVector}[1]{\hat{\bfx}_#1}
\newcommand{\observerEstimatedStateVectorElement}[2]{\hat{x}_{#1 #2}}
\newcommand{\observerEstimatedStateVectorDerivative}[1]{\dot{\hat{\bfx}}_#1}
\newcommand{\observerEstimatedOutput}[1]{\hat{y}_#1}

\newcommand{\observerGainVector}[1]{\bfl_#1}
\newcommand{\observerGainVectorElement}[2]{l_{#1 #2}}
\newcommand{\observerBandwidth}{\omega_o}
\newcommand{\particularObserverBandwidth}[1]{\omega_{o #1}}

\newcommand{\observationErrorVector}[1]{\tilde{\bfx}_#1}
\newcommand{\observationErrorVectorDerivative}[1]{\dot{\tilde{\bfx}}_#1}

\newcommand{\observationErrorAggregatedSubvector}[1]{\tilde{\bfchi}_#1}

\newcommand{\observerStateAggregatedVector}{\bfchi}
\newcommand{\observerStateEstimateAggregatedVector}{\hat{\bfchi}}
\newcommand{\observationErrorAggregatedVector}{\tilde{\bfchi}}
\newcommand{\observationErrorAggregatedVectorDerivative}{\dot{\tilde{\bfchi}}}
\newcommand{\observatorOutputObservationError}{\tilde{\bfy}}
\newcommand{\observerGainAggregatedMatrix}[1]{\bfL_#1}

\newcommand{\environmentalDampingMatrixGlobal}{\bfDelta_\eta}
\newcommand{\environmentalDampingMatrixLocal}{\bfDelta}
\newcommand{\environmentalDampingElementLocal}[1]{\rho_#1}
\newcommand{\environmentalDampingElement}{\rho}

\newcommand{\adrcGainMatrixGlobal}{\bfK_\eta}
\newcommand{\adrcGainMatrixLocal}{\bfK}
\newcommand{\adrcGainMatrixElementLocal}[1]{k_#1}
\newcommand{\adrcGainMatrixAngularElement}{k_o}

\newcommand{\configurationErrorVector}{\bfe}
\newcommand{\configurationModuloErrorVector}{\bfe_{2\pi}}
\newcommand{\configurationErrorDomain}{\mathcal{Q}_e}

\newcommand{\orientationError}{\bfe_o}
\newcommand{\orientationErrorElement}[1]{e_#1}
\newcommand{\auxiliaryOrientationError}{\bar{\bfe}_a}
\newcommand{\auxiliaryOrientationErrorElement}[1]{e_{#1 a}}
\newcommand{\auxiliaryOrientationErrorDerivative}{\dot{\bar{\bfe}}_a}
\newcommand{\exendedAuxiliaryOrientationError}{\bfe_a}
\newcommand{\extendedAuxiliaryOrientationErrorDerivative}{\dot{\bfe}_a}
\newcommand{\auxiliaryOrientationErrorPitch}{e_{\pitch a}}
\newcommand{\auxiliaryOrientationErrorYaw}{e_{\yaw a}}
\newcommand{\rollError}{e_\phi}
\newcommand{\rollErrorDerivative}{\dot{e}_\phi}

\newcommand{\positionError}{\bfe_p}
\newcommand{\reducedPositionError}{\bar{\bfe}_p}
\newcommand{\positionErrorVectorElement}[1]{e_#1}
\newcommand{\positionErrorDerivative}{\dot{\bfe}_p}

\newcommand{\vfoLongitudinalGain}{k_p}
\newcommand{\vfoOrientationGainMatrix}{\bfK_a}
\newcommand{\vfoOrientationGainPitch}{k_\pitch}
\newcommand{\vfoOrientationGainYaw}{k_\yaw}
\newcommand{\rollControllerGain}{k_\phi}
\newcommand{\longitudinalCompensationCoefficient}{\delta_p}
\newcommand{\angularCompensationCoefficient}{\delta_o}


\newcommand{\desiredPositionDerivative}{\dot{\bfeta}_{pd}}
\newcommand{\desiredPositionDerivativeUpperBound}{\bar{m}_1}
\newcommand{\desiredPositionSecondDerivative}{\ddot{\bfeta}_{pd}}
\newcommand{\reducedDesiredPositionDerivative}{\dot{\bar{\bfeta}}_{pd}}
\newcommand{\desiredXPosition}{x_d}
\newcommand{\desiredYPosition}{y_d}
\newcommand{\desiredZPosition}{z_d}
\newcommand{\desiredXPositionDerivative}{\dot{x}_d}
\newcommand{\desiredYPositionDerivative}{\dot{y}_d}
\newcommand{\desiredZPositionDerivative}{\dot{z}_d}

\newcommand{\reducedAuxiliaryOrientation}{\bar{\bfeta}_{oa}}
\newcommand{\reducedAuxiliaryOrientationDerivative}{\dot{\bar{\bfeta}}_{oa}}
\newcommand{\reducedAuxiliaryOrientationDerivativeEstimate}{\hat{\dot{\bar{\bfeta}}}_{oa}}
\newcommand{\reducedAuxiliaryOrientationDerivativeObservationError}{\tilde{\dot{\bar{\bfeta}}}_{oa}}

\newcommand{\auxiliaryYawDerivativeErrorFunction}{f_\yaw}
\newcommand{\auxiliaryPitchDerivativeErrorFunction}{f_\pitch}
\newcommand{\auxiliaryPitchDerivative}{\dot{\pitch}_a}
\newcommand{\auxiliaryYawDerivative}{\dot{\yaw}_a}
\newcommand{\auxiliaryPitchDerivativeObservationError}{\tilde{\dot{\pitch}}_a}
\newcommand{\auxiliaryYawDerivativeObservationError}{\tilde{\dot{\yaw}}_a}

\newcommand{\convergenceVectorField}{\bfh}
\newcommand{\longitudinalConvergenceVectorField}{\bfh_p}
\newcommand{\longitudinalConvergenceVectorFieldEstimate}{\hat{\bfh}_p}
\newcommand{\longitudinalConvergenceVectorFieldEstimateDerivative}{\dot{\hat{\bfh}}_p}
\newcommand{\longitudinalConvergenceVectorFieldEstimateDerivativeEstimate}{\hat{\dot{\hat{\bfh}}}_p}
\newcommand{\longitudinalConvergenceVectorFieldFeedforward}{\bfh_p^{ff}}
\newcommand{\longitudinalConvergenceVectorFieldFeedback}{\bfh_p^{fb}}
\newcommand{\longitudinalConvergenceVectorFieldCompensation}{\bfh_p^{dc}}
\newcommand{\longitudinalConvergenceVectorFieldFeedforwardDerivative}{\dot{\bfh}_p^{ff}}
\newcommand{\longitudinalConvergenceVectorFieldFeedbackDerivative}{\dot{\bfh}_p^{fb}}
\newcommand{\longitudinalConvergenceVectorFieldCompensationDerivative}{\dot{\bfh}_p^{dc}}
\newcommand{\longitudinalConvergenceVectorFieldFeedforwardDerivativeEstimate}{\hat{\dot{\bfh}}_p^{ff}}
\newcommand{\longitudinalConvergenceVectorFieldFeedbackDerivativeEstimate}{\hat{\dot{\bfh}}_p^{fb}}
\newcommand{\longitudinalConvergenceVectorFieldCompensationDerivativeEstimate}{\hat{\dot{\bfh}}_p^{dc}}
\newcommand{\longitudinalConvergenceVectorFieldCompensationEstimate}{\hat{\bfh}_p^{dc}}
\newcommand{\longitudinalConvergenceVectorFieldCompensationEstimateDerivative}{\dot{\hat{\bfh}}_p^{dc}}
\newcommand{\longitudinalConvergenceVectorFieldCompensationEstimateDerivativeEstimate}{\hat{\dot{\hat{\bfh}}}_p^{dc}}
\newcommand{\reducedLongitudinalConvergenceVectorField}{\bar{\bfh}_p}
\newcommand{\angularConvergenceVectorField}{\bfh_o}
\newcommand{\angularConvergenceVectorFieldEstimate}{\hat{\bfh}_o}
\newcommand{\angularConvergenceVectorFieldFeedforward}{\bfh_o^{ff}}
\newcommand{\angularConvergenceVectorFieldFeedforwardEstimate}{\hat{\bfh}_o^{ff}}
\newcommand{\angularConvergenceVectorFieldFeedback}{\bfh_o^{fb}}
\newcommand{\angularConvergenceVectorFieldCompensation}{\bfh_o^{dc}}
\newcommand{\angularConvergenceVectorFieldCompensationEstimate}{\hat{\bfh}_o^{dc}}
\newcommand{\convergenceVectorFieldElement}[1]{h_#1}
\newcommand{\convergenceVectorFieldEstimateElement}[1]{\hat{h}_#1}
\newcommand{\convergenceVectorFieldEstimateDerivativeElement}[1]{\dot{\hat{h}}_#1}
\newcommand{\convergenceVectorFieldEstimateDerivativeEstimateElement}[1]{\hat{\dot{\hat{h}}}_#1}
\newcommand{\longitudinalConvergenceVectorFieldDerivative}{\dot{\bfh}_p}
\newcommand{\longitudinalConvergenceVectorFieldDerivativeEstimate}{\hat{\dot{\bfh}}_p}
\newcommand{\longitudinalConvergenceVectorFieldDerivativeError}{\tilde{\dot{\bfh}}_p}
\newcommand{\longitudinalConvergenceVectorFieldEstimateDerivativeError}{\tilde{\dot{\hat{\bfh}}}_p}
\newcommand{\longitudinalModifiedConvergenceVectorFieldEstimateDerivativeError}{\tilde{\dot{\hat{\bfh}}}_p^*}
\newcommand{\longitudinalConvergenceVectorFieldEstimateDerivativeErrorAuxiliaryFunction}{f_h}
\newcommand{\convergenceVectorFieldElementDerivative}[1]{\dot{h}_#1}
\newcommand{\convergenceVectorFieldElementDerivativeEstimate}[1]{\hat{\dot{h}}_#1}
\newcommand{\convergenceVectorFieldElementDerivativeError}[1]{\tilde{\dot{h}}_#1}
\newcommand{\convergenceVectorFieldEstimateDerivativeErrorElement}[1]{\tilde{\dot{\hat{h}}}_#1}
\newcommand{\modifiedConvergenceVectorFieldEstimateDerivativeErrorElement}[1]{\tilde{\dot{\hat{h}}}_#1^*}
\newcommand{\modifiedConvergenceVectorField}{\bfh^*}
\newcommand{\longitudinalModifiedConvergenceVectorField}{\bfh_p^*}
\newcommand{\angularModifiedConvergenceVectorField}{\bfh_o^*}
\newcommand{\modifiedConvergenceVectorFieldElement}[1]{h^*_#1}
\newcommand{\modifiedConvergenceVectorFieldEstimate}{\hat{\bfh}^*}
\newcommand{\modifiedConvergenceVectorFieldEstimateElement}[1]{\hat{h}^*_#1}
\newcommand{\modifiedConvergenceVectorFieldEstimateDerivativeElement}[1]{\dot{\hat{h}}^*_#1}
\newcommand{\modifiedConvergenceVectorFieldEstimateDerivative}{\dot{\hat{\bfh}}^*}
\newcommand{\modifiedConvergenceVectorFieldEstimateDerivativeEstimate}{\hat{\dot{\hat{\bfh}}}^*}
\newcommand{\longitudinalModifiedConvergenceVectorFieldEstimate}{\hat{\bfh}_p^*}
\newcommand{\longitudinalModifiedConvergenceVectorFieldEstimateDerivative}{\dot{\hat{\bfh}}_p^*}
\newcommand{\longitudinalModifiedConvergenceVectorFieldEstimateDerivativeEstimate}{\hat{\dot{\hat{\bfh}}}_p^*}
\newcommand{\angularModifiedConvergenceVectorFieldEstimate}{\hat{\bfh}_o^*}
\newcommand{\modifiedConvergenceVectorFieldElementEstimate}[1]{\hat{h}^*_#1}
\newcommand{\modifiedConvergenceVectorFieldElementEstimateDerivativeEstimate}[1]{\hat{\dot{\hat{h}}}^*_#1}


\newcommand{\desiredYawAngle}{\yaw_d}
\newcommand{\desiredPitchAngle}{\pitch_d}
\newcommand{\desiredRollAngle}{\phi_d}
\newcommand{\auxiliaryPitchAngle}{\pitch_a}
\newcommand{\auxiliaryPitchAngleDerivative}{\dot{\pitch}_a}
\newcommand{\auxiliaryPitchAngleDerivativeEstimate}{\hat{\dot{\pitch}}_a}
\newcommand{\auxiliaryPitchAngleDerivativeError}{\tilde{\dot{\pitch}}_a}
\newcommand{\auxiliaryYawAngle}{\yaw_a}
\newcommand{\auxiliaryBoundedYawAngle}{\yaw_a^*}
\newcommand{\auxiliaryYawAngleDerivative}{\dot{\yaw}_a}
\newcommand{\auxiliaryYawAngleDerivativeEstimate}{\hat{\dot{\yaw}}_a}
\newcommand{\auxiliaryYawAngleDerivativeError}{\tilde{\dot{\yaw}}_a}

\newcommand{\auxiliaryDesiredPitchDifference}{\varepsilon_\pitch}
\newcommand{\auxiliaryDesiredYawDifference}{\varepsilon_\yaw}
\newcommand{\auxiliaryDesiredYawDifferenceFunction}{f_{\varepsilon\yaw}}
\newcommand{\auxiliaryDesiredPitchDifferenceFunction}{f_{\varepsilon\pitch}}

\newcommand{\auxiliaryDesiredYawDifferenceUltimateBound}{r_{\varepsilon\yaw}^\infty}
\newcommand{\auxiliaryDesiredPitchDifferenceUltimateBound}{r_{\varepsilon\pitch}^\infty}

\newcommand{\observationErrorSubsystem}{\bfSigma_1}
\newcommand{\velocityErrorSubsystem}{\bfSigma_2}
\newcommand{\auxiliaryOrientationErrorSubsystem}{\bfSigma_3}
\newcommand{\positionErrorSubsystem}{\bfSigma_4}

\newcommand{\observationErrorStateMatrix}{\bfH_\chi}
\newcommand{\observationErrorDomain}{\mathcal{D}_X}
\newcommand{\observationErrorDisturbanceDomain}{\mathcal{D}_D}
\newcommand{\observationErrorPrimaryBound}{r_X}
\newcommand{\observationErrorTerminalBound}{r_\chi}
\newcommand{\observationErrorDisturbancePrimaryBound}{r_{\dot{D}}}
\newcommand{\observationErrorDisturbanceTerminalBound}{r_{\dot{d}}}
\newcommand{\totalDisturbanceBound}{r_d}
\newcommand{\observationErrorLyapunovFunction}{V_\zeta}
\newcommand{\observationErrorLyapunovFunctionDerivative}{\dot{V}_\zeta}
\newcommand{\observationErrorAlphaOne}{\alpha_{1\zeta}}
\newcommand{\observationErrorAlphaTwo}{\alpha_{2\zeta}}
\newcommand{\observationErrorMajorizationCoefficient}{\nu_\zeta}
\newcommand{\observationErrorChiFunction}{\chi_\zeta}
\newcommand{\observationErrorBetaFunction}{\beta_\zeta}
\newcommand{\observationErrorGammaFunction}{\gamma_\zeta}
\newcommand{\observationErrorUltimateBound}{r_{\chi}^\infty}
\newcommand{\velocityErrorUltimateBound}{r_{\epsilon}^\infty}

\newcommand{\velocityErrorStateMatrix}{\bfH_\epsilon}
\newcommand{\velocityErrorPerturbationVector}{\bfdelta_\epsilon}
\newcommand{\velocityErrorDomain}{\mathcal{D}_E}
\newcommand{\velocityErrorDisturbanceDomain}{\mathcal{D}_{\delta E}}
\newcommand{\velocityErrorModifiedDisturbanceDomain}{\mathcal{D}_{D}^*}
\newcommand{\velocityErrorPrimaryBound}{r_E}
\newcommand{\velocityErrorModifiedDisturbancePrimaryBound}{r_{D^*}}
\newcommand{\velocityErrorModifiedDisturbanceTerminalBound}{r_{d^*}}
\newcommand{\velocityErrorTerminalBound}{r_\epsilon}
\newcommand{\velocityErrorDisturbancePrimaryBound}{r_{\delta E}}
\newcommand{\velocityErrorDisturbanceTerminalBound}{r_{\delta\epsilon}}
\newcommand{\velocityErrorLyapunovFunction}{V_\epsilon}
\newcommand{\velocityErrorLyapunovFunctionDerivative}{\dot{V}_\epsilon}
\newcommand{\velocityErrorAlphaOne}{\alpha_{1\epsilon}}
\newcommand{\velocityErrorAlphaTwo}{\alpha_{2\epsilon}}
\newcommand{\velocityErrorMajorizationCoefficient}{\nu_\epsilon}
\newcommand{\velocityErrorChiFunction}{\chi_\epsilon}
\newcommand{\velocityErrorChiFunctionOne}{\chi_{1\epsilon}}
\newcommand{\velocityErrorChiFunctionTwo}{\chi_{2\epsilon}}
\newcommand{\velocityErrorBetaFunction}{\beta_\epsilon}
\newcommand{\velocityErrorGammaFunction}{\gamma_\epsilon}
\newcommand{\velocityErrorGammaFunctionOne}{\gamma_{1\epsilon}}
\newcommand{\velocityErrorGammaFunctionTwo}{\gamma_{2\epsilon}}

\newcommand{\auxiliaryOrientationPerturbationMatrix}{\bfH_\epsilon}
\newcommand{\auxiliaryOrientationPerturbationVector}{\bfdelta_{eA}}
\newcommand{\auxiliaryOrientationDomain}{\mathcal{D}_{eA}}
\newcommand{\auxiliaryOrientationDisturbanceDomain}{\mathcal{D}_{\delta e A}}
\newcommand{\auxiliaryOrientationObservationErrorDomain}{\auxiliaryOrientationDisturbanceDomain}
\newcommand{\auxiliaryOrientationVelocityErrorDomain}{\mathcal{D}_{\epsilon A}}
\newcommand{\auxiliaryOrientationFFDomain}{\mathcal{D}_{\eta A}}
\newcommand{\auxiliaryOrientationPrimaryBound}{r_{eA}}
\newcommand{\auxiliaryOrientationTerminalBound}{r_{ea}}
\newcommand{\auxiliaryOrientationUltimateBound}{r_{ea}^\infty}
\newcommand{\auxiliaryOrientationDisturbancePrimaryBound}{r_{\delta e A}}
\newcommand{\auxiliaryOrientationDisturbanceTerminalBound}{r_{\delta ea}}
\newcommand{\auxiliaryOrientationObservationPrimaryBound}{r_{\tilde{E}}}
\newcommand{\auxiliaryOrientationVelocityErrorPrimaryBound}{r_{E}}
\newcommand{\auxiliaryOrientationFFPrimaryBound}{r_{\eta A}}
\newcommand{\auxiliaryOrientationObservationTerminalBound}{r_{\tilde{e}}}
\newcommand{\auxiliaryOrientationVelocityErrorTerminalBound}{r_{e}}
\newcommand{\auxiliaryOrientationFFTerminalBound}{r_{\eta a}}
\newcommand{\auxiliaryOrientationLyapunovFunction}{V_a}
\newcommand{\auxiliaryOrientationLyapunovFunctionDerivative}{\dot{V}_a}
\newcommand{\auxiliaryOrientationAlphaOne}{\alpha_{1a}}
\newcommand{\auxiliaryOrientationAlphaTwo}{\alpha_{2a}}
\newcommand{\auxiliaryOrientationMajorizationCoefficient}{\nu_a}
\newcommand{\auxiliaryOrientationChiFunction}{\chi_a}
\newcommand{\auxiliaryOrientationChiFunctionOne}{\chi_{1a}}
\newcommand{\auxiliaryOrientationChiFunctionTwo}{\chi_{2a}}
\newcommand{\auxiliaryOrientationChiFunctionThree}{\chi_{3a}}
\newcommand{\auxiliaryOrientationBetaFunction}{\beta_a}
\newcommand{\auxiliaryOrientationGammaFunction}{\gamma_a}
\newcommand{\auxiliaryOrientationGammaFunctionOne}{\gamma_{1a}}
\newcommand{\auxiliaryOrientationGammaFunctionTwo}{\gamma_{2a}}
\newcommand{\auxiliaryOrientationGammaFunctionThree}{\gamma_{3a}}

\newcommand{\positionErrorStateMatrix}{\bfH_p}
\newcommand{\positionErrorPerturbationVector}{\bfdelta_{p}}
\newcommand{\positionErrorPerturbationObservationErrorVector}{\bfdelta_{ep}}
\newcommand{\positionErrorPerturbationVelocityErrorVector}{\bfdelta_{\epsilon p}}
\newcommand{\positionErrorDomain}{\mathcal{D}_{eP}}
\newcommand{\positionErrorDisturbanceDomain}{\mathcal{D}_{\delta p}}
\newcommand{\positionErrorPerturbationObservationErrorDomain}{\mathcal{D}_{\delta e p}}
\newcommand{\positionErrorPerturbationVelocityErrorDomain}{\mathcal{D}_{\delta \epsilon p}}
\newcommand{\positionErrorPrimaryBound}{r_{eP}}
\newcommand{\positionErrorTerminalBound}{r_{ep}}
\newcommand{\positionErrorUltimateBound}{\positionalControlError}
\newcommand{\positionErrorDisturbancePrimaryBound}{r_{\delta P}}
\newcommand{\positionErrorDisturbanceTerminalBound}{r_{\delta p}}
\newcommand{\positionErrorPerturbationObservationErrorPrimaryBound}{r_{\delta e P}}
\newcommand{\positionErrorPerturbationObservationErrorTerminalBound}{r_{\delta e p}}
\newcommand{\positionErrorPerturbationVelocityErrorPrimaryBound}{r_{\delta \epsilon P}}
\newcommand{\positionErrorPerturbationVelocityErrorTerminalBound}{r_{\delta \epsilon p}}
\newcommand{\positionErrorPerturbationOrientationVectorTerminalBound}{r_{a p}}
\newcommand{\positionErrorLyapunovFunction}{V_p}
\newcommand{\positionErrorLyapunovFunctionDerivative}{\dot{V}_p}
\newcommand{\positionErrorAlphaOne}{\alpha_{1p}}
\newcommand{\positionErrorAlphaTwo}{\alpha_{2p}}
\newcommand{\positionErrorMajorizationCoefficient}{\nu_p}
\newcommand{\positionErrorChiFunction}{\chi_p}
\newcommand{\positionErrorChiFunctionOne}{\chi_{1p}}
\newcommand{\positionErrorChiFunctionTwo}{\chi_{2p}}
\newcommand{\positionErrorChiFunctionThree}{\chi_{3p}}
\newcommand{\positionErrorBetaFunction}{\beta_p}
\newcommand{\positionErrorGammaFunction}{\gamma_p}
\newcommand{\positionErrorGammaFunctionOne}{\gamma_{1p}}
\newcommand{\positionErrorGammaFunctionTwo}{\gamma_{2p}}
\newcommand{\positionErrorGammaFunctionThree}{\gamma_{3p}}
\newcommand{\positionErrorResidueVector}{\bfr}
\newcommand{\positionErrorResidueAngle}{\alpha}
\newcommand{\positionErrorResidueAngleStateMatrix}{\bfH_\alpha}
\newcommand{\positionErrorDirectionVector}{\bfgamma}

\newcommand{\controlError}{\varepsilon_\eta}
\newcommand{\positionalControlError}{\varepsilon_p}
\newcommand{\transformedObservationErrorStateMatrix}{{\bfH_\zeta}}
\newcommand{\observationErrorMap}{\phi_\zeta}
\newcommand{\transformedObservationError}{\pmb{\zeta}}
\newcommand{\transformedObservationErrorDerivative}{\dot{\pmb{\zeta}}}
\newcommand{\observationErrorTransformationMatrix}{\pmb{L}_\zeta}
\newcommand{\transformedObservationErrorGainMatrix}{\pmb{L}_\omega}
\newcommand{\transformedObservationErrorGainMatrixSingle}{\pmb{W}_\omega}
\newcommand{\transformedObservationErrorDomain}{\mathcal{D}_\zeta}
\newcommand{\transformedObservationErrorPrimaryBound}{r_Z}
\newcommand{\transformedObservationErrorTerminalBound}{r_\zeta}
\newcommand{\transformedObservationErrorLyapunovEquationSolution}{\pmb{P}_\zeta}
\newcommand{\observationErrorIdentityMatrix}{\bfI_\chi}


\blfootnote{This work was supported by the statutory grant No. 0211/SBAD/0911.}



\nomenclature{\small $\bfeta$}{\small configuration of the vehicle}
\nomenclature{\small $\bfeta_p,\bfeta_o$}{\small position and orientation of the vehicle}

\nomenclature{\small $\bfnu$}{\small pseudovelocities of the vehicle}
\nomenclature{\small $\bfnu_p,\bfnu_o$}{\small longitudinal and angular pseudovelocities of the vehicle}

\nomenclature{\small $\bfJ$}{\small Jacobian matrix of the velocity transformation}
\nomenclature{\small $\bfR$}{\small rotation matrix between frames \{G\} and \{B\}}
\nomenclature{\small $\bfT$}{\small angular-velocity transformation matrix between frames \{G\} and \{B\}}

\nomenclature{\small $\bfM_{\eta},\bfM$}{\small inertia matrix, respectively, in frame \{G\} and \{B\}}
\nomenclature{\small $\bfmu_{\eta},\bfmu$}{\small combined dynamic phenomena, respectively, in frame \{G\} and \{B\}}
\nomenclature{\small $\bftau_{\eta}^*,\bftau^*$}{\small external perturbation, respectively, in frame \{G\} and \{B\}}
\nomenclature{\small $\bftau_{\eta}, \bftau$}{\small generalized input forces, respectively, in frame \{G\} and \{B\}}

\nomenclature{\small $\pathPointSet$}{\small set of positions included in reference path}
\nomenclature{\small $\desiredRollAngle,\desiredPitchAngle,\desiredYawAngle$}{\small desired roll, pitch, and yaw angles}
\nomenclature{\small $\xi_d$}{\small bi-valued factor determining desired motion strategy}
\nomenclature{\small $\bfnu_d$}{\small desired pseudovelocities}

\nomenclature{\small $\bfe$}{\small tracking error}
\nomenclature{\small $\bfe_p,\bfe_o$}{\small positional and angular tracking error}
\nomenclature{\small $\bfe_{2\pi}$}{\small tracking error with component $e_\psi$ limited to range $[0,2\pi)$}

\nomenclature{\small $\bfnu_c$}{\small commanded pseudovelocities}
\nomenclature{\small $\bfnu_{pc},\bfnu_{oc}$}{\small longitudinal and angular commanded pseudovelocities}
\nomenclature{\small $\bfG$}{\small simplified Jacobian matrix of the velocity transformation}
\nomenclature{\small $\bfh$}{\small convergence vector field}
\nomenclature{\small $\bfh_p,\bfh_o$}{\small longitudinal and angular part of convergence vector field}
\nomenclature{\small $\bfh^*$}{\small modified convergence vector field}
\nomenclature{\small $\bfh_p^*,\bfh_o^*$}{\small longitudinal and angular part of modified convergence vector field}
\nomenclature{\small $\theta_a, \psi_a$}{\small auxiliary pitch and yaw angles}
\nomenclature{\small $k_p,k_\theta,k_\psi$}{\small gains of the VFO controller}
\nomenclature{\small $s_i$}{\small level-surface value, $i\in\{1,2\}$}
\nomenclature{\small $\bfvartheta_\bot, \bfvartheta_i$}{\small tangential and normal unit vectors of the particular level-surface, $i\in\{1,2\}$}

\nomenclature{\small $\bfGamma_\eta, \bfGamma$}{\small actuation matrix, respectively, in $\globalCoordinateSystem$ and $\localCoordinateSystem$}

\nomenclature{\small $\hat{\bfB}$}{\small prescribed rough estimate of matrix $\bfM_\eta^{-1}$}
\nomenclature{\small $\bfd^*$}{\small modified total disturbance vector}
\nomenclature{\small $\bfepsilon$}{\small commanded-velocity tracking error in frame \{G\}}
\nomenclature{\small $\bfepsilon_p,\bfepsilon_o$}{\small longitudinal and angular commanded-velocity tracking errors in \{G\}}
\nomenclature{\small $\epsilon_i$}{\small $i$-th component of vector $\bfepsilon$ ($i\in\{1,...,6\}$) or vector $\bfepsilon_{\nu}$ ($i\in\{u,...,r\}$)}
\nomenclature{\small $\bfd$}{\small total disturbance containing the feedforward term}
\nomenclature{\small $\bfK_\eta, \bfK$}{\small gain matrix of the ADR controller, respectively, in $\globalCoordinateSystem$ and $\localCoordinateSystem$}

\nomenclature{\small $\bfx_i$}{\small extended state of the $i$th degree of freedom, $i\in\{1,...,6\}$}
\nomenclature{\small $\bfeta_c$}{\small commanded configuration}
\nomenclature{\small $\bfl_i$}{\small gain of ESO for the $i$th degree of freedom, $i\in\{1,...,6\}$}
\nomenclature{\small $\omega_{oi}$}{\small bandwidth of ESO for the $i$th degree of freedom, $i\in\{1,...,6\}$}

\nomenclature{\small $\omega_o$}{\small minimal value of $\omega_{oi}$, $i\in\{1,..,6\}$}
\nomenclature{\small $\bfchi$}{\small accumulated vector of the extended states $\bfx_i$ for all $i\in\{1,...,6\}$}
\nomenclature{\small $\bfe_a$}{\small auxiliary orientation error}
\nomenclature{\small $\bfK_a$}{\small gain matrix of the angular part of VFO controller}
\nomenclature{\small $\bfDelta_\eta,\bfDelta$}{matrix of linear damping coefficients, respectively, in $\globalCoordinateSystem$ and $\localCoordinateSystem$}

\nomenclature{\small $\nu_i$}{\small majorization constant, $i\in\{\chi, \epsilon, a, p\}$}
\nomenclature{\small $\longitudinalCompensationCoefficient, \angularCompensationCoefficient$}{\small drift compensation coefficients of the VFO controller}
\nomenclature{\small $\psi_a^*$}{\small auxiliary yaw angle limited to range $[0,2\pi)$}
\nomenclature{\small $u_d$}{\small velocity profile along reference path}


\nomenclature{\small $\bfepsilon_\nu$}{\small commanded-velocity tracking error in frame \{B\}}
\nomenclature{\small $\bfepsilon_1,\bfepsilon_2$}{\small longitudinal and angular commanded-velocity tracking errors in \{B\}}

\nomenclature{\small $\bar{(\cdot)}$}{\small dimensionally-reduced vector $(\cdot)$ (with some components removed)}
\nomenclature{\small $\hat{(\cdot)}$}{\small estimate of variable $(\cdot)$}
\nomenclature{\small $\tilde{(\cdot)}$}{\small observation/estimation error of variable $(\cdot)$}

\setlength{\nomitemsep}{-0.1cm}
\begin{framed}
  \color{black}
	\printnomenclature
  \color{black}
\end{framed}


\section{Introduction}
The interest in automatic control of Unmanned Aerial Vehicles (UAVs) and Autonomous Underwater Vehicles (AUVs) has been gradually growing in recent years, and resulted in a large amount of the designed control structures for various types of objects, such as multicopters \cite{SHAO19,RAFF10,NGUY18,HONG16}, airships \cite{ZHU13,ZHEN19,SUN15,AZIN08}, fixed-wing planes \cite{KANG09} or underwater exploring robots \cite{CAHA14,BECH17,PENG17}. Looking throughout the control architectures proposed in the literature, one can observe that the controllers intended for mobile robots are most frequently focused on solving one (or more) of three classical motion tasks, i.e., path-following \cite{AZIN14,YAO18,REIS19,GUO19,ZHEN19,XU19}, trajectory tracking \cite{BECH17,ABDE10} or set-point stabilization \cite{AZIN08,PAIV06}. All of the aforementioned control tasks have been solved utilizing various control methods, like backstepping \cite{AZIN08,AZIN14}, model predictive control \cite{KANG09}, sliding mode control \cite{YANG15} or active disturbance rejection control \cite{ZHU13,HONG16}, each having specific properties suited for a particular application.

The movement of spatial vehicles can be performed with three different motion strategies. The least common omnidirectional motion \cite{BRES16,MART16} can be applied only to fully-actuated vehicles that can generate substantial forces in every Degree of Freedom (DoF). Actuation systems implemented on the aerial/underwater vehicles are usually capable of executing large control signals only in some distinguished directions, making it necessary to use one of the other available motion strategies. Alternative approaches to the omnidirectional movement are, for example, the unicycle-like and the torpedo-like motion strategies, both relying on a motion realization along with some privileged directions, aligned with a thrust generated by the main actuators.
In the unicycle-like motion, a vehicle movement is decomposed into the planar, and fully decoupled vertical motions \cite{ZHEN19,REIS19,ZHU13}, while in the torpedo-like approach a control object moves directly in the 3d space  \cite{BECH17,MICH19,GUO19,SGOR19}. A specific actuation strategy utilized for the vehicles that are usually moving in a torpedo-like manner allows a precise realization of the given control task with fewer control signals, comparing to the ones using the unicycle-like strategy.

 In this paper, we focus on the development of the control structure guaranteeing a realization of the path following motion task in the torpedo-like strategy by a spatially moving vehicle. In this research, the path is defined as a cross-section of two surfaces described by non-parametrized equations \cite{LAKO17,SGOR19,YAO18,NGUY18}. On the contrary to the classical, parametrized way of defining the reference path \cite{REIS19,GUO19,ZHEN19}, the utilized approach does not impose restrictive constraints on the initial conditions and does not demand a calculation of the shortest distance between a vehicle body and the path. Analytical calculation of the aforementioned path-to-vehicle distance, although straightforward in the case of simple examples (e.g. linear or circular paths), is in general non-trivial for more complex paths with a varying curvature and may be hard to obtain numerically in each control sample with a satisfying control-loop frequency.

 The cascaded control structure presented in this article consists of the inner dynamic-level controller responsible for tracking of the commanded velocities, calculated by the outer kinematic-level controller in a way to attract the vehicle towards the reference path with a prescribed orientation. The dynamic-level control loop is designed according to Active Disturbance Rejection (ADR) method \cite{HAN09,GAO06,LAKO17ADRC, BAI17} introducing a feedforward control from a total disturbance. The effort made to compensate the total disturbance makes the presented control algorithm robust to external forces (caused by winds, currents, etc.) and to parametric uncertainties of a vehicle mathematical model. Due to the use of an error-domain Extended State Observer (ESO) \cite{MICH16,MADO19}, being a part of the ADR controller, the whole control structure has got output-feedback characteristics, which means that it only needs to measure the vehicle position and attitude to achieve expected control quality. The output-feedback property is practically desirable, because measuring all of the state vector elements, including longitudinal velocities, may be expensive (when measured by high-quality sensors) or computationally challenging when estimated upon the visual data.  The outer kinematic-level controller uses Vector Field Orientation (VFO) methodology, introduced initially for wheeled robots in \cite{MICH10}, and developed recently for the 3D vehicle motion \cite{MICH19,LAKO17}. The kinematic controller aims to calculate a vector of commanded velocities, which, while being correctly followed by a dynamic controller, guarantee an accurate following of the reference path.

 This paper is a substantial extension of the conference articles \cite{LAKO17} and \cite{LAKO18}, providing a more detailed description and a formal analysis of the VFO-ADR control structure satisfying the non-parametrized path-following in a torpedo-like motion strategy of the vehicle moving in a 3D space. Comparing to the results presented in \cite{LAKO17}, the VFO controller is extended by a transversal drift compensation component, causing a significant increase in a path-following control performance for the case of underactuated vehicles. \color{black} Work \cite{LAKO18} is a preliminary conference proposition of the VFO-ADR path following controller, which is extend in this article with the more detailed derivation, theoretical analysis, and more thorough simulation verification.

 It is worth emphasizing that, \color{black} unlike the most common cascade control systems designed for UAV/AUV-type objects that are based on the decoupling of longitudinal and angular subsystems, we propose to use the approach utilized in the control of nonholonomic vehicles that decouples system kinematics from system dynamics (see also \cite{WIIG18}). Many solutions considering the ADR-based control of spatial mobile vehicles, for example \cite{BAI17,ZHU13}, demand to measure the whole state vector. On the contrary to these methods, the ADR controller presented in this article uses the error-domain architecture of ESO and only needs the information about the system configuration. The use of the ADR method in the inner-loop controller guarantees also the robustness of the proposed control structure to the external disturbances and significant model uncertainties.  According to the Input-to-State Stability (ISS) procedure \cite{KHAL02,SONT96} conducted in the multi-input approach \cite{ISID17,PENG17}, application of the VFO-ADR control structure to the dynamics of the considered vehicle results in the bounded control errors and a possible reduction of positional errors to arbitrarily small magnitudes. The theoretical analysis is followed by the simulation verification performed in Matlab/Simulink environment.

 \textit{Notation.} For the sake of the notational conciseness, we will use the assignments $\sin \alpha \equiv \textrm{sin} \alpha$,
 $\cos \alpha \equiv \textrm{cos} \alpha$, $\tan \alpha \equiv \textrm{tan} \alpha$, and $\lsShort := \ls$. The nabla operator was treated as
 a column vector $\nabla = [\partial/\partial x \ \partial/\partial y \ \partial/\partial z]^\top$ and the set $\nonNegativeRealNumbers=\{x\in\realNumbers:x\geq0\}$. In order not to make the equations excessively long, we have ommited the signal parameters in longer formulas.

\color{black}


\section{Preliminaries}

\subsection{Mathematical model of a vehicle}
\label{subsec:model_mat}
\begin{figure}[htpb]
	\centering
	\includegraphics[width=0.3\textwidth]{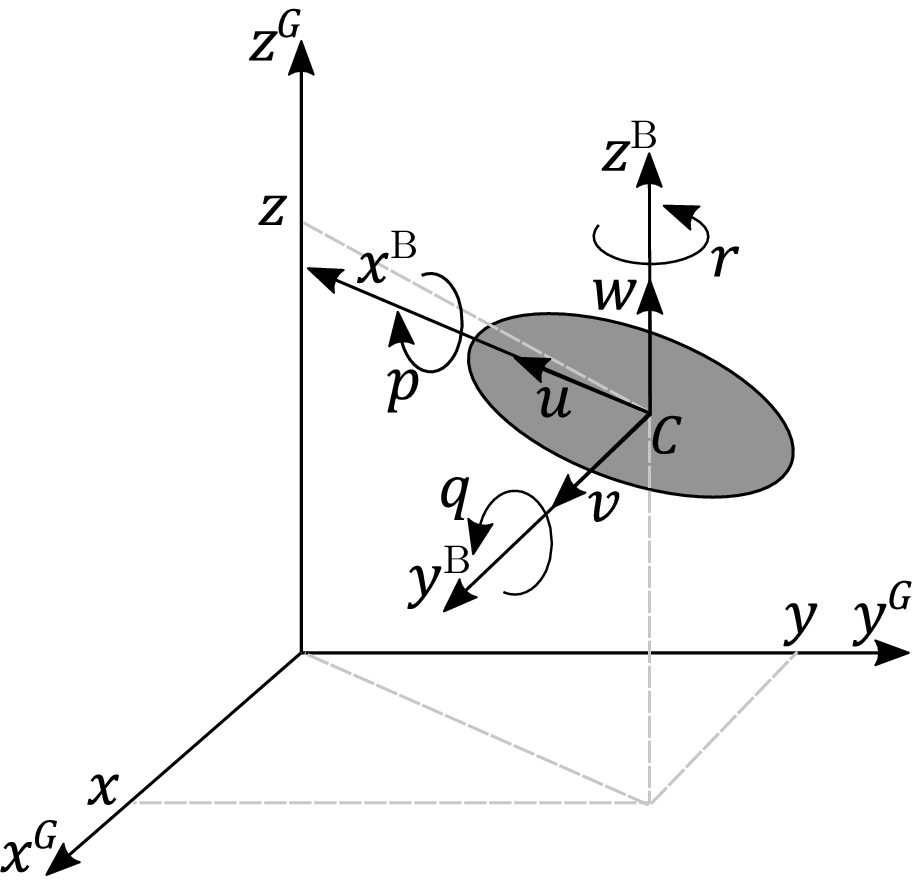}
	\caption{A rigid body in the 3D space.}
	\label{fig:RigidBody}
\end{figure}

To describe the position and attitude of a rigid-body vehicle in a 3D space, the global frame $\globalCoordinateSystem$ and the local (body-fixed) frame $\localCoordinateSystem$ need to be introduced. The origin of $\localCoordinateSystem$ is placed in the robot mass center $\massCenter$, while axis $\xLocalAxis$ is aligned with the direction of a main thrust provided by the onboard actuation system. The configuration vector is represented as
\begin{align}
	\label{eq:etadef}
		\configurationVector = \begin{bmatrix} \positionVector \\ \orientationVector \end{bmatrix} \triangleq \begin{bmatrix} \left[\xPosition \ \yPosition \ \zPosition\right]^\top \\  \left[\roll \ \pitch \ \yaw\right]^\top \end{bmatrix}\in\realNumbers^3\times[-\pi,\pi)\times\left(-\frac{\pi}{2},\frac{\pi}{2}\right)\times\realNumbers,
\end{align}
where $\positionVector\triangleq\left[\xPosition \ \yPosition \ \zPosition\right]^\top$ describes the position of the $\localCoordinateSystem$ origin in a global frame, while $\orientationVector\triangleq\left[\roll \ \pitch \ \yaw\right]^\top$ is a vector of RPY (Roll, Pitch, Yaw) Euler angles describing the vehicle attitude. The vector of pseudovelocities in the local frame
\begin{align}
	\label{eq:nudef}
	\localVelocityVector = \begin{bmatrix} \localLongitudinalVelocityVector \\ \localAngularVelocityVector \end{bmatrix} \triangleq \begin{bmatrix} \left[\xLocalVelocity \ \yLocalVelocity \ \zLocalVelocity\right]^\top \\ \left[\rollLocalVelocity \ \pitchLocalVelocity \ \yawLocalVelocity\right]^\top \end{bmatrix}\in\realNumbers^6
\end{align}
consists of the subvector of longitudinal velocities $\localLongitudinalVelocityVector\triangleq\left[\xLocalVelocity \ \yLocalVelocity \ \zLocalVelocity\right]^\top$ and the subvector of angular velocities $\localAngularVelocityVector\triangleq\left[\rollLocalVelocity \ \pitchLocalVelocity \ \yawLocalVelocity\right]^\top$. The graphical representation of a 3D rigid body, together with the axes of $\globalCoordinateSystem$ and $\localCoordinateSystem$, chosen elements of the configuration vector
\eqref{eq:etadef} and the vector of pseudovelocities \eqref{eq:nudef} is illustrated in Fig.  \ref{fig:RigidBody}.

Kinematics of the considered rigid vehicle, i.e., the velocity transformation between the local and global frames is expressed by
\begin{align}
	\label{eq:KinDef}
	\configurationVectorDerivative = \velocityTransformationMatrix(\orientationVector)\localVelocityVector,
\end{align}
where $\velocityTransformationMatrix(\orientationVector)=\blkdiag\{\rotationMatrix(\orientationVector), \ \angularVelocityTransformationMatrix(\orientationVector)\}$ is the Jacobian matrix dependent  on the rotation matrix $\rotationMatrix(\orientationVector)$ and the angular velocity transformation matrix $\angularVelocityTransformationMatrix(\orientationVector)$ that are, respectively, defined as
\begin{align}
	\rotationMatrix\left(\orientationVector\right) &= \begin{bmatrix} \cos\yaw\cos\pitch & \cos\yaw\sin\roll\sin\pitch-\cos\roll\sin\yaw & \cos\yaw\cos\roll\sin\pitch+\sin\roll\sin\yaw \\ \sin\yaw\cos\pitch & \sin\yaw\sin\roll\sin\pitch+\cos\roll\cos\yaw & \sin\yaw\cos\roll\sin\pitch-\sin\roll\cos\yaw \\ -\sin\pitch & \sin\roll\cos\pitch & \cos\roll\cos\pitch \end{bmatrix}
  \label{eq:Rdef}, \\
	\angularVelocityTransformationMatrix\left(\orientationVector\right) &= \begin{bmatrix} 1 & \sin\roll\tan\pitch & \cos\roll\tan\pitch \\ 0 & \cos\roll & -\sin\roll \\ 0 & {\sin\roll}/{\cos\pitch} & {\cos\roll}/{\cos\pitch} \end{bmatrix}.
  \label{eq:velocityTransformationMatrix}
\end{align}
\begin{cremark}
	\label{rem:1}
	To prevent the vehicle configuration running into the  singular points of transformation $\angularVelocityTransformationMatrix(\orientationVector)$, the pitch angle is restricted in the sense that
	$\forall_{t\geq0} \  \pitch(t)\in\left(-\frac{\pi}{2},\frac{\pi}{2}\right)$, see \eqref{eq:etadef}.
\end{cremark}

According to work \cite{FOSS99}, general equations of the 6 DoF rigid body dynamics are respectively expressed in the global and local frames as
\begin{align}
	\label{DynDefG}
	&\globalCoordinateSystem: \ \massMatrixGlobal(\configurationVector)\configurationVectorSecondDerivative+\dynamicPhenomenaVectorGlobal(\configurationVector,\configurationVectorDerivative)+\perturbationsVectorGlobal = \actuationMatrixGlobal(\configurationVector)\controlSignalGlobal, \\
	\label{DynDefB} &\localCoordinateSystem: \ \massMatrixLocal\localVelocityVectorDerivative + \dynamicPhenomenaVectorLocal(\configurationVector,\localVelocityVector) + \perturbationsVectorLocal = \actuationMatrixLocal\controlSignalLocal.
\end{align}
where $\massMatrixGlobal$, $\massMatrixLocal$ are the inertia matrices, $\dynamicPhenomenaVectorGlobal$, $\dynamicPhenomenaVectorLocal$ represent the vectors accumulating the gravitational, Coriolis, centripetal and restoring forces together with the influence of environmental damping terms, $\controlSignalGlobal$, $\controlSignalLocal$ are the control signal vectors (corresponding to the forces and torques along/around particular axes), while $\perturbationsVectorGlobal$, $\perturbationsVectorLocal$ refer to the combined external disturbances and the unmodeled dynamic phenomena.  Relations between the matrices and vectors occurring in the equations of robot dynamics \eqref{DynDefG}-\eqref{DynDefB} are as follows: $\massMatrixLocal = \velocityTransformationMatrix^\top\massMatrixGlobal\velocityTransformationMatrix$,  $\dynamicPhenomenaVectorLocal= \velocityTransformationMatrix^\top\left(\dynamicPhenomenaVectorGlobal+\massMatrixGlobal\velocityTransformationMatrixDerivative\localVelocityVector\right)$, $\perturbationsVectorLocal = \velocityTransformationMatrix^\top\perturbationsVectorGlobal$ and $\controlSignalLocal = \velocityTransformationMatrix^\top\controlSignalGlobal$. Input matrices
\begin{align}
    \label{eq:gamma}
    &\actuationMatrixLocal \triangleq \diag\{1, \ \actuationMatrixLocalElement{v}, \ \actuationMatrixLocalElement{w}, \ \actuationMatrixLocalElement{p}, \ 1, \ 1\}, \quad \ \actuationMatrixLocalElement{v},\actuationMatrixLocalElement{w},\color{black}\actuationMatrixLocalElement{p}\color{black}\in\{0,1\}
\end{align}
and $\actuationMatrixGlobal = \velocityTransformationMatrix^{-\top}\actuationMatrixLocal\velocityTransformationMatrix^\top$ determine the type of an actuation system mounted on the vehicle in such a manner, that for $\actuationMatrixLocalElement{i}=0$ and $i\in\{v,w,p\}$, the control signal in a particular axis cannot be generated, causing the non-actuation of the corresponding DoF. A selected form of matrix $\actuationMatrixLocal$ implies that the surge, pitch, and yaw axes are certainly actuated, and thus the privileged direction of motion should be described along with them.
Within this research, we assume the anisotropy of the robot actuation system (see \cite{MICH19}) even in the fully actuated case, implying that even if the forces along $\yLocalAxis$, $\zLocalAxis$, and around $\xLocalAxis$ axes can be generated, they only fulfill the compensation function due to the lower capabilities of the thrust-generating actuators in these directions. The aforementioned actuator distribution justifies the selection of a torpedo-like motion strategy described in the Introduction (see also \cite{MICH19,BECH17}).

The overall state-space equations of the considered rigid body vehicle (assuming the invertability of $\massMatrixLocal$, see \cite{FOSS99}) are represented by a set of differential equations
\begin{align}
	\label{FullKinDyn}
	\begin{cases}
		\configurationVectorDerivative = \velocityTransformationMatrix(\orientationVector)\localVelocityVector, \\
		\localVelocityVectorDerivative = -\massMatrixLocal^{-1}\left[\dynamicPhenomenaVectorLocal(\configurationVector,\localVelocityVector)+\perturbationsVectorLocal\right]+\massMatrixLocal^{-1}\actuationMatrixLocal\controlSignalLocal.
	\end{cases}
\end{align}
The dynamical system \eqref{FullKinDyn} will be further considered under following assumptions:
\begin{assumption}
	\label{ass:measurableConfiguration}
	Configuration vector $\configurationVector(t)$ is the only measurable signal that can be used to design a control law.
\end{assumption}
\begin{assumption}
	The vehicle dynamics included in \eqref{FullKinDyn} is structurally and parametrically uncertain.
\end{assumption}
\begin{assumption}
	\label{ass:limitedDisturbance}
	The elements of vector $\perturbationsVectorLocal(t)$ are bounded Lipschitz functions.
\end{assumption}
\begin{assumption}
	\label{ass:rollActuation}
	We analyze a subclass of systems described by \eqref{FullKinDyn}, restricted to the objects where the roll axis is actuated, implying that $\actuationMatrixLocalElement{p}=1$, see \eqref{eq:gamma}.
\end{assumption}

\subsection{Reference path definition}

According to the definition introduced in \cite{LAKO17} and \cite{SGOR19}, the positional reference path can be considered as a set of reference points
\begin{align}
    \label{eq:2.6}
    \color{black} \pathPointSet \triangleq \{\positionVector \in \admissiblePositionSet : \levelSurface{1}(\positionVector) = \levelSurface{2}(\positionVector) = 0 \}, \color{black}
\end{align}
where $\admissiblePositionSet$ is a domain of admissible robot positions.  Set $\pathPointSet$ is interpreted as a cross-section of the surfaces described with a pair of non-parametric equations \color{black} $\levelSurface{1}(\positionVector)=0$ and $\levelSurface{2}(\positionVector)=0$. \color{black} Besides the geometrical curve definition itself, we define the reference velocity along the path
\begin{align}
    \label{eq:us}
        \pathVelocityProfile \triangleq \pathVelocityDirection\pathVelocityMagnitude, \hspace{0.2cm} \pathVelocityMagnitude = const > 0, \hspace{0.2cm} \pathVelocityDirection \in \{-1,+1\},
\end{align}
where $\pathVelocityDirection$ determines the movement strategy ($\pathVelocityDirection= -1$ for backward and $\pathVelocityDirection= +1$ for forward movement). Due to a non-zero value of the constant $\pathVelocityMagnitude$, the velocity $\pathVelocityProfile$ impose a persistently exciting reference motion for a vehicle.
\begin{cremark}
	The reference velocity $\pathVelocityProfile$ introduced in \eqref{eq:us} is assumed to be constant for the sake of simplicity. In more general case, it can depend on the position $\positionVector$.
\end{cremark}

Consistently with \cite{LAKO17}, let us introduce some assumptions concerning surfaces $\levelSurface{1}(\positionVector)$ and $\levelSurface{2}(\positionVector)$, which will be necessary for the control design purposes.
\begin{assumption}
	\label{ass:3}
	For all admissible points $\positionVector \in \admissiblePositionSet$, a gradient of the $j$-th level-surface, i.e.,  $\levelSurfaceGradient{j}(\positionVector) \triangleq$ $ [\partial \levelSurface{j}/\partial \xPosition \ \partial \levelSurface{j}/\partial \yPosition \ \partial \levelSurface{j}/\partial \zPosition]^\top$ has a finite, and strictly positive norm:
	$\forall \positionVector \in \admissiblePositionSet, \ \levelSurfaceGradientLowerBound{j}<\|\levelSurfaceGradient{j}(\positionVector)\|<\levelSurfaceGradientUpperBound{j},$
	for $\levelSurfaceGradientLowerBound{j},\levelSurfaceGradientUpperBound{j}>0$ and  $j\in\{1,2\}$.
\end{assumption}
\begin{assumption}
	\label{ass:4}
	For all admissible points $\positionVector\in\admissiblePositionSet$, the derivatives of the $j$-th level-surface with respect to the arguments up to the order 3 are bounded, and the Hessian $\levelSurfaceHessianWithArgument{j}\in \realNumbers^{3\times3}$ has a finite norm, that is, $\forall\positionVector\in\admissiblePositionSet, \ \module{\levelSurfaceHessianWithArgument{j}} < \levelSurfaceHessianUpperBound{j},$ for $\levelSurfaceHessianUpperBound{j}>0$ and $j\in\{1,2\}$.
\end{assumption}
\begin{assumption}
	\label{ass:nonCollinearity}
	For all admissible positions $\positionVector \in \admissiblePositionSet$, the gradients of functions $\levelSurface{1}(\positionVector)$, $\levelSurface{2}(\positionVector)$ are not collinear:
	$\forall~\positionVector~\in~\admissiblePositionSet, \  \levelSurfaceGradient{1}(\positionVector) \neq k\levelSurfaceGradient{2}(\positionVector), \hspace{0.2cm} k \in \realNumbers.$
	%
\end{assumption}

To define the reference pitch and yaw angles along the path, we have to introduce unit vectors normal to the particular path-describing surfaces in the form
\begin{align}
    \label{eq:2.8}
    \pathNormalUnitVector{j}(\positionVector) = \begin{bmatrix} \pathNormalUnitVectorElement{j}{x}(\positionVector) \\ \pathNormalUnitVectorElement{j}{y}(\positionVector) \\ \pathNormalUnitVectorElement{j}{z}(\positionVector) \end{bmatrix} \triangleq -\frac{\levelSurfaceGradient{j}(\positionVector)}{\|\levelSurfaceGradient{j}(\positionVector)\|}, \ \textrm{for} \ j\in\{1,2\},
\end{align}
and a unit vector tangential to both surfaces represented by
\begin{align}
	\label{eq:2.8_2}
	\pathTangentialUnitVector(\positionVector) = \begin{bmatrix} \pathTangentialUnitVectorElement{x}(\positionVector) \\ \pathTangentialUnitVectorElement{y}(\positionVector) \\ \pathTangentialUnitVectorElement{z}(\positionVector) \end{bmatrix} \triangleq \pathMotionDirection \left[\pathNormalUnitVector{1}(\positionVector)\times \pathNormalUnitVector{2}(\positionVector)\right],
\end{align}
where $\pathMotionDirection\in\{-1,1\}$ describes the desired movement direction along the path.
In the case when the vehicle position $\positionVector\not\in\pathPointSet$, the values of $\levelSurface{1}(\positionVector)$, $\levelSurface{2}(\positionVector)$ are not zero, and $\pathTangentialUnitVector(\positionVector)$ is not necessarily aligned with the reference path. Surface $\levelSurface{j}(\positionVector)=k, k\in\realNumbers$ will henceforth be called a \textit{level-surface} on the level $k$.
\begin{cremark}
	\label{rem:5}
	To avoid the situation, when the reference motion goes through the singularity point described in the Remark \ref{rem:1}, the level-surface equations should be designed in a way to guarantee
	\begin{align}
		\label{eq:barvarthetabotlim}
		\pathReducedTangentialUnitVectorLowerBound<\|\pathReducedTangentialUnitVector(\color{black}\positionVector\color{black})\|, \quad \pathReducedTangentialUnitVectorLowerBound>0,
	\end{align}
	\color{black} for some positive constant $\pathReducedTangentialUnitVectorLowerBound$, \color{black} position $\positionVector\in\pathPointSet$\color{black}, and $\pathReducedTangentialUnitVector(\color{black}\positionVector\color{black})\triangleq[\pathTangentialUnitVectorElement{x} \ \pathTangentialUnitVectorElement{y}]^\top$ being a projection of $\pathTangentialUnitVector(\color{black}\positionVector\color{black})$ onto the $\{\xGlobalAxis\yGlobalAxis\}$ plane.
\end{cremark}

Due to the underactuation of the considered vehicle along $\yLocalAxis$, $\zLocalAxis$, and around $\xLocalAxis$ axes, see \eqref{eq:gamma}, and the fact that the main propulsion is acting in the surge direction, the reference attitude of $\xLocalAxis$ axis should be aligned with the reference path, implying the reference yaw, pitch, and roll angles to be defined, respectively, as
\begin{align}
    \label{eq:2.10}
    \desiredYawAngle(\color{black}\positionVector\color{black}) &\triangleq \atantwo(\pathVelocityDirection\pathTangentialUnitVectorElement{y}(\color{black}\positionVector\color{black}), \hspace{0.1cm} \pathVelocityDirection\pathTangentialUnitVectorElement{x}(\color{black}\positionVector\color{black})) \in [-\pi,\pi), \\
    \label{eq:2.11}
    \desiredPitchAngle(\color{black}\positionVector\color{black}) &\triangleq \atan(-\pathTangentialUnitVectorElement{z}(\color{black}\positionVector\color{black})/ \beta_1(\color{black}\positionVector\color{black})) \in\left(-\frac{\pi}{2},\frac{\pi}{2}\right), \\ 
		\desiredRollAngle(\color{black}\positionVector\color{black}) & \triangleq 0, \label{eq:2.123}
\end{align}
\color{black} when $\positionVector\in\pathPointSet$, and \color{black}
where
$\atantwo: \realNumbers\times\realNumbers\rightarrow[-\pi,\pi)$ is a four-quadrant inverse tangent function including the bi-valued variable $\pathVelocityDirection$ introduced in \eqref{eq:us}, while
\begin{align}
	\beta_1(\color{black}\positionVector\color{black})=\pathTangentialUnitVectorElement{x}(\color{black}\positionVector\color{black})\cos\desiredYawAngle(\color{black}\positionVector\color{black})+\pathTangentialUnitVectorElement{y}(\color{black}\positionVector\color{black})\sin\desiredYawAngle(\color{black}\positionVector\color{black}).
\end{align}

\begin{cremark}
	\label{rem:nonbanked}
	The motion strategy considered within this article is assumed to have a non-banked characteristic (see \cite{FOSS99}), implying the $\yLocalAxis$ axis to be parallel to the $\xGlobalAxis\yGlobalAxis$ plane, and resulting in the fixed value of the reference roll angle $\desiredRollAngle\equiv0$ introduced in \eqref{eq:2.123}.
\end{cremark}

\subsection{Formulation of control objectives}

Let the path-following errors be accumulated in the vector
\begin{align}
    \label{eq:2.12}
    \configurationErrorVector(\configurationVector) = \left[\begin{array}{c}
     \positionError(\positionVector) \\ \hdashline[2pt/2pt]  \orientationError(\configurationVector)
    \end{array}\right] \triangleq \left[\begin{array}{c}
    \levelSurface{1}(\positionVector) \\
    \levelSurface{2}(\positionVector) \\  \hdashline[2pt/2pt]
		\desiredRollAngle-\roll \\
		\desiredPitchAngle(\positionVector) - \pitch \\
    \desiredYawAngle(\positionVector) - \yaw
    \end{array}\right] \in  \configurationErrorDomain,
\end{align}
where $\configurationErrorDomain=\realNumbers^2 \times [-\pi,\pi)^2 \times\realNumbers$. The values of each level-surface, $\levelSurface{1}(\positionVector)$ or $\levelSurface{2}(\positionVector)$,
correspond to some non-Euclidean measure of the signed distance between the origin of $\localCoordinateSystem$ and the particular level-surface. The control objective of the path-following task is to find an output-feedback control law $\controlSignalLocal = \controlSignalLocal(\configurationVector)$ that when applied into \eqref{FullKinDyn}, guarantees \color{black} ultimate \color{black} boundedness of the error
\begin{align}
	\label{eq:e2pidef}
	\configurationModuloErrorVector(\configurationVector)\triangleq\begin{bmatrix}\positionError(\positionVector) \\ \orientationError(\configurationVector) \  \modulo \ 2\pi\end{bmatrix}
\end{align}
in the sense that
\begin{align}
	\label{eq:controlobjective}
	&\exists T \in [0,\infty) : \forall t \geq T \ \|\configurationModuloErrorVector(\configurationVector(t))\|\leq\controlError ,  \\
	\label{eq:controlobjective2}
	&\exists T \in [0,\infty) : \forall t \geq T \ \|\positionError(\positionVector(t))\|\leq\positionalControlError ,
\end{align}
for some upper bounds $\controlError, \positionalControlError > 0$, where $\positionalControlError$ can be made arbitrarily small.
The modulo $2\pi$ operation utilized in \eqref{eq:e2pidef} on angular error vector $\orientationError(\configurationVector)$ is introduced to remove the effects of unlimited domain of the yaw angle error $\orientationErrorElement{\yaw}\triangleq\desiredYawAngle - \yaw\in\realNumbers$ introduced in \eqref{eq:2.12}.


\section{The design of VFO-ADR controller}

The kinematic-level VFO controller presented in this article uses the values of velocity error estimates calculated by ESO. As a consequence, we will start with the presentation of the dynamic-level ADR controller explaining the procedure of estimating required signals, and continue with the description of the VFO path-following controller.

\subsection{Dynamic-level ADR controller}
\label{subsec:adrc}

The main task of the inner dynamic-level controller is to calculate the values of the generalized forces applied to the vehicle body, aggregated within the control signal vector $\controlSignalLocal$ introduced in the equation \eqref{DynDefB}. Obtained values of the control signals should guarantee tracking of the commanded velocities
\begin{align}
\label{etacfromnuc}
	\commandedConfigurationVectorDerivative(t) = \begin{bmatrix} \commandedPositionVectorDerivative(t) \\ \commandedOrientationVectorDerivative(t) \end{bmatrix} =  \velocityTransformationMatrix(\orientationVector(t))\commandedVelocityVectorLocal(t),
\end{align}
with sufficient accuracy. The vector of commanded velocities $\commandedConfigurationVectorDerivative(t)$ expressed in frame $\globalCoordinateSystem$ is calculated using the Jacobian matrix introduced in \eqref{eq:KinDef} and  the commanded pseudovelocity vector
\begin{align}
\label{nucdef}
	\commandedVelocityVectorLocal = \begin{bmatrix} \longitudinalCommandedVelocityVectorLocal \\ \angularCommandedVelocityVectorLocal \end{bmatrix} \triangleq \begin{bmatrix} \left[\commandedXLocalVelocity \ \commandedYLocalVelocity \ \commandedZLocalVelocity\right]^\top \\ \left[\commandedRollLocalVelocity \ \commandedPitchLocalVelocity \ \commandedYawLocalVelocity\right]^\top \end{bmatrix}\in\realNumbers^6
\end{align}
expressed in the body-fixed frame.
Vector $\commandedVelocityVectorLocal$ is computed online by the VFO kinematic (outer-loop) controller described in Section \ref{subsec:vfo}.

The velocity tracking error in the global coordinate system is defined as
\begin{align}
	\label{epsilondef}
	 \velocityErrorVector(t) \eqdef \commandedConfigurationVectorDerivative(t)-\configurationVectorDerivative(t) = \begin{bmatrix}\longitudinalVelocityErrorVector(t) \\ \angularVelocityErrorVector(t)\end{bmatrix}= \begin{bmatrix} \left[\velocityErrorVectorElement{1}(t) \ \velocityErrorVectorElement{2}(t) \ \velocityErrorVectorElement{3}(t)\right]^\top \\ \left[\velocityErrorVectorElement{4}(t) \ \velocityErrorVectorElement{5}(t) \ \velocityErrorVectorElement{6}(t)\right]^\top \end{bmatrix} ,
\end{align}
while its dynamics can be derived upon \eqref{DynDefG} and \eqref{DynDefB} in the form
\begin{align}
	\velocityErrorVectorDerivative = \commandedConfigurationVectorSecondDerivative - \configurationVectorSecondDerivative &= \commandedConfigurationVectorSecondDerivative +\massMatrixGlobal^{-1}\left[\dynamicPhenomenaVectorGlobal+\perturbationsVectorGlobal\right] \nonumber \\
	&-\massMatrixGlobal^{-1}\actuationMatrixGlobal\controlSignalGlobal+\globalInputMatrixEstimate\actuationMatrixGlobal\controlSignalGlobal - \globalInputMatrixEstimate\actuationMatrixGlobal\controlSignalGlobal \nonumber \\
	&=\totalDisturbanceVector(\commandedConfigurationVectorSecondDerivative,\configurationVector,\configurationVectorDerivative,\controlSignalGlobal,\perturbationsVectorGlobal)-\globalInputMatrixEstimate\actuationMatrixGlobal\controlSignalGlobal \nonumber \\
	&= \totalDisturbanceVector(\commandedConfigurationVectorSecondDerivative,\configurationVector,\configurationVectorDerivative,\controlSignalGlobal,\perturbationsVectorGlobal)-\velocityTransformationMatrix\localInputMatrixEstimate\actuationMatrixLocal\velocityTransformationMatrix^\top\controlSignalGlobal.
	\label{eq:dotepsilon}
\end{align}
Referring to the commonly used nomenclature utilized in the ADR-focused research (e.g. \cite{HAN09,LAKO17ADRC}), vector $\totalDisturbanceVector(\cdot)\triangleq[\totalDisturbanceVectorElement{1}(\cdot) \ ... \ \totalDisturbanceVectorElement{6}(\cdot)]^\top$ will be treated as the so-called \textit{total disturbance} of system (\ref{eq:dotepsilon}), while $\globalInputMatrixEstimate\in\realNumbers^{6\times6}$, and  $\localInputMatrixEstimate\triangleq\diag\{\localInputMatrixEstimateDiagonalElement{1},...,\localInputMatrixEstimateDiagonalElement{6}\}\in\realNumbers^{6\times6}$ are, respectively, the rough approximations of the $\massMatrixGlobal^{-1}$ and $\massMatrixLocal^{-1}$ matrices.  The mutual relation between the the inverse inertia matrices in global, and body-fixed coordinate frames has the form $\massMatrixGlobal^{-1}=\velocityTransformationMatrix\massMatrixLocal^{-1}\velocityTransformationMatrix^\top$, thus, the corresponding relation between their estimates can be written down as $\globalInputMatrixEstimate=\velocityTransformationMatrix\localInputMatrixEstimate\velocityTransformationMatrix^\top$.

%
A specific representation of input signals influencing the dynamical system represented by \eqref{DynDefG} and \eqref{DynDefB} results in the possibility of designing the controls $\controlSignalLocal, \ \controlSignalGlobal$ in the same way for the underactuated vehicle as for the fully-actuated one. Particular elements of the control vector, associated with the non-actuated axes, will be later cut out by matrices $\actuationMatrixLocal, \ \actuationMatrixGlobal$ implying the applied control vectors to be in the form $\actuationMatrixLocal\controlSignalLocal, \ \actuationMatrixGlobal\controlSignalGlobal$ for $\localCoordinateSystem$ and $\globalCoordinateSystem$ frames, respectively.
\begin{cremark}
   A lack of control signals in the non-actuated axes results in the presence of a transversal drift affecting the system dynamic behavior and lowering the path-following precision. This phenomenon can be suppressed with a properly designed kinematic-level controller compensating the transversal-drift.
\end{cremark}
According to the velocity error dynamics derived in \eqref{eq:dotepsilon}, we propose the following nominal control law for generalized forces
\begin{align}
	\controlSignalGlobal &\eqdef \globalInputMatrixEstimate^{-1}\left[\totalDisturbanceVectorEstimate + \adrcGainMatrixGlobal\velocityErrorVectorEstimate\right] \nonumber \\
	&= \velocityTransformationMatrix^{-\top}\localInputMatrixEstimate^{-1}\velocityTransformationMatrix^{-1}\left[\totalDisturbanceVectorEstimate+\adrcGainMatrixGlobal\velocityErrorVectorEstimate\right], \label{tauetadef}
\end{align}
where $\totalDisturbanceVectorEstimate$ is an estimate of the total disturbance, $\velocityErrorVectorEstimate$ is an estimate of the velocity tracking error, and
\color{black}
	$\adrcGainMatrixGlobal=\velocityTransformationMatrix\adrcGainMatrixLocal\velocityTransformationMatrix^{-1}\in\realNumbers^{6\times6}$
denotes a gain matrix of the ADR controller with a constant matrix
\begin{align}
	\adrcGainMatrixLocal=\diag\{\adrcGainMatrixElementLocal{1},...,\adrcGainMatrixElementLocal{6}\}\succ0
	\label{eq:adrcLocalGainMatrix}
\end{align}
corresponding to the controller gains expressed in local reference frame $\localCoordinateSystem$.
\color{black}
\begin{cremark}
	According to Assumption \ref{ass:measurableConfiguration}, $\configurationVector$ is the only measurable signal that we can use in the design of feedback control. Vectors $\totalDisturbanceVector$ and $\velocityErrorVector$ are dependent not only on the configuration vector, but also on its derivative, therefore in \eqref{tauetadef} we utilized the estimates $\totalDisturbanceVectorEstimate$ and $\velocityErrorVectorEstimate$ which will be calculated by the extended state observer presented in the latter part of this subsection.
	%
\end{cremark}
According to the mutual relations of the particular model components presented in the description of \eqref{DynDefG} and \eqref{DynDefB}, we can transform the control signals represented in frame $\globalCoordinateSystem$ to the local coordinate system $\localCoordinateSystem$ expressing them as
\begin{align}
	\controlSignalLocal &= \velocityTransformationMatrix^\top\controlSignalGlobal \eq{tauetadef} \localInputMatrixEstimate^{-1}\velocityTransformationMatrix^{-1}\left[\totalDisturbanceVectorEstimate + \adrcGainMatrixGlobal\,\velocityErrorVectorEstimate\right] \nonumber \\
	&= \left[ \controlSignalLocalAlongX \ \controlSignalLocalAlongY \ \controlSignalLocalAlongZ \ \controlSignalLocalAroundX \ \controlSignalLocalAroundY \ \controlSignalLocalAroundZ\right]^\top.
	\label{controllawdef}
\end{align}

We propose to obtain the estimates $\velocityErrorVectorEstimate$ and $\totalDisturbanceVectorEstimate$, utilized in \eqref{tauetadef} and \eqref{controllawdef}, using the high-gain linear ESO (see \cite{MICH19,KHAL08}). The vector of an extended state associated with the particular degree of freedom can be defined as
\begin{align}
\label{exidef}
	\extendedStateVector{i} = \begin{bmatrix} \extendedStateVectorElement{1}{i} \\ \extendedStateVectorElement{2}{i} \\ \extendedStateVectorElement{3}{i} \end{bmatrix} \eqdef \begin{bmatrix} \commandedConfigurationVectorElement{i} - \configurationVectorElement{i} \\ \velocityErrorVectorElement{i} \\ \totalDisturbanceVectorElement{i} \end{bmatrix}\in\realNumbers^3, \quad i\in\{1,...,6\},
\end{align}
where $\commandedConfigurationVector(t) =  [\commandedConfigurationVectorElement{1}(t)  \dots \commandedConfigurationVectorElement{6}(t)]^\top \triangleq \configurationVector(0) + \int_0^t \velocityTransformationMatrix({\orientationVector}(\xi))\commandedVelocityVectorLocal(\xi)d\xi$ and $\configurationVectorElement{i}$ corresponds to the $i$-th element of the configuration vector \eqref{eq:etadef}.
The state-space equations \color{black} associated with the $i$-th extended state vector \eqref{exidef} are calculated upon relation \eqref{epsilondef} and its derivative \eqref{eq:dotepsilon}, \color{black} and are expressed by
\begin{align}
  \label{xidef}
  \begin{cases}
    \extendedStateVectorDerivative{i} = \underbrace{\begin{bmatrix} 0 & 1 & 0 \\ 0 & 0 & 1 \\ 0 & 0 & 0 \end{bmatrix}}_{\observerStateMatrix}\extendedStateVector{i}+\underbrace{\begin{bmatrix} 0 \\ -1 \\ 0 \end{bmatrix}}_{\observerInputRowVector}\cutoffMatrix{i}\velocityTransformationMatrix\localInputMatrixEstimate\actuationMatrixLocal\velocityTransformationMatrix^\top\controlSignalGlobal + \begin{bmatrix} 0 \\ 0 \\ 1 \end{bmatrix}\totalDisturbanceVectorElementDerivative{i} \\
    \observerOutput{i} = \begin{bmatrix} 1 & 0 & 0 \end{bmatrix}\extendedStateVector{i} = \observerOutputMatrix\extendedStateVector{i},
  \end{cases}
\end{align}
where
\begin{align}
  \cutoffMatrix{i} = [\cutoffMatrixElement{i}{1} \ ... \ \cutoffMatrixElement{i}{6}]\in\realNumbers^6: \ \cutoffMatrixElement{i}{j} \triangleq \begin{cases} 1, \ \textrm{when} \  i=j \\
  0, \ \textrm{when} \ i\neq j
  \end{cases}
\end{align}
is a cutting out vector, selecting the appropriate element of the control input vector.

According to equation \eqref{xidef}, we propose a linear observer for a single degree of freedom
\begin{align}
	\label{LESOdef}
	\observerEstimatedStateVectorDerivative{i}  = \observerStateMatrix\observerEstimatedStateVector{i} + \observerInputRowVector\cutoffMatrix{i}\velocityTransformationMatrix\localInputMatrixEstimate\actuationMatrixLocal\velocityTransformationMatrix^\top \controlSignalGlobal + \observerGainVector{i}[(\commandedConfigurationVectorElement{i} - \configurationVectorElement{i}) - \observerEstimatedStateVectorElement{1}{i}],
\end{align}
where  $\observerEstimatedStateVector{i}\triangleq[\observerEstimatedStateVectorElement{1}{i} \ \observerEstimatedStateVectorElement{2}{i} \ \observerEstimatedStateVectorElement{3}{i}]^\top$ is an estimate of the extended state vector \eqref{exidef}, while
\begin{align}
	\label{eq:observergainrule}
	\observerGainVector{i} = [\observerGainVectorElement{1}{i} \ \observerGainVectorElement{2}{i} \ \observerGainVectorElement{3}{i}]^\top \eqdef [3\particularObserverBandwidth{i} \ 3\particularObserverBandwidth{i}^2 \ \particularObserverBandwidth{i}^3]^\top, \ \particularObserverBandwidth{i}>0
\end{align}
is a vector of observer gains. Tuning strategy of the observer gains expressed by  \eqref{eq:observergainrule} is based on the choice of a single parameter value  ($\particularObserverBandwidth{i}$), interpreted as a bandwidth pulsation of ESO (see \cite{GAO06}).

\subsection{Kinematic-level VFO controller}
\label{subsec:vfo}

The outer-loop controller is designed according to the VFO methodology \cite{MICH10,MICH19} and results in the calculation of the commanded velocity vector $\commandedVelocityVectorLocal$, see \eqref{nucdef}, satisfying the desired torpedo-like motion strategy along the reference path. For the design purposes, we are going to keep
\begin{align}
	\label{eq:zalozenia}
	\roll(t)=0,
\end{align}
%
%
according to the postulated non-banked motion (see Remark \ref{rem:nonbanked}). As a consequence,
the roll angle error introduced in \eqref{eq:2.12} should satisfy $\forall t\geq0 \ \rollError(t)=\desiredRollAngle(t)-\roll(t)=0$. Stabilization of the roll angle is obtained with the use of an auxiliary controller designed according to the vehicle kinematics \eqref{eq:KinDef} as
\begin{align}
\label{pcdef}
	\commandedRollLocalVelocity(t) \eqdef \rollControllerFunction(\rollError(t),\cdot) - \sin\roll(t)\tan\pitch(t)\,\commandedPitchLocalVelocity(t) - \cos\roll(t)\tan\pitch(t)\,\commandedYawLocalVelocity(t),
\end{align}
where $\rollControllerFunction(\rollError(t),\cdot)$ is a feedback function that should guarantee the error $\rollError(t)$ to converge to zero. At this step, we are not choosing any particular function $\rollControllerFunction(\rollError,\cdot)$ -  we will introduce and discuss one in the section considering the simulation/experimental results of the proposed control structure.

\begin{assumption}
	\label{ass:roll}
	According to a fixed value of the desired roll angle \eqref{eq:2.123} and a proper selection of the $\rollControllerFunction(\rollError,\cdot)$ function, we assume that  the controlled object fulfills the non-banked motion strategy, satisfying
	\begin{align}
		\label{eq:rollPostulate}
		\forall t>T, \roll(t)=0
	\end{align}
	for some finite time $T$.
\end{assumption}

According to the torpedo-like motion philosophy, the components of commanded velocity vector $\commandedVelocityVectorLocal$ associated with the non-privileged longitudinal axes are fixed at the zero values, i.e., $\commandedYLocalVelocity \eqdef 0$ and $\commandedZLocalVelocity\eqdef 0$, implying that the velocities
\begin{align}
	\label{eq:localUnderactuatedVelocities}
	\yLocalVelocity(t)=\velocityErrorVectorElement{v}(t), \quad \zLocalVelocity(t)=\velocityErrorVectorElement{w}(t),
\end{align}
where $\velocityErrorVectorElement{v}(t)$ and $\velocityErrorVectorElement{w}(t)$ are the elements of the velocity tracking error vector in the body-fixed frame as
\begin{align}
		\label{eq:epsilonnudef}
		\velocityErrorVectorLocal&=[\velocityErrorVectorElement{u} \ \velocityErrorVectorElement{v} \ \velocityErrorVectorElement{w} \ \velocityErrorVectorElement{p} \ \velocityErrorVectorElement{q} \ \velocityErrorVectorElement{r}]^\top \nonumber \\
		&\triangleq\commandedVelocityVectorLocal-\localVelocityVector=\velocityTransformationMatrix^{-1}(\orientationVector)\commandedConfigurationVectorDerivative-\velocityTransformationMatrix^{-1}(\orientationVector)\configurationVectorDerivative=\velocityTransformationMatrix^{-1}(\orientationVector)\velocityErrorVector.
\end{align}

Application of the VFO controller proposed in \cite{LAKO17} to the underactuated vehicle will not provide a satisfactory control performance (especially in not actuated degrees of freedom), thus we propose a modification of the kinematic-level controller to deal with a transversal-drift appearing in the non-actuated axes. The VFO kinematic control law will be derived assumming that the postulate \eqref{eq:rollPostulate} is satisfied.
Upon definition \eqref{epsilondef} and design assumption \eqref{eq:zalozenia}, the vehicle kinematics \eqref{eq:KinDef} can be rewritten in a simplified form
\begin{align}
	&\begin{bmatrix} \xPositionDerivative \\ \yPositionDerivative \\ \zPositionDerivative \\ \pitchDerivative \\ \yawDerivative \end{bmatrix} = \begin{bmatrix} \cos\yaw\cos\pitch & 0 & 0 \\ \sin\yaw\cos\pitch & 0 & 0 \\ -\sin\pitch & 0 & 0 \\ 0 & 1 & 0 \\ 0 & 0 & \frac{1}{\cos\pitch} \end{bmatrix} \begin{bmatrix} \commandedXLocalVelocity \\ \commandedPitchLocalVelocity \\ \commandedYawLocalVelocity \end{bmatrix} - \begin{bmatrix} \velocityErrorVectorElement{1} \\ \velocityErrorVectorElement{2} \\ \velocityErrorVectorElement{3} \\ \velocityErrorVectorElement{5} \\ \velocityErrorVectorElement{6} \\ \end{bmatrix}\ \nonumber \\ & \Leftarrow \ \reducedConfigurationVectorDerivative = \simplifiedVelocityTransformationMatrix(\reducedConfigurationVector)\reducedCommandedVelocityVectorLocal-\reducedVelocityErrorVector,
	\label{KinDefMod}
\end{align}
where $\reducedConfigurationVector=[\xPosition \ \yPosition \ \zPosition \ \pitch \ \yaw]^\top\in\realNumbers^5, \ \reducedCommandedVelocityVectorLocal=[\commandedXLocalVelocity \ \commandedPitchLocalVelocity \ \commandedYawLocalVelocity]^\top\in\realNumbers^3, \ \textrm{and} \ \reducedVelocityErrorVector=[\velocityErrorVectorElement{1} \ \velocityErrorVectorElement{2} \ \velocityErrorVectorElement{3} \ \velocityErrorVectorElement{5} \ \velocityErrorVectorElement{6}]^\top\in\realNumbers^5$ are reduced versions of the original vectors. According to the assumptions of the VFO methodology described in \cite{MICH10}, the vehicle should move along the integral curves determined by the so-called \textit{convergence vector field}, marked as
\begin{align}
\label{hdef}
	&\convergenceVectorField(\reducedConfigurationVector,t) = \begin{bmatrix} \longitudinalConvergenceVectorField(\reducedConfigurationVector,t) \\ \angularConvergenceVectorField(\reducedConfigurationVector,t) \end{bmatrix}\in\realNumbers^5, \ \textrm{where}\\ &\longitudinalConvergenceVectorField = \begin{bmatrix} \convergenceVectorFieldElement{\xPosition} \\ \convergenceVectorFieldElement{\yPosition} \\ \convergenceVectorFieldElement{\zPosition} \end{bmatrix}\in\realNumbers^3, \ \angularConvergenceVectorField = \begin{bmatrix} \convergenceVectorFieldElement{\pitch} \\ \convergenceVectorFieldElement{\yaw} \end{bmatrix}\in\realNumbers^2.
	\nonumber
\end{align}
In the analyzed case, a properly designed vector field $\convergenceVectorField$ should guarantee that $\lim_{t\rightarrow\infty}[\convergenceVectorField(t)-\reducedConfigurationVectorDerivative(t)]=0$. According to formula (\ref{KinDefMod}) and form (\ref{hdef}), the limit case can be rewritten as a postulate
\begin{align}
	\label{eq:lim}
  \begin{bmatrix} \convergenceVectorFieldElement{\xPosition} - \commandedXLocalVelocity\cos\pitch \cos\yaw +\velocityErrorVectorElement{1} \\ \convergenceVectorFieldElement{\yPosition} - \commandedXLocalVelocity\cos\pitch \sin\yaw +\velocityErrorVectorElement{2} \\ \convergenceVectorFieldElement{\zPosition} + \commandedXLocalVelocity\sin\pitch +\velocityErrorVectorElement{3} \\ \convergenceVectorFieldElement{\pitch} - \commandedPitchLocalVelocity +\velocityErrorVectorElement{5} \\ \convergenceVectorFieldElement{\yaw} - \frac{1}{\cos\pitch}\commandedYawLocalVelocity +\velocityErrorVectorElement{6}
  \end{bmatrix} = \begin{bmatrix} 0 \\ 0 \\ 0 \\ 0 \\ 0 \end{bmatrix}.
\end{align}
Solving \eqref{eq:lim} with respect to $\reducedCommandedVelocityVectorLocal$, results in the nominal forms of pseudovelocities
\begin{align}
    \reducedCommandedNominalVelocityVectorLocal &= \begin{bmatrix}   \commandedXNominalLocalVelocity & \commandedPitchNominalLocalVelocity & \commandedYawNominalLocalVelocity \end{bmatrix}^\top \nonumber \\ &\triangleq \begin{bmatrix} (\convergenceVectorFieldElement{\xPosition}+\velocityErrorVectorElement{1})\cos\pitch \cos\yaw+(\convergenceVectorFieldElement{\yPosition}+\velocityErrorVectorElement{2})\cos\pitch \sin\yaw - (\convergenceVectorFieldElement{\zPosition}+\velocityErrorVectorElement{3})\sin\pitch \\ \convergenceVectorFieldElement{\pitch}+\velocityErrorVectorElement{5} \\ (\convergenceVectorFieldElement{\yaw}+\velocityErrorVectorElement{6}) \cos\pitch  \end{bmatrix}  \nonumber \\
		&=\begin{bmatrix} \modifiedConvergenceVectorFieldElement{x}\cos\pitch \cos\yaw+\modifiedConvergenceVectorFieldElement{y}\cos\pitch \sin\yaw - \modifiedConvergenceVectorFieldElement{z}\sin\pitch \\ \modifiedConvergenceVectorFieldElement{\pitch} \\ \modifiedConvergenceVectorFieldElement{\yaw} \cos\pitch  \end{bmatrix},
		\label{nucbardefNominal}
\end{align}
where $\modifiedConvergenceVectorField(\convergenceVectorField,\velocityErrorVector) = [{\longitudinalModifiedConvergenceVectorField}^\top \ {\angularModifiedConvergenceVectorField}^\top]^{\top} = [\modifiedConvergenceVectorFieldElement{\xPosition} \ \modifiedConvergenceVectorFieldElement{\yPosition} \ \modifiedConvergenceVectorFieldElement{\zPosition} \ \modifiedConvergenceVectorFieldElement{\pitch} \ \modifiedConvergenceVectorFieldElement{\yaw} ]^{\top}\triangleq\convergenceVectorField+\reducedVelocityErrorVector$ is a modified convergence vector field. The desired output-feedback characteristics of the final control structure implies the inability of taking velocity measurements and makes the direct use of the velocity tracking error $\velocityErrorVector$ in the controller equations impossible. To calculate the applied commanded velocities  $\reducedCommandedVelocityVectorLocal$, we propose to utilize the modified convergence vector field
\begin{align}
	\label{eq:hstardef}
	\modifiedConvergenceVectorFieldEstimate(\convergenceVectorField,\velocityErrorVectorEstimate) = \begin{bmatrix} \longitudinalModifiedConvergenceVectorFieldEstimate(\longitudinalConvergenceVectorField,\longitudinalVelocityErrorVectorEstimate) \\ \angularModifiedConvergenceVectorFieldEstimate(\angularConvergenceVectorField,\angularVelocityErrorVectorEstimate,\color{black}\longitudinalVelocityErrorVectorEstimateDerivative\color{black}) \end{bmatrix} 
\end{align}
that takes the estimates of the velocity tracking error $\velocityErrorVectorEstimate$ in the place of unavailable vector $\velocityErrorVector$.
Let us first define the longitudinal part of the modified convergence vector field as
\begin{align}
	\label{eq:hpstardef}
	\longitudinalModifiedConvergenceVectorFieldEstimate(\longitudinalConvergenceVectorField,\longitudinalVelocityErrorVectorEstimate) \triangleq \longitudinalConvergenceVectorField + \longitudinalCompensationCoefficient\longitudinalVelocityErrorVectorEstimate,
\end{align}
where $\longitudinalCompensationCoefficient\in[0,1)$ is the design parameter. According to the so-called \textit{cautious compensation} method, utilized for ground vehicles in \cite{MICH10skid}, an introduction of $\longitudinalCompensationCoefficient$ will make the control algorithm more robust to the possible overcompensation of velocity tracking errors caused by a non-zero estimation errors of the extended  states $\extendedStateVector{i}$.

To complete the definition of a modified convergence vector field from \eqref{eq:hpstardef}, we define the longitudinal part of the convergence vector field as follows
\begin{align}
	\label{eq:hpdef}
	\longitudinalConvergenceVectorField(\positionVector)\triangleq \pathVelocityProfile\pathTangentialUnitVector(\positionVector)+\vfoLongitudinalGain[\levelSurface{1}(\positionVector)\pathNormalUnitVector{1}(\positionVector)+\levelSurface{2}(\positionVector)\pathNormalUnitVector{2}(\positionVector)],
\end{align}
where $\vfoLongitudinalGain>0$ is a design parameter of the VFO kinematic controller, whereas \color{black} the orthogonal \color{black} unit vectors, $\pathTangentialUnitVector$ and $\pathNormalUnitVector{j}$, were defined in \eqref{eq:2.8_2} and \eqref{eq:2.8}, respectively. \color{black} The first component of \eqref{eq:hpdef} is a feedforward term computed along a reference path, while the second component is a feedback term that attracts a vehicle towards a reference path. \color{black}

According to equations \eqref{eq:lim} and \eqref{nucbardefNominal}, the postulated values of yaw and pitch angles should satisfy the so-called \textit{orienting conditions} (see \cite{MICH19}):
\begin{align}
	&\yaw - \atantwoc(\pathVelocityDirection\modifiedConvergenceVectorFieldElementEstimate{y}, \  \pathVelocityDirection\modifiedConvergenceVectorFieldElementEstimate{x}) = 0, \quad \textrm{and} \\
	&\pitch - \atan\left(\frac{-\modifiedConvergenceVectorFieldElementEstimate{z}}{\modifiedConvergenceVectorFieldElementEstimate{x}\cos\yaw+\modifiedConvergenceVectorFieldElementEstimate{y}\sin\yaw}\right) = 0,
\end{align}
where $\atantwoc(\cdot,\cdot):\realNumbers\times\realNumbers\rightarrow\realNumbers$ is a continous version of a four-quadrant function $\atantwo(\cdot,\cdot):\realNumbers\times\realNumbers\rightarrow[-\pi,\pi)$, and is described in detail in \cite{MICH10}.
Referring to the vehicle kinematics \eqref{eq:KinDef}, the dynamics of the Euler angles is described with a set of differential equations, making the instantaneous satisfaction of orienting conditions impossible. Due to this fact, we introduce an auxiliary angular error
\begin{align}
	\label{eq:3.22}
	\auxiliaryOrientationError(\positionVector,\longitudinalVelocityErrorVectorEstimate) &\triangleq \begin{bmatrix}\auxiliaryOrientationErrorPitch(\positionVector,\longitudinalVelocityErrorVectorEstimate) \\ \auxiliaryOrientationErrorYaw
	(\positionVector,\longitudinalVelocityErrorVectorEstimate)
	\end{bmatrix} \nonumber \\
	&\triangleq \begin{bmatrix} \auxiliaryPitchAngle(\positionVector,\longitudinalVelocityErrorVectorEstimate) - \pitch \\ \auxiliaryYawAngle(\positionVector,\longitudinalVelocityErrorVectorEstimate) - \yaw \end{bmatrix} \in \left[-\pi,\pi\right)\times\realNumbers,
\end{align}
where $\auxiliaryPitchAngle$ and $\auxiliaryYawAngle$ are the elements of the auxiliary orientation vector defined as
\begin{align}
		&\reducedAuxiliaryOrientation(\positionVector,\longitudinalVelocityErrorVectorEstimate) = \begin{bmatrix} \auxiliaryPitchAngle(\positionVector,\longitudinalVelocityErrorVectorEstimate) \\
    \auxiliaryYawAngle(\positionVector,\longitudinalVelocityErrorVectorEstimate) \end{bmatrix} \nonumber
		\\ &\triangleq \begin{bmatrix} \atan\left(\frac{-\modifiedConvergenceVectorFieldElementEstimate{z}(\positionVector,\longitudinalVelocityErrorVectorEstimate)}{\modifiedConvergenceVectorFieldElementEstimate{x}(\positionVector,\longitudinalVelocityErrorVectorEstimate)\cos\auxiliaryYawAngle+\modifiedConvergenceVectorFieldElementEstimate{y}(\positionVector,\longitudinalVelocityErrorVectorEstimate)\sin\auxiliaryYawAngle}\right) \\ \atantwoc(\pathVelocityDirection\modifiedConvergenceVectorFieldElementEstimate{y}(\positionVector,\longitudinalVelocityErrorVectorEstimate), \ \pathVelocityDirection\modifiedConvergenceVectorFieldElementEstimate{x}(\positionVector,\longitudinalVelocityErrorVectorEstimate))  \end{bmatrix}\in\mathcal{Q},
		 \label{eq:3.20}
\end{align}
for $\mathcal{Q}=(-\pi/2,\pi/2)\times\realNumbers$.
\begin{cremark}
	The auxiliary orientation vector defined by \eqref{eq:3.20} expresses the orientation of a longitudinal part of modified convergence vector field $\longitudinalModifiedConvergenceVectorFieldEstimate$ defined by \eqref{eq:hpstardef}-\eqref{eq:hpdef}. When the vehicle is on a reference path and the velocity tracking error $\velocityErrorVector=\zeroMatrix$ - the auxiliary orientation corresponds to the desired orientation determined by \eqref{eq:2.10}-\eqref{eq:2.11}.
\end{cremark}
\color{black}
According to the derivations presented in \ref{app:angularModifiedConvergenceVectorFieldEstimate},
we introduce the angular part of modified convergence vector field \eqref{eq:hstardef} in a form
\begin{align}
	\label{eq:angularModifiedConvergenceVectorFieldEstimate}
	\angularModifiedConvergenceVectorFieldEstimate(\configurationVector, \reducedAngularVelocityErrorVectorEstimate, \longitudinalVelocityErrorVectorEstimateDerivative) = \reducedAuxiliaryOrientationDerivativeEstimate(\configurationVector,\longitudinalVelocityErrorVectorEstimate,\longitudinalVelocityErrorVectorEstimateDerivative)+\vfoOrientationGainMatrix\auxiliaryOrientationError(\positionVector,\longitudinalVelocityErrorVectorEstimate) + \angularCompensationCoefficient\reducedAngularVelocityErrorVectorEstimate,
\end{align}
where $\reducedAuxiliaryOrientationDerivativeEstimate(\cdot) \triangleq [\auxiliaryPitchAngleDerivativeEstimate(\cdot) \ \auxiliaryYawAngleDerivativeEstimate(\cdot)]^\top$ is an estimate of the reduced auxiliary orientation derivative,  $\vfoOrientationGainMatrix = \diag\{\vfoOrientationGainPitch, \ \vfoOrientationGainYaw\}: \vfoOrientationGainPitch,\vfoOrientationGainYaw>0$ is a gain matrix, and $\reducedAngularVelocityErrorVectorEstimate = [\velocityErrorVectorElementEstimate{5} \ \velocityErrorVectorElementEstimate{6}]^{\top}$, while $\angularCompensationCoefficient\in[0,1]$ is a design parameter introduced to allow the cautious compensation of the angular velocity tracking errors.

\color{black}

The reduced commanded local pseudovelocity vector \eqref{nucbardefNominal} is recalculated with the longitudinal and angular parts of the modified convergence vector field described respectively in \eqref{eq:hpstardef} and  \eqref{eq:angularModifiedConvergenceVectorFieldEstimate}, resulting in the final form
\begin{align}
    \reducedCommandedVelocityVectorLocal &= \begin{bmatrix}   \commandedXLocalVelocity & \commandedPitchLocalVelocity & \commandedYawLocalVelocity \end{bmatrix}^\top \nonumber \\
		&=\begin{bmatrix} \modifiedConvergenceVectorFieldEstimateElement{x}\cos\pitch \cos\yaw+\modifiedConvergenceVectorFieldEstimateElement{y}\cos\pitch \sin\yaw - \modifiedConvergenceVectorFieldEstimateElement{z}\sin\pitch \\ \modifiedConvergenceVectorFieldEstimateElement{\pitch} \\ \modifiedConvergenceVectorFieldEstimateElement{\yaw} \cos\pitch  \end{bmatrix}
		\label{nucbardef}
\end{align}
that only requires the information about the configuration $\configurationVector(t)$ (see Assumption \ref{ass:measurableConfiguration}).

\begin{figure*}[ht!]
	\centering
	\includegraphics[width=0.8\textwidth]{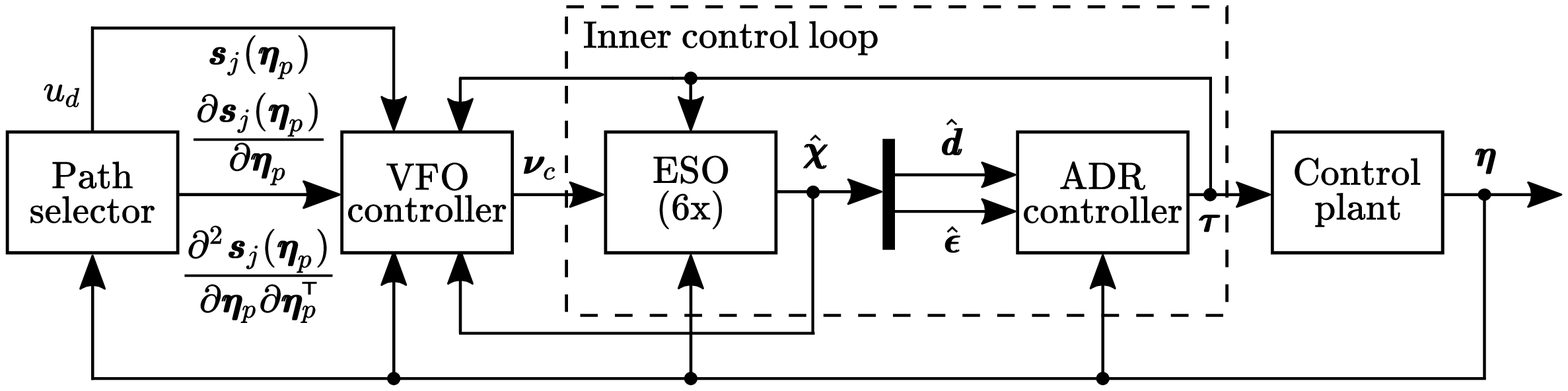}
	\caption{Block diagram of the  VFO-ADR control system for the underactuated vehicle. }
	\label{fig:algorytm}
\end{figure*}

A block diagram of the proposed control structure is presented in Fig.  \ref{fig:algorytm}.

\begin{cremark}
	\label{rem8}
	The elements of auxiliary orientation $\reducedAuxiliaryOrientation$ defined in \eqref{eq:3.20}, together with their derivatives presented in \eqref{eq:auxiliaryPitchDerivative} and \eqref{eq:auxiliaryYawDerivative} are well determined if only ${\modifiedConvergenceVectorFieldElementEstimate{x}}^2+{\modifiedConvergenceVectorFieldElementEstimate{y}}^2 \neq 0$. When the aforementioned relation is satisfied, not only a proper definition of the auxiliary orientation and its derivative is guaranteed, but also the domain $\auxiliaryPitchAngle\in(-\frac{\pi}{2}, \  \frac{\pi}{2})$ is preserved, in accordance with the pitch angle constraint described in the Remark \ref{rem:1}. A non-zero value of the velocity profile $\pathVelocityProfile$ introduced in \eqref{eq:us} guarantees a persistent excitation of the VFO controller, causing that the situation when ${\modifiedConvergenceVectorFieldElementEstimate{x}}^2+{\modifiedConvergenceVectorFieldElementEstimate{y}}^2 = 0$ may potentially appear only in the transient stage (however, it is very rare and non-attracting). To prevent the indeterminacy of auxiliary angles, we propose to freeze the values of $\reducedAuxiliaryOrientation$ and $\reducedAuxiliaryOrientationDerivative$ in a previous state as long as ${\modifiedConvergenceVectorFieldElementEstimate{x}}^2+{\modifiedConvergenceVectorFieldElementEstimate{y}}^2 < \varepsilon$, for sufficiently small, non-zero value of $\varepsilon$.
\end{cremark}
\color{black}

\renewcommand{\labelenumi}{P\theenumi.}

\section{The main result and stability analysis}
\label{sec:stability}

\subsection{Error domain dynamics of particular subsystems}

To show the boundedness of the path-following errors introduced in \eqref{eq:2.12}, particular subsystems of the proposed control structure have to be analyzed in an appropriate order. This subsection is focused on the definition of these subsystems,  and the derivation of their dynamic equations.

Let us address first the observation errors associated with the extended state observer. According to dynamics \eqref{xidef}, and ESO equations \eqref{LESOdef}, we define a combined observation error
\begin{equation}
    \label{eq:ObsErrorVect}
    \observationErrorAggregatedVector \triangleq \begin{bmatrix} \observatorOutputObservationError \\ \velocityErrorVectorObservationError \\ \totalDisturbanceVectorObservationError \end{bmatrix} \triangleq \observerStateAggregatedVector-\observerStateEstimateAggregatedVector = \begin{bmatrix} \observerAggregatedOutput-\observerAggregatedOutputEstimate \\ \velocityErrorVector - \velocityErrorVectorEstimate \\ \totalDisturbanceVector - \totalDisturbanceVectorEstimate \end{bmatrix}\in\realNumbers^{18},
\end{equation}
where $\observerAggregatedOutput = [\observerOutput{1} ... \observerOutput{6}]^\top \eqtext{\eqref{exidef}} [\extendedStateVectorElement{1}{1}...\extendedStateVectorElement{1}{6}]^\top$ is the combined observer output (see \eqref{xidef}). The dynamics of $\observationErrorAggregatedVector$ can be written down as
\begin{align}
\label{eq:sigma1}
    \observationErrorAggregatedVectorDerivative(t) &= -\underbrace{\begin{bmatrix} \observerGainAggregatedMatrix{1} & \pmb{-\identityMatrix} & \zeroMatrix \\
                                    \observerGainAggregatedMatrix{2} & \zeroMatrix & \pmb{-\identityMatrix} \\
                                    \observerGainAggregatedMatrix{3} & \zeroMatrix & \zeroMatrix\end{bmatrix}}_{\observationErrorStateMatrix}\observationErrorAggregatedVector(t) +\underbrace{\begin{bmatrix} \zeroMatrix \\ \zeroMatrix \\ \pmb{\identityMatrix}  \end{bmatrix}}_{\observationErrorIdentityMatrix}\totalDisturbanceVectorDerivative(t),
\end{align}
where $\observerGainAggregatedMatrix{i}\triangleq\diag\{\observerGainVectorElement{i}{1} \ ... \ \observerGainVectorElement{i}{6}\}, \ i\in\{1,2,3\}$ are the observer gain matrices resulting from \eqref{eq:observergainrule}, while $\zeroMatrix$ and $\identityMatrix$ are, respectively, the zero and identity matrices of the appropriate dimensions.

To address dynamics of the velocity error, we introduce a modified disturbance vector
%
\begin{align}
  \label{eq:modifiedTotalDisturbance}
  \modifiedTotalDisturbanceVector &= [\modifiedTotalDisturbanceVectorElement{1} \ \modifiedTotalDisturbanceVectorElement{2} \ \modifiedTotalDisturbanceVectorElement{3} \ \modifiedTotalDisturbanceVectorElement{4} \ \modifiedTotalDisturbanceVectorElement{5} \ \modifiedTotalDisturbanceVectorElement{6}]^\top \nonumber \\
  &\triangleq \totalDisturbanceVector + \massMatrixGlobal^{-1}\environmentalDampingMatrixGlobal\commandedConfigurationVectorDerivative-\massMatrixGlobal^{-1}\environmentalDampingMatrixGlobal\configurationVectorDerivative = \totalDisturbanceVector + \color{black}\massMatrixGlobal^{-1}\color{black}\environmentalDampingMatrixGlobal\velocityErrorVector,
\end{align}
where the component $\massMatrixGlobal^{-1}\environmentalDampingMatrixGlobal\configurationVectorDerivative$ corresponds to the linear part of the environmental damping (see \cite{FOSS99}) included in vector $\dynamicPhenomenaVectorGlobal$ from \eqref{DynDefG}, while $- \massMatrixGlobal^{-1}\environmentalDampingMatrixGlobal\commandedConfigurationVectorDerivative$ allows us to describe $\modifiedTotalDisturbanceVector$ as a function of $\totalDisturbanceVector$ and $\velocityErrorVector$. Matrix $\environmentalDampingMatrixGlobal = \velocityTransformationMatrix\environmentalDampingMatrixLocal\velocityTransformationMatrix^{-1}$, where
\begin{align}
  \label{eq:dampingMatrix}
  \environmentalDampingMatrixLocal \triangleq\textrm{diag}\{\environmentalDampingElementLocal{u},\environmentalDampingElementLocal{v},\environmentalDampingElementLocal{w},\environmentalDampingElementLocal{p},\environmentalDampingElementLocal{q},\environmentalDampingElementLocal{r}\}, \ \environmentalDampingElementLocal{i}>0,
\end{align}
includes the positive damping coefficients in the body-related coordinate system $\localCoordinateSystem$. It is worth noting that only the linear part of the environmental damping model was excluded from $\totalDisturbanceVector$, while the higher-order terms still reside in vector $\modifiedTotalDisturbanceVector$.

\begin{cremark}
	\label{rem:dbound}
  Upon Assumptions \ref{ass:limitedDisturbance} and \ref{ass:4}, Remark \ref{rem8}, and equations \eqref{tauetadef}, \eqref{etacfromnuc}, and \eqref{nucbardef}, we may claim that the disturbance $\totalDisturbanceVector$, defined in \eqref{eq:dotepsilon}, its derivative $\totalDisturbanceVectorDerivative$, and the modified total disturbance $\modifiedTotalDisturbanceVector$ from \eqref{eq:modifiedTotalDisturbance}, are bounded in some compact sets, in particular
	\begin{align}
    \label{eq:rem9}
  	\sup_{t\geq0}\module{\totalDisturbanceVectorDerivative(t)}\leq\observationErrorDisturbanceTerminalBound, \quad
    \sup_{t\geq0}\module{\modifiedTotalDisturbanceVector(t)}\leq\velocityErrorModifiedDisturbanceTerminalBound 
  \end{align}
	for $\totalDisturbanceVectorDerivative\in\observationErrorDisturbanceDomain\subset\realNumbers^6, \ \modifiedTotalDisturbanceVector\in\velocityErrorModifiedDisturbanceDomain\subset\realNumbers^6,$ where  $\observationErrorDisturbanceTerminalBound,\velocityErrorModifiedDisturbanceTerminalBound>0$ are some finite upper bounds determined by a compact set of the vehicle's state ($\configurationVector,\configurationVectorDerivative$).
	%
\end{cremark}

After transformations presented in \ref{app:velocityErrorDynamics}, the velocity error dynamics described firstly in \eqref{eq:dotepsilon} can be expressed as
\begin{align}
    \velocityErrorVectorDerivative
    &= -\velocityTransformationMatrix\underbrace{\left(\actuationMatrixLocal\adrcGainMatrixLocal+(\identityMatrix-\actuationMatrixLocal)\massMatrixLocal^{-1}\environmentalDampingMatrixLocal\right)}_{\velocityErrorStateMatrix}\velocityTransformationMatrix^{-1}\velocityErrorVector \nonumber \\
    &+\velocityTransformationMatrix\actuationMatrixLocal\adrcGainMatrixLocal\velocityTransformationMatrix^{-1}\velocityErrorVectorObservationError+\velocityTransformationMatrix\actuationMatrixLocal\velocityTransformationMatrix^{-1}\totalDisturbanceVectorObservationError+\velocityTransformationMatrix(\identityMatrix-\actuationMatrixLocal)\velocityTransformationMatrix^{-1}\modifiedTotalDisturbanceVector
    \label{eq:velocityTrackingErrorDynamics}
\end{align}
where $\adrcGainMatrixLocal$ is the constant gain matrix defined in \eqref{eq:adrcLocalGainMatrix}.

\begin{cremark}
  Matrix $\velocityErrorStateMatrix$ introduced in \eqref{eq:velocityTrackingErrorDynamics} is a convex combination of the matrices $\adrcGainMatrixLocal$ and $\massMatrixLocal^{-1}\environmentalDampingMatrixLocal$ that are both positive definite, thus according to \cite{CHAR95},  matrix $\velocityErrorStateMatrix$ is positive definite itself.
\end{cremark}

\color{black}
\begin{cremark}
  \label{rem:velocityErrorStateMatrix}
  A constant matrix $\massMatrixLocal$, associated with a vehicle which center of mass is located in the origin of a local coordinate frame (see \cite{FOSS99}), leads to the diagonal matrix
  \begin{equation}
    \velocityErrorStateMatrix = \diag\{\adrcGainMatrixElementLocal{1}, \ \actuationMatrixLocalElement{v}\adrcGainMatrixElementLocal{2}+(1-\actuationMatrixLocalElement{v})\frac{\environmentalDampingElementLocal{v}}{m}, \ \actuationMatrixLocalElement{w}\adrcGainMatrixElementLocal{3}+(1-\actuationMatrixLocalElement{w})\frac{\environmentalDampingElementLocal{w}}{m}, \ \adrcGainMatrixElementLocal{4}, \ \adrcGainMatrixElementLocal{5}, \ \adrcGainMatrixElementLocal{6}\},
    \label{eq:58}
  \end{equation}
  where $m$ is a total vehicle mass. In the case when some of the axes are underactuated, i.e. $\actuationMatrixLocalElement{i}=0, \ i\in\{v,w\}$, the corresponding controller gains associated with underactuated axes vanish in matrix $\velocityErrorStateMatrix$.
\end{cremark}
\color{black}

Now, let us turn to the angular part of the kinematic subsystem. The equation of roll error dynamics, calculated upon \eqref{eq:KinDef}, \eqref{eq:epsilonnudef}, \eqref{epsilondef} and \eqref{pcdef}, has the form
\begin{align}
  \label{eq:ephidyn}
  \rollErrorDerivative(t)= -\rollControllerFunction(\rollError(t),\cdot)+\velocityErrorVectorElement{1}(t).
\end{align}
Under Assumption \ref{ass:roll}, the dynamics of an auxiliary error vector  \eqref{eq:3.22} can be derived using equations \eqref{eq:KinDef}, \eqref{eq:epsilonnudef} in the form
\begin{align}
  \auxiliaryOrientationErrorDerivative&=\begin{bmatrix} \auxiliaryPitchDerivative-\commandedPitchLocalVelocity+\velocityErrorVectorElement{5} \\
	\auxiliaryYawAngleDerivative-\frac{1}{\cos\pitch}\commandedYawLocalVelocity+\velocityErrorVectorElement{6}\end{bmatrix} \nonumber \\
  &\eqtext{\eqref{eq:angularModifiedConvergenceVectorFieldEstimate}, \eqref{nucbardef}} \begin{bmatrix} -\vfoOrientationGainPitch\auxiliaryOrientationErrorPitch+\angularCompensationCoefficient\velocityErrorVectorElementObservationError{\pitch} + (1-\angularCompensationCoefficient)\velocityErrorVectorElement{\pitch}+\auxiliaryPitchDerivativeObservationError \\
  -\vfoOrientationGainYaw\auxiliaryOrientationErrorYaw+\angularCompensationCoefficient\velocityErrorVectorElementObservationError{\yaw}+(1-\angularCompensationCoefficient)\velocityErrorVectorElement{\yaw}+\auxiliaryYawDerivativeObservationError \end{bmatrix} \nonumber \\
  &= -\vfoOrientationGainMatrix\auxiliaryOrientationError+\angularCompensationCoefficient\reducedAngularVelocityErrorVectorObservationError+(1-\angularCompensationCoefficient)\reducedAngularVelocityErrorVector+\reducedAuxiliaryOrientationDerivativeObservationError
  \label{eq:auxiliaryOrientationErrorDynamics}
\end{align}
where $\reducedAngularVelocityErrorVector \triangleq [\velocityErrorVectorElement{5} \ \velocityErrorVectorElement{6}]^\top$ is the dimensionally reduced angular velocity error, $\reducedAngularVelocityErrorVectorObservationError \triangleq [\velocityErrorVectorObservationErrorElement{5} \ \velocityErrorVectorObservationErrorElement{6}]^\top$ is the observation inaccuracy of the reduced vector of the angular velocity error, while
\begin{align}
  \reducedAuxiliaryOrientationDerivativeObservationError = \begin{bmatrix} \auxiliaryPitchDerivativeObservationError \\ \auxiliaryYawDerivativeObservationError \end{bmatrix} \triangleq \begin{bmatrix}  \auxiliaryPitchDerivative-\auxiliaryPitchAngleDerivativeEstimate \\ \auxiliaryYawDerivative - \auxiliaryYawAngleDerivativeEstimate \end{bmatrix}
\end{align}
correspond to the difference between the nominal derivatives of auxiliary angles from \eqref{eq:auxiliaryYawDerivative}-\eqref{eq:auxiliaryPitchDerivative} and the applied derivatives of auxiliary angles utilized in \eqref{eq:angularConvergenceVectorFieldEstimate}. More detailed representation of the particular elements of $\reducedAuxiliaryOrientationDerivativeObservationError$ is provided in \ref{app:reducedAuxiliaryOrientationDerivativeObservationError}.

The dynamics of positional error $\positionError$ will be written down separately for its particular components. Knowing that the gradient of the particular level-surfaces $\levelSurfaceGradient{j}$ and the vector $\pathTangentialUnitVector$, tangential to the cross-section of level surfaces (see \eqref{eq:2.8_2}) are perpendicular to each other, relation
\begin{align}
  &\levelSurfaceGradientTranspose{j}\pathVelocityProfile\frac{\pathMotionDirection(\levelSurfaceGradient{1}\times\levelSurfaceGradient{2})}{\module{\levelSurfaceGradient{1}\times\levelSurfaceGradient{2}}} \nonumber \\
  &\eqtext{\eqref{eq:hpstardef}, \eqref{eq:2.8}}\levelSurfaceGradientTranspose{j}\longitudinalModifiedConvergenceVectorFieldEstimate+\levelSurfaceGradientTranspose{j}\vfoLongitudinalGain \levelSurface{1}\frac{\levelSurfaceGradient{1}}{\module{\levelSurfaceGradient{1}}}+\levelSurfaceGradientTranspose{j}\vfoLongitudinalGain \levelSurface{2}\frac{\levelSurfaceGradient{2}}{\module{\levelSurfaceGradient{2}}} \nonumber \\
  &- \levelSurfaceGradientTranspose{j}\longitudinalCompensationCoefficient\longitudinalVelocityErrorVectorEstimate=0
  \label{eq:levelSurfaceDynamicsRelation}
\end{align}
is always satisfied.
As a consequence, we can express the level-surface dynamics as
\begin{align}
  \levelSurfaceDerivative{1}&=\levelSurfaceGradientTranspose{1}\positionVectorDerivative \nonumber \\
  &=\levelSurfaceGradientTranspose{1}(\commandedPositionVectorDerivative-\longitudinalVelocityErrorVector)-\levelSurfaceGradientTranspose{1}\longitudinalModifiedConvergenceVectorFieldEstimate-\vfoLongitudinalGain \levelSurface{1}\module{\levelSurfaceGradient{1}} \nonumber \\
  &-\vfoLongitudinalGain \levelSurface{2}\cos\angleBetweenlevelSurfaceGradients\module{\levelSurfaceGradient{1}}+\levelSurfaceGradientTranspose{1}\longitudinalCompensationCoefficient\longitudinalVelocityErrorVectorEstimate,
  \label{eq:s1dot}  \\
  \levelSurfaceDerivative{2}
  &=\levelSurfaceGradientTranspose{2}(\commandedPositionVectorDerivative-\longitudinalVelocityErrorVector)-\levelSurfaceGradientTranspose{2}\longitudinalModifiedConvergenceVectorFieldEstimate-\vfoLongitudinalGain \levelSurface{2}\module{\levelSurfaceGradient{2}} \nonumber \\
  &-\vfoLongitudinalGain \levelSurface{1}\cos\angleBetweenlevelSurfaceGradients\module{\levelSurfaceGradient{2}} + \levelSurfaceGradientTranspose{2}\longitudinalCompensationCoefficient\longitudinalVelocityErrorVectorEstimate,
  \label{eq:s2dot}
\end{align}
where $\angleBetweenlevelSurfaceGradients\angle(\levelSurfaceGradient{1},\levelSurfaceGradient{2})\neq k\pi, \ k\in\integerNumbers$ corresponds to the angle between gradients of level-surfaces and its value is constrained according to Assumption \ref{ass:nonCollinearity}.

\subsection{Statement of the main result}
\begin{proposition}
  \label{prop:1}
  Under Assumptions \ref{ass:measurableConfiguration} - \ref{ass:roll}, \color{black} recalling \eqref{eq:rem9}, and postulating that $\environmentalDampingElementLocal{v}=\environmentalDampingElementLocal{w}=:\environmentalDampingElement$ is large enough (see \eqref{eq:dampingMatrix})\color{black}, the output-feedback VFO-ADR cascaded control law resulting from a combination of the inner-loop ADR controller \eqref{controllawdef} and the outer-loop controller, consisting of the VFO part \eqref{nucbardef} and the auxiliary roll stabilizer \eqref{pcdef}, applied into the uncertain and underactuated vehicle represented by \eqref{FullKinDyn}, locally \color{black}(for sufficiently small initial errors $\|\observationErrorAggregatedVector(0)\|, \ \|\velocityErrorVector(0)\|, \ \|\auxiliaryOrientationError(0)\|, \ \|\positionError(0)\|$) \color{black} guarantees satisfaction of control objectives \eqref{eq:controlobjective}-\eqref{eq:controlobjective2} with ultimate upper bounds  $\controlError=\controlError(\observerBandwidth,\adrcGainMatrixLocal,\vfoOrientationGainMatrix,\vfoLongitudinalGain,\angularCompensationCoefficient,\longitudinalCompensationCoefficient,r_{\dot{d}},r_{d^*})$ and $\positionalControlError=\positionalControlError(\observerBandwidth,\adrcGainMatrixLocal,\vfoOrientationGainMatrix,\vfoLongitudinalGain,\angularCompensationCoefficient,\longitudinalCompensationCoefficient,r_{\dot{d}},r_{d^*})$, \color{black} by taking $\adrcGainMatrixLocal=\textrm{blkdiag}\left\{\frac{\environmentalDampingElement}{m}\identityMatrix, \ \adrcGainMatrixAngularElement\identityMatrix\right\}$, where $\identityMatrix\in\realNumbers^{3\times3}$, and $k_o>0$.
  \color{black} Moreover, the upper bound $\positionalControlError(\observerBandwidth,\adrcGainMatrixLocal,\vfoOrientationGainMatrix,\vfoLongitudinalGain,\angularCompensationCoefficient,\longitudinalCompensationCoefficient,r_{\dot{d}},r_{d^*})$ can be made arbitrarily small, for sufficiently small upper bounds $r_{\dot{d}},r_{d^*}$ introduced in \eqref{eq:rem9}, by increasing the gain $\vfoLongitudinalGain$ and using sufficiently large $\minEigenvalue{\vfoOrientationGainMatrix}$ and $\observerBandwidth$.
\end{proposition}
\color{black}
\begin{cremark}
  The relation $\environmentalDampingElementLocal{v}=\environmentalDampingElementLocal{w}=:\environmentalDampingElement$, postulated in Proposition \ref{prop:1}, is satisfied for a vehicle with a body symmetric with respect to axis $\xLocalAxis$. This conservative assumption, together with the conservative postulate about sufficiently small initial errors and upper bounds of disturbances, and the selected tuning strategy for $\adrcGainMatrixLocal$, all have been introduced only for stability analysis purposes. According to the simulation results presented in Section \ref{sec:simulations}, the above conservative postulates can be relaxed in practice without loosing the terminal boundedness stated in \eqref{eq:controlobjective}-\eqref{eq:controlobjective2}. 
  \label{rem:adrcTuning}
\end{cremark}
\color{black}

\subsection{Proof of Proposition \ref{prop:1}}
\color{black}
The analysis
is divided into a set of shorter Lemmas corresponding the subsystems of observation error \eqref{eq:sigma1}, velocity error \eqref{eq:velocityTrackingErrorDynamics}, auxiliary orientation error \eqref{eq:auxiliaryOrientationErrorDynamics}, and position error \eqref{eq:s1dot}-\eqref{eq:s2dot}. The analysis will be carried out using the ISS method presented in \cite{SONT96,ISID17}.
\begin{lemma}
  \label{lem:1}
  Observation error subsystem \eqref{eq:sigma1} is locally ISS with respect to the perturbation $\totalDisturbanceVectorDerivative(\cdot)\in\observationErrorDisturbanceDomain$, see \eqref{eq:rem9}, implying the ultimate boundedness of $\observationErrorAggregatedVector$, i.e., $\lsShort\module{\observationErrorAggregatedVector(t)}\leq\observationErrorUltimateBound(\observerBandwidth,\observationErrorDisturbanceTerminalBound)$, where
  \begin{align}
    \observationErrorUltimateBound(\observerBandwidth,\observationErrorDisturbanceTerminalBound) = \max\left\{\frac{1}{\observerBandwidth^2},1\right\}\frac{2\lambda_{\textrm{max}}^{3/2}(\transformedObservationErrorLyapunovEquationSolution)}{\lambda_{\textrm{min}}^{1/2}(\transformedObservationErrorLyapunovEquationSolution)\observationErrorMajorizationCoefficient\observerBandwidth}\observationErrorDisturbanceTerminalBound,
  \end{align}
  $\observationErrorMajorizationCoefficient\in(0,1)$, and  $\transformedObservationErrorLyapunovEquationSolution=\transformedObservationErrorLyapunovEquationSolution^\top\succ0$. Thus, the non-negative function $\observationErrorUltimateBound(\observerBandwidth,\observationErrorDisturbanceTerminalBound)$ is bounded for any $\observerBandwidth>0$, and $\observationErrorUltimateBound(\observerBandwidth,\observationErrorDisturbanceTerminalBound)\rightarrow0$ as $\observerBandwidth\rightarrow\infty$, for any sufficiently small upper bound $\observationErrorDisturbanceTerminalBound$ introduced in \eqref{eq:rem9}.
\end{lemma}

\begin{lemma}
  \label{lem:2}
    Velocity error subsystem \eqref{eq:velocityTrackingErrorDynamics} is locally ISS with respect to the perturbations $\velocityErrorPerturbationVector\triangleq\left[\totalDisturbanceVectorObservationError^\top \ \velocityErrorVectorObservationError^\top \right]^\top\in\velocityErrorDisturbanceDomain$ and $\modifiedTotalDisturbanceVector\in\velocityErrorModifiedDisturbanceDomain$ implying the ultimate boundedness of $\velocityErrorVector$, i.e., $\lsShort\module{\velocityErrorVector(t)}\leq\velocityErrorUltimateBound(\adrcGainMatrixLocal,\observationErrorUltimateBound,\velocityErrorModifiedDisturbanceTerminalBound)$ where
    \begin{align}
      \velocityErrorUltimateBound(\adrcGainMatrixLocal,\observationErrorUltimateBound,\velocityErrorModifiedDisturbanceTerminalBound) = \frac{1+\maxEigenvalue{\adrcGainMatrixLocal}}{\velocityErrorMajorizationCoefficient\minEigenvalue{\velocityErrorStateMatrix}}\color{black} r_{\chi}^\infty
      \color{black}+ \frac{\maxEigenvalue{\identityMatrix-\actuationMatrixLocal}}{\velocityErrorMajorizationCoefficient\minEigenvalue{\velocityErrorStateMatrix}}\velocityErrorModifiedDisturbanceTerminalBound,
      \label{eq:lem2}
    \end{align}
     while $r_{\chi}^\infty=r_{\chi}^\infty(\observerBandwidth,\observationErrorDisturbanceTerminalBound)$ has been introduced in Lemma \ref{lem:1}. According to right-hand side of \eqref{eq:lem2} and observing that $\velocityErrorStateMatrix=\adrcGainMatrixLocal$ for $\adrcGainMatrixLocal=\blkdiag\{\frac{\rho}{m}\identityMatrix,k_o\identityMatrix\}\succ0$ and $\rho:=\environmentalDampingElementLocal{v}=\environmentalDampingElementLocal{w}>0$, the non-negative function  $\velocityErrorUltimateBound(\adrcGainMatrixLocal,\observationErrorUltimateBound,\velocityErrorModifiedDisturbanceTerminalBound)$ is bounded, upon Lemma \ref{lem:1}, for any sufficiently small upper bounds $\observationErrorDisturbanceTerminalBound$ and $\velocityErrorModifiedDisturbanceTerminalBound$ introduced in \eqref{eq:rem9}. Moreover, $\velocityErrorUltimateBound(\adrcGainMatrixLocal,0,\velocityErrorModifiedDisturbanceTerminalBound)\equiv0$ if the vehicle is fully actuated (i.e., if $\actuationMatrixLocal=\identityMatrix$); otherwise $\velocityErrorUltimateBound(\adrcGainMatrixLocal,\observationErrorUltimateBound,\velocityErrorModifiedDisturbanceTerminalBound)$ can be made sufficiently small by taking $\observerBandwidth$ large enough and if $\minEigenvalue{\adrcGainMatrixLocal}\equiv\minEigenvalue{\velocityErrorStateMatrix}=\min\{\frac{\rho}{m},k_o\}$ is large enough as well.
\end{lemma}

\begin{lemma}
  \label{lem:3}
  Auxiliary orientation error subsystem \eqref{eq:auxiliaryOrientationErrorDynamics} is locally ISS with respect to the perturbations $\reducedAngularVelocityErrorVectorObservationError \in \auxiliaryOrientationObservationErrorDomain, \ \reducedAngularVelocityErrorVector \in \auxiliaryOrientationVelocityErrorDomain, \ \textrm{and} \ \reducedAuxiliaryOrientationDerivativeObservationError \in\auxiliaryOrientationFFDomain$, implying the ultimate boundedness of $\auxiliaryOrientationError$, i.e., $\lsShort\module{\auxiliaryOrientationError(t)}\leq\auxiliaryOrientationUltimateBound(\vfoOrientationGainMatrix,\angularCompensationCoefficient,\observationErrorUltimateBound,\velocityErrorUltimateBound)$ where
  \begin{align}
    \auxiliaryOrientationUltimateBound(\vfoOrientationGainMatrix,\angularCompensationCoefficient,\observationErrorUltimateBound,\velocityErrorUltimateBound) = \frac{\angularCompensationCoefficient+\bar{f}}{\auxiliaryOrientationMajorizationCoefficient\minEigenvalue{\vfoOrientationGainMatrix}}\observationErrorUltimateBound+\frac{1-\angularCompensationCoefficient}{\auxiliaryOrientationMajorizationCoefficient\minEigenvalue{\vfoOrientationGainMatrix}}\velocityErrorUltimateBound
    \label{eq:lem3}
  \end{align}
  while $r_{\chi}^\infty=r_{\chi}^\infty(\observerBandwidth,\observationErrorDisturbanceTerminalBound)$ and $\velocityErrorUltimateBound=\velocityErrorUltimateBound(\adrcGainMatrixLocal,\observationErrorUltimateBound,\velocityErrorModifiedDisturbanceTerminalBound)$ have been introduced, respectively, in Lemma 1 and Lemma 2, $\angularCompensationCoefficient\in[0,1]$, and $\bar{f}>0$ is some bounding function for any compact set of a vehicle state. According to the right-hand side of \eqref{eq:lem3}, and upon Lemmas 1-2, the non-negative function $\auxiliaryOrientationUltimateBound(\vfoOrientationGainMatrix,\angularCompensationCoefficient,\observationErrorUltimateBound,\velocityErrorUltimateBound)$ is bounded for any sufficiently small upper bounds $\observationErrorDisturbanceTerminalBound$ and $\velocityErrorModifiedDisturbanceTerminalBound$ introduced in \eqref{eq:rem9}. Moreover, $\auxiliaryOrientationUltimateBound(\vfoOrientationGainMatrix,\angularCompensationCoefficient,\observationErrorUltimateBound\velocityErrorUltimateBound)\rightarrow0$ as $\observationErrorUltimateBound\rightarrow0$ and $\minEigenvalue{\vfoOrientationGainMatrix}\rightarrow\infty$, for any finite $\velocityErrorUltimateBound$ and any finite upper bounds $\observationErrorDisturbanceTerminalBound$ and $\velocityErrorModifiedDisturbanceTerminalBound$. Additionally, if one selects $\angularCompensationCoefficient=1$, then $\auxiliaryOrientationUltimateBound(\vfoOrientationGainMatrix,1,\observationErrorUltimateBound\velocityErrorUltimateBound)\rightarrow0$ as $\observationErrorUltimateBound\rightarrow0$ for any finite $\minEigenvalue{\vfoOrientationGainMatrix}>0$ and any finite upper bounds $\observationErrorDisturbanceTerminalBound$ and $\velocityErrorModifiedDisturbanceTerminalBound$.
\end{lemma}

\begin{lemma}
  \label{lem:4}
    Dynamics of the positional error $\positionError$ (cf. \eqref{eq:2.12}), represented by the equations \eqref{eq:s1dot}-\eqref{eq:s2dot} is locally ISS with respect to perturbations $\positionErrorPerturbationObservationErrorVector\triangleq[\auxiliaryOrientationError^\top \ \longitudinalVelocityErrorVectorObservationError^\top]^\top\in\positionErrorPerturbationObservationErrorDomain$, $\positionErrorPerturbationVelocityErrorVector\triangleq[\auxiliaryOrientationError^\top \ \longitudinalVelocityErrorVector^\top]^\top\in\positionErrorPerturbationVelocityErrorDomain$, and $\auxiliaryOrientationError\in\auxiliaryOrientationDomain$, implying the ultimate boundedness of $\positionError$, i.e., $\lsShort\module{\positionError(t)}\leq\positionErrorUltimateBound(\vfoLongitudinalGain,\longitudinalCompensationCoefficient,\observationErrorUltimateBound,\velocityErrorUltimateBound,\auxiliaryOrientationUltimateBound)$, where
    \begin{align}
      \positionErrorUltimateBound(\vfoLongitudinalGain,\longitudinalCompensationCoefficient,\observationErrorUltimateBound,\velocityErrorUltimateBound,\auxiliaryOrientationUltimateBound) = \frac{m^*}{k_p}&\Bigg[\frac{l_1(\longitudinalCompensationCoefficient,\velocityErrorUltimateBound,\auxiliaryOrientationUltimateBound)}{m_1(\velocityErrorUltimateBound,\auxiliaryOrientationUltimateBound)} \nonumber \\
      &+\frac{l_2(\longitudinalCompensationCoefficient,\observationErrorUltimateBound,\auxiliaryOrientationUltimateBound)}{m_2(\observationErrorUltimateBound,\auxiliaryOrientationUltimateBound)}+\frac{l_3(\auxiliaryOrientationUltimateBound)}{m_3(\auxiliaryOrientationUltimateBound)}\Bigg],
      \label{eq:lem4}
    \end{align}
    where $m^*>0$ is some constant, $l_i(\cdot), \ m_i(\cdot), \ i\in\{1,2,3\}$ are some polynomial functions of their arguments, and $m_i(\cdot)$ are positive for their arguments being sufficiently small, while $r_{\chi}^\infty=r_{\chi}^\infty(\observerBandwidth,\observationErrorDisturbanceTerminalBound)$, $\velocityErrorUltimateBound=\velocityErrorUltimateBound(\adrcGainMatrixLocal,\observationErrorUltimateBound,\velocityErrorModifiedDisturbanceTerminalBound)$ and $\auxiliaryOrientationUltimateBound=\auxiliaryOrientationUltimateBound(\vfoOrientationGainMatrix,\angularCompensationCoefficient,\observationErrorUltimateBound,\velocityErrorUltimateBound)$ have been introduced in Lemmas 1-3 (thus, $\positionErrorUltimateBound(\vfoLongitudinalGain,\longitudinalCompensationCoefficient,\observationErrorUltimateBound,\velocityErrorUltimateBound,\auxiliaryOrientationUltimateBound) = \positionErrorUltimateBound(\observerBandwidth,\adrcGainMatrixLocal,\vfoOrientationGainMatrix,\vfoLongitudinalGain,\angularCompensationCoefficient,\longitudinalCompensationCoefficient,\observationErrorDisturbanceTerminalBound,\velocityErrorModifiedDisturbanceTerminalBound)$). For sufficiently small upper-bounding functions $\observationErrorUltimateBound$, $\velocityErrorUltimateBound$, and $\auxiliaryOrientationUltimateBound$, which can be obtained by selecting $\observerBandwidth$ and $\minEigenvalue{\vfoOrientationGainMatrix}$ large enough and if $\rho>0$ is sufficiently large (according to Lemmas 1-3), the right-hand side of \eqref{eq:lem4} is non-negative and bounded. Moreover, under the above conditions, $\positionErrorUltimateBound(\observerBandwidth,\adrcGainMatrixLocal,\vfoOrientationGainMatrix,\vfoLongitudinalGain,\angularCompensationCoefficient,\longitudinalCompensationCoefficient,\observationErrorDisturbanceTerminalBound,\velocityErrorModifiedDisturbanceTerminalBound)\rightarrow0$ as $\vfoLongitudinalGain\rightarrow\infty$.
     %
     %
\end{lemma}

\begin{lemma}
  \label{lem:5}
  In the VFO-ADR control system, the angular error $\orientationError \ \textrm{mod} \  2\pi$ (cf. \eqref{eq:2.12}-\eqref{eq:e2pidef}) is ultimately upper bounded as follows: $\lsShort\|\orientationError(t) \ \textrm{mod} \ 2\pi\|_1\leq2\auxiliaryOrientationUltimateBound+r_o^\infty$, where $r_o^\infty>0$ is some bounding function, whereas $\auxiliaryOrientationUltimateBound=\auxiliaryOrientationUltimateBound(\vfoOrientationGainMatrix,\angularCompensationCoefficient,\observationErrorUltimateBound,\velocityErrorUltimateBound)$ has been introduced in Lemma 3. As a consequence, upon the form of \eqref{eq:e2pidef} and the result formulated in Lemma 4, the error $\configurationModuloErrorVector$ is ultimately bounded, i.e., $\lsShort\module{\configurationModuloErrorVector(t)}\leq\controlError(\positionErrorUltimateBound,\auxiliaryOrientationUltimateBound)$ where
  \begin{align}
    \controlError(\positionErrorUltimateBound,\auxiliaryOrientationUltimateBound) = \sqrt{2}\positionErrorUltimateBound+2\auxiliaryOrientationUltimateBound+r_o^\infty,
    \label{eq:lem5}
  \end{align}
  while $\positionErrorUltimateBound=\positionErrorUltimateBound(\vfoLongitudinalGain,\longitudinalCompensationCoefficient,\observationErrorUltimateBound,\velocityErrorUltimateBound,\auxiliaryOrientationUltimateBound)$ has been introduced in Lemma 4. Taking into account the arguments of functions $\positionErrorUltimateBound$ and $\auxiliaryOrientationUltimateBound$, the bounding function $\controlError(\positionErrorUltimateBound,\auxiliaryOrientationUltimateBound)=\controlError(\observerBandwidth,\adrcGainMatrixLocal,\vfoOrientationGainMatrix,\vfoLongitudinalGain,\angularCompensationCoefficient,\longitudinalCompensationCoefficient,\observationErrorDisturbanceTerminalBound,\velocityErrorModifiedDisturbanceTerminalBound)$, and it cannot be made arbitrarily small.
  %
\end{lemma}

According to Lemmas \ref{lem:1}-\ref{lem:5}, the control objectives \eqref{eq:controlobjective}-\eqref{eq:controlobjective2} are satisfied, leading to the conclusions formulated in Proposition \ref{prop:1}.

\subsection{Proofs of Lemmas 1-5}
\renewcommand*{\proofname}{Proof of Lemma 1}

\begin{proof}
  According to Remark \ref{rem:dbound}, admissible domains of the observation error \eqref{eq:ObsErrorVect} and the perturbing input of system \eqref{eq:sigma1} can be defined as
  \begin{align}
    \observationErrorAggregatedVector\in\observationErrorDomain, \ \totalDisturbanceVectorDerivative\in\observationErrorDisturbanceDomain, \label{eq:chidomains}
  \end{align}
  where $\observationErrorDomain\triangleq\{\observationErrorAggregatedVector\in\realNumbers^{18}:\module{\observationErrorAggregatedVector}<\observationErrorPrimaryBound\}$, $\observationErrorDisturbanceDomain\triangleq\{\totalDisturbanceVectorDerivative\in\realNumbers^6:\module{\totalDisturbanceVectorDerivative}<\observationErrorDisturbancePrimaryBound\}$, and $\observationErrorPrimaryBound, \ \observationErrorDisturbancePrimaryBound > 0$. \color{black} In order to show the ultimate boundedness of $\observationErrorAggregatedVector(t)$
  , we follow the idea utilized for a SISO system in \cite{KHAL08}, and introduce the transformation
  \begin{align}
    \observationErrorAggregatedVector \triangleq \observationErrorTransformationMatrix\transformedObservationError:  \transformedObservationErrorDomain\rightarrow\observationErrorDomain,
    \label{eq:observationErrorTransformation}
  \end{align}
  where $\observationErrorTransformationMatrix \triangleq \blkdiag\{\transformedObservationErrorGainMatrixSingle^{-2},\transformedObservationErrorGainMatrixSingle^{-1},\identityMatrix\}$ for $\transformedObservationErrorGainMatrixSingle \triangleq \diag\{\particularObserverBandwidth{1},...,\particularObserverBandwidth{6}\}\succ0$ and $\transformedObservationErrorDomain\triangleq\{\transformedObservationError\in\realNumbers^{18}:\module{\transformedObservationError}<\module{\observationErrorTransformationMatrix^{-1}}\observationErrorPrimaryBound=:\transformedObservationErrorPrimaryBound\}$. Utilizing transformation \eqref{eq:observationErrorTransformation}, and substituting particular values of observer gains from \eqref{eq:observergainrule} into $\observationErrorStateMatrix$, we can rewrite \eqref{eq:sigma1} as
  \begin{align}
    \transformedObservationErrorDerivative &= -\observationErrorTransformationMatrix^{-1}\observationErrorStateMatrix\observationErrorTransformationMatrix\transformedObservationError+\observationErrorTransformationMatrix^{-1}\observationErrorIdentityMatrix\totalDisturbanceVectorDerivative \nonumber \\
    &= -\transformedObservationErrorGainMatrix\transformedObservationErrorStateMatrix\transformedObservationError+\observationErrorIdentityMatrix\totalDisturbanceVectorDerivative,
    \label{eq:transformedObservationErrorDerivative}
  \end{align}
  where $\transformedObservationErrorGainMatrix = \blkdiag\{\transformedObservationErrorGainMatrixSingle,\transformedObservationErrorGainMatrixSingle,\transformedObservationErrorGainMatrixSingle\}$, while
  \begin{align}
     \transformedObservationErrorStateMatrix = \begin{bmatrix} 3\identityMatrix & -\identityMatrix & \zeroMatrix \\
      3\identityMatrix & \zeroMatrix & -\identityMatrix \\
      \identityMatrix & \zeroMatrix & \zeroMatrix\end{bmatrix}.
  \end{align}

  Let us introduce a positive definite function $\observationErrorLyapunovFunction\triangleq\transformedObservationError^\top\transformedObservationErrorLyapunovEquationSolution\transformedObservationError, \ \observationErrorLyapunovFunction:\transformedObservationErrorDomain\rightarrow\nonNegativeRealNumbers$, such that $\transformedObservationErrorLyapunovEquationSolution=\transformedObservationErrorLyapunovEquationSolution^\top\succ0$ is a solution of Lyapunov equation $\transformedObservationErrorStateMatrix^\top\transformedObservationErrorGainMatrix^\top\transformedObservationErrorLyapunovEquationSolution+\transformedObservationErrorLyapunovEquationSolution\transformedObservationErrorGainMatrix\transformedObservationErrorStateMatrix = \transformedObservationErrorGainMatrix$. Function $\observationErrorLyapunovFunction$ satisfies  $\observationErrorAlphaOne(\module{\transformedObservationError})\leq\observationErrorLyapunovFunction(\transformedObservationError)\leq\observationErrorAlphaTwo(\module{\transformedObservationError})$, for the limiting functions $\observationErrorAlphaOne(\module{\transformedObservationError})\triangleq\minEigenvalue{\transformedObservationErrorLyapunovEquationSolution}\module{\transformedObservationError}^2\in\kappaFunction$, and $\observationErrorAlphaTwo(\module{\transformedObservationError})\triangleq\maxEigenvalue{\transformedObservationErrorLyapunovEquationSolution}\module{\transformedObservationError}^2\in\kappaFunction.$
  A time derivative of $\observationErrorLyapunovFunction$ can be assessed as
  \begin{align}
    \observationErrorLyapunovFunctionDerivative&=\transformedObservationErrorDerivative^\top\transformedObservationErrorLyapunovEquationSolution\transformedObservationError+\transformedObservationError^\top\transformedObservationErrorLyapunovEquationSolution\transformedObservationErrorDerivative  \nonumber \\
    &= -\transformedObservationError^\top(\transformedObservationErrorStateMatrix^\top\transformedObservationErrorGainMatrix^\top\transformedObservationErrorLyapunovEquationSolution+\transformedObservationErrorLyapunovEquationSolution\transformedObservationErrorGainMatrix\transformedObservationErrorStateMatrix)\transformedObservationError + 2\transformedObservationError^\top\transformedObservationErrorLyapunovEquationSolution\observationErrorIdentityMatrix \totalDisturbanceVectorDerivative \nonumber \\
    &=  -\transformedObservationError^\top\transformedObservationErrorGainMatrix\transformedObservationError+2\transformedObservationError^\top\transformedObservationErrorLyapunovEquationSolution\observationErrorIdentityMatrix \totalDisturbanceVectorDerivative \nonumber \\
    &\leq -(1-\observationErrorMajorizationCoefficient)\observerBandwidth\module{\transformedObservationError}^2+ \module{\transformedObservationError}\left(2\maxEigenvalue{\transformedObservationErrorLyapunovEquationSolution}\module{\totalDisturbanceVectorDerivative}-\observationErrorMajorizationCoefficient\observerBandwidth\module{\transformedObservationError}\right), \label{eq:vchi}
  \end{align}
  where $\observationErrorMajorizationCoefficient\in(0,1)$ is a majorization constant and $\observerBandwidth = \min_{i\in\{1, ... ,6\}}\{\particularObserverBandwidth{i}\}$ is the lowest eigenvalue of matrix $\transformedObservationErrorGainMatrix$. According to \eqref{eq:vchi},
  \begin{align}
      \observationErrorLyapunovFunctionDerivative\leq-(1-\observationErrorMajorizationCoefficient)\observerBandwidth\module{\transformedObservationError}^2,
  \end{align}
  for
  \begin{align}
      \module{\transformedObservationError}\geq\observationErrorChiFunction\left(\|{\totalDisturbanceVectorDerivative}\|\right), \label{eq:Vchiogr}
  \end{align}
  where
  \begin{align}
  	\observationErrorChiFunction(\|{\totalDisturbanceVectorDerivative}\|) = \frac{{2\maxEigenvalue{\transformedObservationErrorLyapunovEquationSolution}\|\totalDisturbanceVectorDerivative}\|}{\observationErrorMajorizationCoefficient\observerBandwidth}\in\kappaFunction.
  \end{align}
  One concludes that subsystem \eqref{eq:sigma1} is locally ISS with respect to perturbation $\totalDisturbanceVectorDerivative$, that is
  \begin{align}
    \forall t\geq 0 \ \module{\transformedObservationError(t)}\leq\max\left\{\observationErrorBetaFunction(\|{\transformedObservationError(0)\|},t),\observationErrorGammaFunction\left(\sup_{t\geq0}\|{\totalDisturbanceVectorDerivative(t)}\|\right)\right\},
    \label{eq:ogrZ}
  \end{align}
  for some $\kappaLambdaFunction$-class function $\observationErrorBetaFunction(\cdot,\cdot)$, where
  \begin{align}
  	\observationErrorGammaFunction = \observationErrorAlphaOne^{-1}\left(\observationErrorAlphaTwo\left(\observationErrorChiFunction\left(\|{\totalDisturbanceVectorDerivative(t)\|}\right)\right)\right) = \sqrt{\frac{\maxEigenvalue{\transformedObservationErrorLyapunovEquationSolution}}{\minEigenvalue{\transformedObservationErrorLyapunovEquationSolution}}} \observationErrorChiFunction\left(\|{\totalDisturbanceVectorDerivative(t)}\|\right).
  \end{align}
  According to \eqref{eq:ogrZ} and \eqref{eq:observationErrorTransformation}, the original observation error $\|\observationErrorAggregatedVector(t)\|$ is also bounded.

  The result of local ISS is fulfilled within the domains $\observationErrorDisturbanceDomain$ and $\transformedObservationErrorDomain$ if the initial condition meets the requirement
  $\module{\transformedObservationError(0)}<\transformedObservationErrorTerminalBound = \observationErrorAlphaTwo^{-1}(\observationErrorAlphaOne(\transformedObservationErrorPrimaryBound))=\sqrt{\minEigenvalue{\transformedObservationErrorLyapunovEquationSolution}/\maxEigenvalue{\transformedObservationErrorLyapunovEquationSolution}}\transformedObservationErrorPrimaryBound,$
  and if the perturbation is upper-bounded by
  \begin{align}
  	\sup_{t\geq0}\|{\totalDisturbanceVectorDerivative(t)\|} < \observationErrorChiFunction^{-1}(\min\{\transformedObservationErrorTerminalBound,\observationErrorChiFunction(\observationErrorDisturbancePrimaryBound)\}) =: \observationErrorDisturbanceTerminalBound
  \end{align}
  (see Remark \ref{rem:dbound}). Formula \eqref{eq:ogrZ} implies the asymptotic relation, i.e.,
  \begin{align}
      \lsShort\module{\transformedObservationError(t)}\leq \observationErrorGammaFunction(\lsShort\|{\totalDisturbanceVectorDerivative(t)}\|)
      <\frac{2\lambda_{\textrm{max}}^{3/2}(\transformedObservationErrorLyapunovEquationSolution)}{\lambda_{\textrm{min}}^{1/2}(\transformedObservationErrorLyapunovEquationSolution)\observationErrorMajorizationCoefficient\observerBandwidth}\observationErrorDisturbanceTerminalBound. \label{eq:zinfty}
  \end{align}
  By referring to transformation \eqref{eq:observationErrorTransformation} and result \eqref{eq:zinfty}, the original aggregated observation error satisfies
  \begin{align}
      \lsShort\module{\observationErrorAggregatedVector(t)}&\leq\maxEigenvalue{\observationErrorTransformationMatrix}\lsShort\module{\transformedObservationError(t)} \nonumber \\
      &< \max\left\{\frac{1}{\observerBandwidth^2},1\right\}\frac{2\lambda_{\textrm{max}}^{3/2}(\transformedObservationErrorLyapunovEquationSolution)}{\lambda_{\textrm{min}}^{1/2}(\transformedObservationErrorLyapunovEquationSolution)\observationErrorMajorizationCoefficient\observerBandwidth}\observationErrorDisturbanceTerminalBound =: r_\chi^\infty.
      \label{eq:chiInfty}
  \end{align}
  Now, the conclusions of Lemma~\ref{lem:1} follow.\color{black}
\end{proof}

\renewcommand*{\proofname}{Proof of Lemma 2}
\begin{proof}
  The norm of modified disturbance $\modifiedTotalDisturbanceVector$ is bounded according to Remark \ref{rem:dbound} and the components of vector \eqref{eq:ObsErrorVect} are bounded according to result \eqref{eq:chiInfty}, thus we can assume the existence of the compact domains
  \begin{align}
    \label{eq:velocityTrackingErrorDomains}
    \velocityErrorVector\in\velocityErrorDomain, \ \velocityErrorPerturbationVector\triangleq\left[\totalDisturbanceVectorObservationError^\top \ \velocityErrorVectorObservationError^\top \right]^\top\in\velocityErrorDisturbanceDomain, \ {\modifiedTotalDisturbanceVector}\in\velocityErrorModifiedDisturbanceDomain,
  \end{align}
  where $\velocityErrorDomain\triangleq\{\velocityErrorVector\in\realNumbers^6:\module{\velocityErrorVector}<\velocityErrorPrimaryBound\}$, $\velocityErrorDisturbanceDomain\triangleq\{\velocityErrorPerturbationVector\in\realNumbers^{12}:\module{\velocityErrorPerturbationVector}<\velocityErrorDisturbancePrimaryBound\}$, $\velocityErrorModifiedDisturbanceDomain \triangleq \{\modifiedTotalDisturbanceVector\in\realNumbers^6:\module{\modifiedTotalDisturbanceVector}<\velocityErrorModifiedDisturbancePrimaryBound\}$, and $\velocityErrorPrimaryBound, \ \velocityErrorDisturbancePrimaryBound, \ \velocityErrorModifiedDisturbancePrimaryBound > 0$. Let us propose the positive definite function in the form $\velocityErrorLyapunovFunction\triangleq\frac{1}{2}\velocityErrorVector^\top\velocityErrorVector, \ \velocityErrorLyapunovFunction:\velocityErrorDomain\rightarrow\nonNegativeRealNumbers$, such that $\velocityErrorAlphaOne(\module{\velocityErrorVector})\leq\velocityErrorLyapunovFunction(\velocityErrorVector)\leq\velocityErrorAlphaTwo(\module{\velocityErrorVector})$, for the limiting functions $\velocityErrorAlphaOne(\module{\velocityErrorVector})=\velocityErrorAlphaTwo(\module{\velocityErrorVector})\triangleq\frac{1}{2}\module{\velocityErrorVector}^2\in\kappaFunction.$
  According to the velocity error dynamics \eqref{eq:velocityTrackingErrorDynamics}, the time derivative of $\velocityErrorLyapunovFunction$ can be assessed as
  \begin{align}
    \velocityErrorLyapunovFunctionDerivative &=  \velocityErrorVector^\top\velocityErrorVectorDerivative \nonumber \\
    &= -\velocityErrorVector^\top\velocityTransformationMatrix\velocityErrorStateMatrix\velocityTransformationMatrix^{-1}\velocityErrorVector+\velocityErrorVector^\top\velocityTransformationMatrix\actuationMatrixLocal\adrcGainMatrixLocal\velocityTransformationMatrix^{-1}\velocityErrorVectorObservationError \nonumber\\
    &+\velocityErrorVector^\top\velocityTransformationMatrix\actuationMatrixLocal\velocityTransformationMatrix^{-1}\totalDisturbanceVectorObservationError+\velocityErrorVector^\top\velocityTransformationMatrix(\identityMatrix-\actuationMatrixLocal)\velocityTransformationMatrix^{-1}\modifiedTotalDisturbanceVector.
    \label{eq:velocityErrorLyapunovFunctionDerivative}
  \end{align}
  \color{black}
  According to Proposition \ref{prop:1} and \eqref{eq:58}, we have $\velocityTransformationMatrix\velocityErrorStateMatrix\velocityTransformationMatrix^{-1}=\velocityErrorStateMatrix$, what allows us to assess the upper bound of \eqref{eq:velocityErrorLyapunovFunctionDerivative} as follows \color{black}
  \begin{align}
    \velocityErrorLyapunovFunctionDerivative&\leq -\minEigenvalue{\velocityErrorStateMatrix}\color{black}\module{\velocityErrorVector}^2+\maxEigenvalue{\adrcGainMatrixLocal}\color{black}\module{\velocityErrorVector}\module{\velocityErrorVectorObservationError} \nonumber\\
    &+\module{\velocityErrorVector}\module{\totalDisturbanceVectorObservationError}+\maxEigenvalue{\identityMatrix-\actuationMatrixLocal}\module{\velocityErrorVector}\module{\modifiedTotalDisturbanceVector} \nonumber\\
    &\leq -(1-\velocityErrorMajorizationCoefficient)\minEigenvalue{\velocityErrorStateMatrix}\color{black}\module{\velocityErrorVector}^2 \nonumber\\
    &+ \module{\velocityErrorVector}\Big[\left(1+\maxEigenvalue{\adrcGainMatrixLocal}\color{black}\right)\module{\velocityErrorPerturbationVector}+\maxEigenvalue{\identityMatrix-\actuationMatrixLocal}\module{\modifiedTotalDisturbanceVector} \nonumber\\
    &-\velocityErrorMajorizationCoefficient\minEigenvalue{\velocityErrorStateMatrix}\color{black}\module{\velocityErrorVector}\Big],
    \label{eq:velocityTrackingErrorLyapunovDerivative}
  \end{align}
  \color{black}
  where $\velocityErrorMajorizationCoefficient\in(0,1)$ is a majorization constant.
  %
  %
  According to \eqref{eq:velocityTrackingErrorLyapunovDerivative}, one can write
  \begin{align}
    \label{eq:velocityTrackingErrorLyapunovDerivativeBound}
    &\velocityErrorLyapunovFunctionDerivative\leq-(1-\velocityErrorMajorizationCoefficient)\minEigenvalue{\velocityErrorStateMatrix}\color{black}\module{\velocityErrorVector}^2
  \end{align}
  for
  \begin{align}
    &\module{\velocityErrorVector}\geq\velocityErrorChiFunctionOne(\module{\velocityErrorPerturbationVector
    })+\velocityErrorChiFunctionTwo(\module{\modifiedTotalDisturbanceVector}),
  \end{align}
  where
  \begin{align}
    \velocityErrorChiFunctionOne(\module{\velocityErrorPerturbationVector}) &= \frac{1+\maxEigenvalue{\adrcGainMatrixLocal}}{\velocityErrorMajorizationCoefficient\minEigenvalue{\velocityErrorStateMatrix}}\module{\velocityErrorPerturbationVector}\in\kappaFunction, \\
    \velocityErrorChiFunctionTwo(\module{\modifiedTotalDisturbanceVector}) &=
    \frac{\maxEigenvalue{\identityMatrix-\actuationMatrixLocal}}{\velocityErrorMajorizationCoefficient\minEigenvalue{\velocityErrorStateMatrix}}\module{\modifiedTotalDisturbanceVector}\in\kappaFunction.
  \end{align}
  As a consequence, the velocity error subsystem is locally ISS with respect to perturbations $\velocityErrorPerturbationVector$ and $\modifiedTotalDisturbanceVector$, and
  \begin{align}
    \forall t\geq0 \ \module{\velocityErrorVector(t)}\leq\max\Big\{\velocityErrorBetaFunction(\|\velocityErrorVector(0)\|,t),
    &\velocityErrorGammaFunctionOne\Big(\sup_{t\geq0}\module{\velocityErrorPerturbationVector(t)}\Big) \nonumber \\
    &+\velocityErrorGammaFunctionTwo\Big(\sup_{t\geq0}\module{\modifiedTotalDisturbanceVector(t)}\Big)\Big\},
    \label{eq:celocityTrackingErrorTimeResponse}
  \end{align}
  for some  $\kappaLambdaFunction$-class function $\velocityErrorBetaFunction(\cdot,\cdot)$, and functions
  \begin{align}
    \velocityErrorGammaFunctionOne(\module{\velocityErrorPerturbationVector}) &=\velocityErrorAlphaOne^{-1}(\velocityErrorAlphaTwo(\velocityErrorChiFunctionOne(\module{\velocityErrorPerturbationVector})))=\velocityErrorChiFunctionOne(\module{\velocityErrorPerturbationVector}) \\
    \velocityErrorGammaFunctionTwo(\module{\modifiedTotalDisturbanceVector}) &=\velocityErrorAlphaOne^{-1}(\velocityErrorAlphaTwo(\velocityErrorChiFunctionTwo(\module{\modifiedTotalDisturbanceVector})))=\velocityErrorChiFunctionTwo(\module{\modifiedTotalDisturbanceVector}).
  \end{align}
  The result of local ISS is valid within the domains \eqref{eq:velocityTrackingErrorDomains} if the initial condition meets the requirement $\module{\velocityErrorVector(0)}<\velocityErrorTerminalBound=\velocityErrorAlphaTwo^{-1}(\velocityErrorAlphaOne(\velocityErrorPrimaryBound)) = \velocityErrorPrimaryBound,$
  and if the perturbations are upper-bounded by
  \begin{align}
    \sup_{t\geq0}\module{\velocityErrorPerturbationVector(t)}&<\velocityErrorChiFunctionOne^{-1}(\min\{\velocityErrorTerminalBound,\velocityErrorChiFunctionOne(\velocityErrorDisturbancePrimaryBound)\}) =: \velocityErrorDisturbanceTerminalBound \\
    \sup_{t\geq0}\module{\modifiedTotalDisturbanceVector(t)}&<\velocityErrorChiFunctionTwo^{-1}(\min\{\velocityErrorTerminalBound,\velocityErrorChiFunctionTwo(\velocityErrorModifiedDisturbancePrimaryBound)\}) =: \velocityErrorModifiedDisturbanceTerminalBound. \label{eq:modifiedConvergenceVectorFieldVector}
  \end{align}
  %
  Formula \eqref{eq:celocityTrackingErrorTimeResponse} implies the asymptotic gain property
  \begin{align}
    \lsShort&\module{\velocityErrorVector(t)} < \velocityErrorGammaFunctionOne(\lsShort\module{\velocityErrorPerturbationVector(t)})+\velocityErrorGammaFunctionTwo(\lsShort\module{\modifiedTotalDisturbanceVector(t)}) \nonumber \\
    &\leq \velocityErrorGammaFunctionOne(\lsShort\module{\observationErrorAggregatedVector(t)})+\velocityErrorGammaFunctionTwo(\lsShort\module{\modifiedTotalDisturbanceVector(t)}) \nonumber \\
    &\leq \frac{1+\maxEigenvalue{\adrcGainMatrixLocal}}{\velocityErrorMajorizationCoefficient\minEigenvalue{\velocityErrorStateMatrix}}\color{black} r_{\chi}^\infty
    \color{black}+ \frac{\maxEigenvalue{\identityMatrix-\actuationMatrixLocal}}{\velocityErrorMajorizationCoefficient\minEigenvalue{\velocityErrorStateMatrix}}\velocityErrorModifiedDisturbanceTerminalBound=:\velocityErrorUltimateBound,
    \label{eq:epsilonsup}
  \end{align}
  which reveals the ultimate boundedness of the velocity error $\|\velocityErrorVector(t)\|$. Result \eqref{eq:epsilonsup} is consistent with \eqref{eq:lem2} and confirms the statements from Lemma~\ref{lem:2}.
  %
\end{proof}

\renewcommand*{\proofname}{Proof of Lemma 3}
\begin{proof}
  The admissible domains for an auxiliary orientation error itself and its perturbing inputs, justified by results \eqref{eq:chiInfty} and \eqref{eq:epsilonsup}, are defined as
  \begin{align}
    \auxiliaryOrientationError\in\auxiliaryOrientationDomain, \ \reducedAngularVelocityErrorVectorObservationError \in \auxiliaryOrientationObservationErrorDomain, \ \reducedAngularVelocityErrorVector \in \auxiliaryOrientationVelocityErrorDomain, \ \reducedAuxiliaryOrientationDerivativeObservationError \in\auxiliaryOrientationFFDomain,
    \label{eq:auxiliaryOrientationErrorDomains}
  \end{align}
  where   $\auxiliaryOrientationDomain\triangleq\{\auxiliaryOrientationError\in[-\pi,\pi)\times\realNumbers:\module{\auxiliaryOrientationError}<\auxiliaryOrientationPrimaryBound\}$,  $\auxiliaryOrientationDisturbanceDomain\triangleq\{\reducedAngularVelocityErrorVectorObservationError\in\realNumbers^2:\module{\reducedAngularVelocityErrorVectorObservationError}<\auxiliaryOrientationObservationPrimaryBound\}$, $\auxiliaryOrientationVelocityErrorDomain \triangleq \{ \reducedAngularVelocityErrorVector\in\realNumbers^2:\module{\reducedAngularVelocityErrorVector}<\auxiliaryOrientationVelocityErrorPrimaryBound\}$, $\auxiliaryOrientationFFDomain\triangleq\{\reducedAuxiliaryOrientationDerivativeObservationError\in\realNumbers^2:\|{\reducedAuxiliaryOrientationDerivativeObservationError}\|<\auxiliaryOrientationFFPrimaryBound\}$, and $\auxiliaryOrientationPrimaryBound, \ \auxiliaryOrientationDisturbancePrimaryBound,$ $ \auxiliaryOrientationVelocityErrorPrimaryBound, \  \auxiliaryOrientationFFPrimaryBound > 0$.
  A positive definite function $\auxiliaryOrientationLyapunovFunction:\auxiliaryOrientationDomain\rightarrow\nonNegativeRealNumbers$ is proposed in the form $\auxiliaryOrientationLyapunovFunction\triangleq\frac{1}{2}\auxiliaryOrientationError^\top\auxiliaryOrientationError$,
  such that $\auxiliaryOrientationAlphaOne(\module{\auxiliaryOrientationError})\leq\auxiliaryOrientationLyapunovFunction(\auxiliaryOrientationError)\leq\auxiliaryOrientationAlphaTwo(\module{\auxiliaryOrientationError})$ for the limiting functions
  $\auxiliaryOrientationAlphaOne(\module{\auxiliaryOrientationError})=\auxiliaryOrientationAlphaTwo(\module{\auxiliaryOrientationError})\triangleq\frac{1}{2}\module{\auxiliaryOrientationError} \in \kappaFunction.$
  The time derivative of function $\auxiliaryOrientationLyapunovFunction$ can be assessed as follows:
  \begin{align}
    \auxiliaryOrientationLyapunovFunctionDerivative&=\auxiliaryOrientationError^\top\auxiliaryOrientationErrorDerivative \nonumber \\
    &\eq{eq:auxiliaryOrientationErrorDynamics} -\auxiliaryOrientationError^\top\vfoOrientationGainMatrix\auxiliaryOrientationError+\auxiliaryOrientationError^\top\angularCompensationCoefficient\reducedAngularVelocityErrorVectorObservationError+\auxiliaryOrientationError^\top(1-\angularCompensationCoefficient)\reducedAngularVelocityErrorVector+\auxiliaryOrientationError^\top\reducedAuxiliaryOrientationDerivativeObservationError \nonumber \\
    &\leq -\minEigenvalue{\vfoOrientationGainMatrix}\module{\auxiliaryOrientationError}^2+\module{\auxiliaryOrientationError}\Big[ \angularCompensationCoefficient\module{\reducedAngularVelocityErrorVectorObservationError} \nonumber \\
    &+(1-\angularCompensationCoefficient)\module{\reducedAngularVelocityErrorVector}+\|{\reducedAuxiliaryOrientationDerivativeObservationError}\|
    \Big] \nonumber \\
    &=  -(1-\auxiliaryOrientationMajorizationCoefficient)\minEigenvalue{\vfoOrientationGainMatrix}\module{\auxiliaryOrientationError}^2+\module{\auxiliaryOrientationError}\Big[ \angularCompensationCoefficient\module{\reducedAngularVelocityErrorVectorObservationError} \nonumber \\
    &+(1-\angularCompensationCoefficient)\module{\reducedAngularVelocityErrorVector}+\|{\reducedAuxiliaryOrientationDerivativeObservationError}\|-\auxiliaryOrientationMajorizationCoefficient\minEigenvalue{\vfoOrientationGainMatrix}\module{\auxiliaryOrientationError}
    \Big],
    \label{eq:Va2}
  \end{align}
  where $\auxiliaryOrientationMajorizationCoefficient \in (0,1)$ is a majorization factor. According to  \eqref{eq:Va2}, one concludes
  \begin{align}
    \label{ea:Vaogr}
    &\auxiliaryOrientationLyapunovFunctionDerivative \leq -(1-\auxiliaryOrientationMajorizationCoefficient)\minEigenvalue{\vfoOrientationGainMatrix}\module{\auxiliaryOrientationError}^2
  \end{align}
  for
  \begin{align}
     &\module{\auxiliaryOrientationError}\geq\auxiliaryOrientationChiFunctionOne(\module{\reducedAngularVelocityErrorVectorObservationError})+\auxiliaryOrientationChiFunctionTwo(\module{\reducedAngularVelocityErrorVector})+\auxiliaryOrientationChiFunctionThree\left(\module{\reducedAuxiliaryOrientationDerivativeObservationError}\right),
  \end{align}
  where
  \begin{align}
    \auxiliaryOrientationChiFunctionOne(\module{\reducedAngularVelocityErrorVectorObservationError}) &= \frac{\angularCompensationCoefficient}{\auxiliaryOrientationMajorizationCoefficient\minEigenvalue{\vfoOrientationGainMatrix}}\module{\reducedAngularVelocityErrorVectorObservationError} \in\kappaFunction, \\
    \auxiliaryOrientationChiFunctionTwo(\module{\reducedAngularVelocityErrorVector}) &= \frac{(1-\angularCompensationCoefficient)}{\auxiliaryOrientationMajorizationCoefficient\minEigenvalue{\vfoOrientationGainMatrix}}\module{\reducedAngularVelocityErrorVector}\in\kappaFunction, \\
    \auxiliaryOrientationChiFunctionThree(\module{\reducedAuxiliaryOrientationDerivativeObservationError}) &= \frac{1}{\auxiliaryOrientationMajorizationCoefficient\minEigenvalue{\vfoOrientationGainMatrix}}\module{\reducedAuxiliaryOrientationDerivativeObservationError}\in\kappaFunction.
  	\label{eq:201}
  \end{align}
  Thus, subsystem describing dynamics of error $\auxiliaryOrientationError$ is locally ISS with respect to the inputs $\reducedAngularVelocityErrorVectorObservationError, \  \reducedAngularVelocityErrorVector \ \textrm{and} \  \reducedAuxiliaryOrientationDerivativeObservationError$. It implies
  \begin{align}
    \forall t\geq 0 \ \module{\auxiliaryOrientationError(t)}&\leq\max\Big\{\auxiliaryOrientationBetaFunction(\|\auxiliaryOrientationError(0)\|,t),\auxiliaryOrientationGammaFunctionOne\Big(\sup_{t\geq0}\|{\reducedAngularVelocityErrorVectorObservationError(t)}\|\Big) \nonumber \\
    &+\auxiliaryOrientationGammaFunctionTwo\Big(\sup_{t\geq0}\|{\reducedAngularVelocityErrorVector(t)}\|\Big)+\auxiliaryOrientationGammaFunctionThree\Big(\sup_{t\geq0}\|{\reducedAuxiliaryOrientationDerivativeObservationError(t)}\|\Big)\Big\}
    \label{eq:ogrea}
  \end{align}
  for some function $\auxiliaryOrientationBetaFunction(\cdot,\cdot)$ of class $\kappaLambdaFunction$, and for
  \begin{align}
    \auxiliaryOrientationGammaFunctionOne(\module{\reducedAngularVelocityErrorVectorObservationError})&=\auxiliaryOrientationAlphaOne^{-1}(\auxiliaryOrientationAlphaTwo(\auxiliaryOrientationChiFunctionOne(\module{\reducedAngularVelocityErrorVectorObservationError(t)})))=\auxiliaryOrientationChiFunctionOne(\module{\reducedAngularVelocityErrorVectorObservationError(t)}), \\
    \auxiliaryOrientationGammaFunctionTwo(\module{\reducedAngularVelocityErrorVector})&=\auxiliaryOrientationAlphaOne^{-1}(\auxiliaryOrientationAlphaTwo(\auxiliaryOrientationChiFunctionTwo(\module{\reducedAngularVelocityErrorVector(t)})))=\auxiliaryOrientationChiFunctionTwo(\module{\reducedAngularVelocityErrorVector(t)}), \\
    \auxiliaryOrientationGammaFunctionThree(\module{\reducedAuxiliaryOrientationDerivativeObservationError})&=\auxiliaryOrientationAlphaOne^{-1}(\auxiliaryOrientationAlphaTwo(\auxiliaryOrientationChiFunctionThree(\module{\reducedAuxiliaryOrientationDerivativeObservationError(t)})))=\auxiliaryOrientationChiFunctionThree(\module{\reducedAuxiliaryOrientationDerivativeObservationError(t)}).
  \end{align}
  The local ISS result is valid within the domains defined in \eqref{eq:auxiliaryOrientationErrorDomains} if the initial auxiliary orientation error satisfies the inequality
  \begin{align}
    \module{\auxiliaryOrientationError(0)}<\auxiliaryOrientationTerminalBound=\auxiliaryOrientationAlphaTwo^{-1}(\auxiliaryOrientationAlphaOne(\auxiliaryOrientationPrimaryBound)) = \auxiliaryOrientationPrimaryBound
  \end{align}
  and if the norms of disturbing vectors are upper bounded by
  \begin{align}
    \sup_{t\geq0}\module{\reducedAngularVelocityErrorVectorObservationError(t)}<\auxiliaryOrientationObservationTerminalBound = \auxiliaryOrientationChiFunctionOne^{-1}(\min\{\auxiliaryOrientationTerminalBound,\auxiliaryOrientationChiFunctionOne(\auxiliaryOrientationObservationPrimaryBound)\}) \\
    \sup_{t\geq0}\module{\reducedAngularVelocityErrorVector(t)}<\auxiliaryOrientationVelocityErrorTerminalBound = \auxiliaryOrientationChiFunctionTwo^{-1}(\min\{\auxiliaryOrientationTerminalBound,\auxiliaryOrientationChiFunctionTwo(\auxiliaryOrientationVelocityErrorPrimaryBound)\}) \\
    \sup_{t\geq0}\module{\reducedAuxiliaryOrientationDerivativeObservationError(t)}<\auxiliaryOrientationFFTerminalBound = \auxiliaryOrientationChiFunctionThree^{-1}(\min\{\auxiliaryOrientationTerminalBound,\auxiliaryOrientationChiFunctionThree(\auxiliaryOrientationFFPrimaryBound)\}).
  \end{align}
  Inequality \eqref{eq:ogrea} implies the asymptotic relation corresponding to \eqref{eq:lem3}, i.e.,
  \begin{align}
    \lsShort&\module{\auxiliaryOrientationError(t)}<\auxiliaryOrientationGammaFunctionOne(\lsShort\module{\reducedAngularVelocityErrorVectorObservationError(t)}) \nonumber\\
    &+\auxiliaryOrientationGammaFunctionTwo(\lsShort\module{\reducedAngularVelocityErrorVector(t)})+\auxiliaryOrientationGammaFunctionThree(\lsShort\module{\reducedAuxiliaryOrientationDerivativeObservationError(t)}) \nonumber \\
    &\leq \auxiliaryOrientationGammaFunctionOne(\lsShort\module{\observationErrorAggregatedVector(t)})+\auxiliaryOrientationGammaFunctionTwo(\lsShort\module{\velocityErrorVector(t)}) \nonumber \\
    &+\auxiliaryOrientationGammaFunctionThree(\sup_{t\geq0}(\abs{\auxiliaryYawDerivativeErrorFunction(t)}\abs{\longitudinalConvergenceVectorFieldEstimateDerivativeErrorAuxiliaryFunction(t)}+\abs{\auxiliaryPitchDerivativeErrorFunction(t)}\abs{\longitudinalConvergenceVectorFieldEstimateDerivativeErrorAuxiliaryFunction(t)})\lsShort\module{\observationErrorAggregatedVector(t)}) \nonumber \\
    &\leq \frac{\angularCompensationCoefficient}{\auxiliaryOrientationMajorizationCoefficient\minEigenvalue{\vfoOrientationGainMatrix}}\color{black}r_\chi^\infty \color{black}
    + \frac{1-\angularCompensationCoefficient}{\auxiliaryOrientationMajorizationCoefficient\minEigenvalue{\vfoOrientationGainMatrix}}\Bigg[\frac{1+\maxEigenvalue{\adrcGainMatrixLocal}}{\velocityErrorMajorizationCoefficient\minEigenvalue{\velocityErrorStateMatrix}}\color{black}r_\chi^\infty \color{black} \nonumber \\
    &+ \frac{\maxEigenvalue{\identityMatrix-\actuationMatrixLocal}}{\velocityErrorMajorizationCoefficient\minEigenvalue{\velocityErrorStateMatrix}}\velocityErrorModifiedDisturbanceTerminalBound\Bigg] + \frac{\bar{f}}{\auxiliaryOrientationMajorizationCoefficient\minEigenvalue{\vfoOrientationGainMatrix}}\color{black}r_\chi^\infty
    \nonumber \\
    \color{black} &= \frac{\angularCompensationCoefficient+\bar{f}}{\auxiliaryOrientationMajorizationCoefficient\minEigenvalue{\vfoOrientationGainMatrix}}\color{black}r_\chi^\infty \color{black}
    + \frac{1-\angularCompensationCoefficient}{\auxiliaryOrientationMajorizationCoefficient\minEigenvalue{\vfoOrientationGainMatrix}}\velocityErrorUltimateBound
     =: \auxiliaryOrientationUltimateBound,
    \label{eq:AuxiliaryOrientationUltimateBound}
  \end{align}
  \color{black}
  where $\bar{f} = \sup_{t\geq0}(\abs{\auxiliaryYawDerivativeErrorFunction(t)}\abs{\longitudinalConvergenceVectorFieldEstimateDerivativeErrorAuxiliaryFunction(t)}+\abs{\auxiliaryPitchDerivativeErrorFunction(t)}\abs{\longitudinalConvergenceVectorFieldEstimateDerivativeErrorAuxiliaryFunction(t)})$\color{black}, while functions $\abs{\longitudinalConvergenceVectorFieldEstimateDerivativeErrorAuxiliaryFunction(t)}, \ \abs{\auxiliaryPitchDerivativeErrorFunction(t)} \ \textrm{and} \ \abs{\auxiliaryYawDerivativeErrorFunction(t)}$ result from the analysis presented in \ref{app:reducedAuxiliaryOrientationDerivativeObservationError}.
  Upon \eqref{eq:AuxiliaryOrientationUltimateBound}, the conclusions stated in Lemma \ref{lem:3} follow. 
\end{proof}

\renewcommand*{\proofname}{Proof of Lemma 4}
\begin{proof}
  The predefined domain for the position error is determined by
  \begin{align}
    \positionError\in\positionErrorDomain,
    \label{eq:positionErrorDomain}
  \end{align}
  where $\positionErrorDomain\triangleq\{\positionError\in\realNumbers^2:\module{\positionError}<\positionErrorPrimaryBound\}$ for $\positionErrorPrimaryBound>0$.
  Upon results \eqref{eq:chiInfty}, \eqref{eq:epsilonsup}, and \eqref{eq:AuxiliaryOrientationUltimateBound}, we can define the domains related to the perturbing vectors as
  \begin{align}
    \label{eq:positionErrorPerturbationDomain1}
    \positionErrorPerturbationObservationErrorVector\in\positionErrorPerturbationObservationErrorDomain, \ \positionErrorPerturbationVelocityErrorVector\in\positionErrorPerturbationVelocityErrorDomain,
    \ \auxiliaryOrientationError\in\auxiliaryOrientationDomain,
  \end{align}
  where
  $\positionErrorPerturbationObservationErrorDomain \triangleq \{\positionErrorPerturbationObservationErrorVector\in[-\pi,\pi)\times\realNumbers^4:\module{\positionErrorPerturbationObservationErrorVector}<\positionErrorPerturbationObservationErrorPrimaryBound\}$,
  $\positionErrorPerturbationVelocityErrorDomain \triangleq \{\positionErrorPerturbationVelocityErrorVector\in[-\pi,\pi)\times\realNumbers^4:\module{\positionErrorPerturbationVelocityErrorVector}<\positionErrorPerturbationVelocityErrorPrimaryBound\}$,
  $\auxiliaryOrientationDomain \triangleq \{ \auxiliaryOrientationError\in[-\pi,\pi)\times\realNumbers:\module{\auxiliaryOrientationError}<\auxiliaryOrientationPrimaryBound\}$, and $\positionErrorPerturbationObservationErrorPrimaryBound, \ \positionErrorPerturbationVelocityErrorPrimaryBound,$ $ \auxiliaryOrientationPrimaryBound > 0$. Vectors assigned to the particular domains in \eqref{eq:positionErrorPerturbationDomain1} are bounded upon \eqref{eq:zinfty}, \eqref{eq:epsilonsup}, and \eqref{eq:AuxiliaryOrientationUltimateBound}. Let us propose the positive definite  function
  \begin{align}
    \positionErrorLyapunovFunction&\triangleq 2\int_{\pmb{0}}^{\positionError}\begin{bmatrix} \frac{\levelSurface{1}}{\module{\levelSurfaceGradient{1}}} \nonumber \\ \frac{\levelSurface{2}}{\module{\levelSurfaceGradient{2}}} \end{bmatrix}^\top d\positionError \\  &=2\int_{\pmb{0}}^{\positionError}\left(\frac{\levelSurface{1}}{\module{\levelSurfaceGradient{1}}}d\levelSurface{1}+\frac{\levelSurface{2}}{\module{\levelSurfaceGradient{2}}}d\levelSurface{2}\right).
    \label{eq:Vp}
  \end{align}
  Under Assumption \ref{ass:3}, there exists a lower-bound of the $\positionErrorLyapunovFunction$ function, i.e.,
  \begin{align}
    &\int_0^{\levelSurface{1}}\frac{\lambda}{\module{\levelSurfaceGradient{1}(\lambda)}}d\lambda+\int_0^{\levelSurface{2}}\frac{\lambda}{\module{\levelSurfaceGradient{2}(\lambda)}}d\lambda \nonumber \\
    &>\int_0^{\levelSurface{1}}\frac{\lambda}{\levelSurfaceGradientUpperBound{1}}d\lambda+\int_0^{\levelSurface{2}}\frac{\lambda}{\levelSurfaceGradientUpperBound{2}}d\lambda= \frac{1}{2\levelSurfaceGradientUpperBound{1}}\levelSurface{1}^2+\frac{1}{2\levelSurfaceGradientUpperBound{2}}\levelSurface{2}^2,
  \end{align}
  implying that the function $\positionErrorLyapunovFunction : \positionErrorDomain\rightarrow\nonNegativeRealNumbers$ satisfies the requirement of being positive definite always, except  $\positionErrorLyapunovFunction=0$ for $\positionError=\zeroMatrix$.
  By referring to Assumption \ref{ass:3}, one may verify that to satisfy the inequality $\positionErrorAlphaOne(\module{\positionError})\leq \positionErrorLyapunovFunction(\positionError)\leq\positionErrorAlphaTwo(\module{\positionError})$, the limiting functions $\positionErrorAlphaOne, \ \positionErrorAlphaTwo$ can be defined as
  \begin{align}
    \label{eq:alphaPosition1}
    \positionErrorAlphaOne(\module{\positionError})&\triangleq \sqrt{\frac{\levelSurfaceGradientUpperBound{1}^2+\levelSurfaceGradientUpperBound{2}^2}{\levelSurfaceGradientUpperBound{1}^2\levelSurfaceGradientUpperBound{2}^2}}\module{\positionError}^2 \in\kappaFunction, \\
    \label{eq:alphaPosition2}
    \positionErrorAlphaTwo(\module{\positionError})&\triangleq\sqrt{\frac{\levelSurfaceGradientLowerBound{1}^2+\levelSurfaceGradientLowerBound{2}^2}{\levelSurfaceGradientLowerBound{1}^2\levelSurfaceGradientLowerBound{2}^2}}\module{\positionError}^2\in\kappaFunction.
  \end{align}
  The upper bound of $\positionErrorLyapunovFunctionDerivative$ has been derived in  \ref{app:A}, and takes the form
  \begin{align}
      \positionErrorLyapunovFunctionDerivative&=\frac{1}{\module{\levelSurfaceGradient{1}}}\levelSurface{1}\levelSurfaceDerivative{1}+\frac{1}{\module{\levelSurfaceGradient{2}}}\levelSurface{2}\levelSurfaceDerivative{2} \nonumber \\
       &\leq  -\vfoLongitudinalGain(1-\positionErrorMajorizationCoefficient)\positionError^\top\minEigenvalue{\positionErrorStateMatrix}\positionError+\module{\positionError}\Big[2(1-\longitudinalCompensationCoefficient)\module{\longitudinalVelocityErrorVector} \nonumber \\
       &+2\longitudinalCompensationCoefficient\module{\longitudinalVelocityErrorVectorObservationError}+2\sqrt{7}\abs{\pathVelocityProfile}\module{\auxiliaryOrientationError}+2\sqrt{7}\longitudinalCompensationCoefficient(\module{\longitudinalVelocityErrorVector}+\module{\longitudinalVelocityErrorVectorObservationError})\module{\auxiliaryOrientationError} \nonumber \\
       &-\vfoLongitudinalGain\module{\positionError}(\minEigenvalue{\positionErrorStateMatrix}\positionErrorMajorizationCoefficient-2\sqrt{7}\module{\auxiliaryOrientationError})\Big],
       \label{eq:positionErrorLyapunovFunctionDerivative}
  \end{align}
  where $\positionErrorMajorizationCoefficient\in(0,1)$ is a majorization constant, $\positionErrorStateMatrix= [1 \ \cos\angleBetweenlevelSurfaceGradients; \  \cos\angleBetweenlevelSurfaceGradients \ 1]$ is a positive-definite matrix  (see Assumption \ref{ass:nonCollinearity}), and $\minEigenvalue{\positionErrorStateMatrix}=(1-\abs{\cos\angleBetweenlevelSurfaceGradients})$. According to the  inequality \eqref{eq:positionErrorLyapunovFunctionDerivative}, one can write
  \begin{align}
    \label{eq:Vsogr}
      &\positionErrorLyapunovFunctionDerivative\leq-\vfoLongitudinalGain\minEigenvalue{\positionErrorStateMatrix}\module{\positionError}^2
    \end{align}
    for
    \begin{align}
       &\module{\positionError}\geq\positionErrorChiFunctionOne(\module{\positionErrorPerturbationVelocityErrorVector})+\positionErrorChiFunctionTwo(\module{\positionErrorPerturbationObservationErrorVector})+\positionErrorChiFunctionThree(\module{\auxiliaryOrientationError}),
  \end{align}
  where the functions
  \begin{align}
    \label{eq:positionErrorChiFunctionOne}
    \positionErrorChiFunctionOne(\module{\positionErrorPerturbationVelocityErrorVector}) &= \frac{2(1-\longitudinalCompensationCoefficient)\module{\positionErrorPerturbationVelocityErrorVector}+2\sqrt{7}\longitudinalCompensationCoefficient\module{\positionErrorPerturbationVelocityErrorVector}^2}{\vfoLongitudinalGain(\minEigenvalue{\positionErrorStateMatrix}\positionErrorMajorizationCoefficient-2\sqrt{7}\module{\positionErrorPerturbationVelocityErrorVector})}, \\
    \label{eq:positionErrorChiFunctionTwo}
    \positionErrorChiFunctionTwo(\module{\positionErrorPerturbationObservationErrorVector})&= \frac{2\longitudinalCompensationCoefficient\module{\positionErrorPerturbationObservationErrorVector}+2\sqrt{7}\longitudinalCompensationCoefficient\module{\positionErrorPerturbationObservationErrorVector}^2}{\vfoLongitudinalGain(\minEigenvalue{\positionErrorStateMatrix}\positionErrorMajorizationCoefficient-2\sqrt{7}\module{\positionErrorPerturbationObservationErrorVector})}, \\
    \label{eq:positionErrorChiFunctionThree}
    \positionErrorChiFunctionThree(\module{\auxiliaryOrientationError})&=\frac{2\sqrt{7}\abs{\pathVelocityProfile}\module{\auxiliaryOrientationError}}{\vfoLongitudinalGain(\minEigenvalue{\positionErrorStateMatrix}\positionErrorMajorizationCoefficient-2\sqrt{7}\module{\auxiliaryOrientationError})},
  \end{align}
  are locally functions of class $\kappaFunction$ with respect to their arguments for $\module{\positionErrorPerturbationVelocityErrorVector}<\positionErrorMajorizationCoefficient\minEigenvalue{\positionErrorStateMatrix}/2\sqrt{7}$, $\module{\positionErrorPerturbationObservationErrorVector}<\positionErrorMajorizationCoefficient\minEigenvalue{\positionErrorStateMatrix}/2\sqrt{7}$, and $\module{\auxiliaryOrientationError}<\positionErrorMajorizationCoefficient\minEigenvalue{\positionErrorStateMatrix}/2\sqrt{7}$. The latter constraints influence the values of constants $\positionErrorPerturbationVelocityErrorPrimaryBound$, $\positionErrorPerturbationObservationErrorPrimaryBound$, and $\auxiliaryOrientationPrimaryBound$,  used in definitions of the domains $\positionErrorPerturbationVelocityErrorDomain$, $\positionErrorPerturbationObservationErrorDomain$ and $\auxiliaryOrientationDomain$ introduced in \eqref{eq:positionErrorPerturbationDomain1}.

  According to \eqref{eq:Vsogr}, the subsystem describing the position error dynamics is locally ISS with respect to the disturbing inputs $\positionErrorPerturbationVelocityErrorVector$, $\positionErrorPerturbationObservationErrorVector$ and $\auxiliaryOrientationError$, that is,
  \begin{align}
    \forall t\geq 0 \ \module{\positionError(t)}&\leq\max\Big\{\positionErrorBetaFunction(\module{\positionError(0)},t),\positionErrorGammaFunctionOne\Big(\sup_{t\geq0}\module{\positionErrorPerturbationObservationErrorVector(t)}\Big) \nonumber \\
    &+\positionErrorGammaFunctionTwo\Big(\sup_{t\geq0}\module{\positionErrorPerturbationVelocityErrorVector(t)}\Big)+\positionErrorGammaFunctionThree\Big(\sup_{t\geq0}\module{\auxiliaryOrientationError(t)}\Big)\Big\} \
  \end{align}
  for some $\positionErrorBetaFunction(\cdot,\cdot)$ function of class $\kappaLambdaFunction$, and for
  \begin{align}
    \positionErrorGammaFunctionOne(\module{\positionErrorPerturbationVelocityErrorVector}) &= \positionErrorAlphaOne^{-1}(\positionErrorAlphaTwo(\positionErrorChiFunctionOne(\module{\positionErrorPerturbationVelocityErrorVector(t)}))) = m^*\positionErrorChiFunctionOne(\module{\positionErrorPerturbationVelocityErrorVector(t)})
    \label{eq:positionErrorGammaFunctionOne}, \\
    \positionErrorGammaFunctionTwo(\module{\positionErrorPerturbationObservationErrorVector}) &= \positionErrorAlphaOne^{-1}(\positionErrorAlphaTwo(\positionErrorChiFunctionTwo(\module{\positionErrorPerturbationObservationErrorVector(t)})))=m^*\positionErrorChiFunctionTwo(\module{\positionErrorPerturbationObservationErrorVector(t)})
    \label{eq:positionErrorGammaFunctionTwo}, \\
    \positionErrorGammaFunctionThree(\module{\auxiliaryOrientationError}) &= \positionErrorAlphaOne^{-1}(\positionErrorAlphaTwo(\positionErrorChiFunctionThree(\module{\auxiliaryOrientationError(t)}))) =m^*\positionErrorChiFunctionThree(\module{\auxiliaryOrientationError(t)}),
    \label{eq:positionErrorGammaFunctionThree}
  \end{align}
  where
  \begin{align}
  m^* =   \frac{\sqrt{2}}{2}\sqrt[4]{\frac{\levelSurfaceGradientUpperBound{1}^2\levelSurfaceGradientUpperBound{2}^2}{\levelSurfaceGradientUpperBound{1}^2+\levelSurfaceGradientUpperBound{2}^2}}\sqrt{\frac{\levelSurfaceGradientLowerBound{1}^2+\levelSurfaceGradientLowerBound{2}^2}{\levelSurfaceGradientLowerBound{1}^2\levelSurfaceGradientLowerBound{2}^2}}.
  \end{align}
  The ISS result is valid within the domains defined in \eqref{eq:positionErrorDomain} if the initial positional error satisfies
  \begin{align}
    \module{\positionError(0)}&<\positionErrorTerminalBound=\positionErrorAlphaTwo^{-1}(\positionErrorAlphaOne(\positionErrorPrimaryBound)) \nonumber \\
    &=\frac{\sqrt{2}}{2}\sqrt[4]{\frac{\levelSurfaceGradientLowerBound{1}^2\levelSurfaceGradientLowerBound{2}^2}{\levelSurfaceGradientLowerBound{1}^2+\levelSurfaceGradientLowerBound{2}^2}}\sqrt{\frac{\levelSurfaceGradientUpperBound{1}^2+\levelSurfaceGradientUpperBound{2}^2}{\levelSurfaceGradientUpperBound{1}^2\levelSurfaceGradientUpperBound{2}^2}}\positionErrorPrimaryBound,
    \label{eq:positionErrorInitialConditions}
  \end{align}
  and if the disturbing terms are bounded by
  \begin{align}
    \sup_{t\geq0}\module{\positionErrorPerturbationObservationErrorVector(t)}&<\positionErrorPerturbationObservationErrorTerminalBound = \positionErrorChiFunctionOne^{-1}(\min\{\positionErrorTerminalBound,\positionErrorChiFunctionOne(\positionErrorPerturbationObservationErrorPrimaryBound)\}), \\
    \sup_{t\geq0}\module{\positionErrorPerturbationVelocityErrorVector(t)}&<\positionErrorPerturbationVelocityErrorTerminalBound = \positionErrorChiFunctionTwo^{-1}(\min\{\positionErrorTerminalBound,\positionErrorChiFunctionTwo(\positionErrorPerturbationVelocityErrorPrimaryBound)\}), \\
    \sup_{t\geq0}\module{\auxiliaryOrientationError(t)}&<\positionErrorPerturbationOrientationVectorTerminalBound =  \positionErrorChiFunctionThree^{-1}(\min\{\positionErrorTerminalBound,\positionErrorChiFunctionThree(\auxiliaryOrientationTerminalBound)\}).
  \end{align}
  %
  The final result concerning the positional error can be expressed by referring to the asymptotic gain property, i.e.,
  \begin{align}
    \lsShort&\module{\positionError(t)}\leq \positionErrorGammaFunctionOne(\lsShort\module{\positionErrorPerturbationObservationErrorVector(t)}) \nonumber \\
    &+\positionErrorGammaFunctionTwo(\lsShort\module{\positionErrorPerturbationVelocityErrorVector(t)})+\positionErrorGammaFunctionThree(\lsShort\module{\auxiliaryOrientationError(t)}) \nonumber \\
    &=\frac{m^*}{\vfoLongitudinalGain}\frac{2(1-\longitudinalCompensationCoefficient)(\sqrt{3}\velocityErrorUltimateBound+\sqrt{2}\auxiliaryOrientationUltimateBound)+2\sqrt{7}\longitudinalCompensationCoefficient\left[(\velocityErrorUltimateBound)^2+(\auxiliaryOrientationUltimateBound)^2\right]}{\minEigenvalue{\positionErrorStateMatrix}\positionErrorMajorizationCoefficient-2\sqrt{7}(\sqrt{3}\velocityErrorUltimateBound+\sqrt{2}\auxiliaryOrientationUltimateBound)} \nonumber \\
    &+ \frac{m^*}{\vfoLongitudinalGain}\frac{2\longitudinalCompensationCoefficient(\sqrt{3}\observationErrorUltimateBound+\sqrt{2}\auxiliaryOrientationUltimateBound)+2\sqrt{7}\longitudinalCompensationCoefficient\left[(\observationErrorUltimateBound)^2+(\auxiliaryOrientationUltimateBound)^2\right]}{\minEigenvalue{\positionErrorStateMatrix}\positionErrorMajorizationCoefficient-2\sqrt{7}(\sqrt{3}\observationErrorUltimateBound+\sqrt{2}\auxiliaryOrientationUltimateBound)} \nonumber \\
    &+\frac{m^*}{\vfoLongitudinalGain}\frac{2\sqrt{7}|\pathVelocityProfile|\auxiliaryOrientationUltimateBound}{\minEigenvalue{\positionErrorStateMatrix}\positionErrorMajorizationCoefficient-2\sqrt{7}\auxiliaryOrientationUltimateBound}=:\positionErrorUltimateBound,
    \label{eq:PositionErrorUltimateBound}
  \end{align}
  which corresponds to \eqref{eq:lem4}. Now, the conclusions of Lemma \ref{lem:4} follow. 
\end{proof}

\renewcommand*{\proofname}{Proof of Lemma 5}
\begin{proof}
  To prove the boundedness of $\configurationModuloErrorVector$, let us first focus on the terminal behaviour of the angular error component introduced in \eqref{eq:e2pidef}.

  According to roll error dynamics \eqref{eq:ephidyn}, and referring to assumption \ref{ass:roll}, function $\rollControllerFunction(\rollError,\cdot)$ should be designed with any robust control method assuring the finite-time convergence of $\rollError(t)\rightarrow0$, despite perturbation $\velocityErrorVectorElement{1}(t)$. Since the result \eqref{eq:epsilonsup} shows that the signal $\velocityErrorVectorElement{1}(t)$ is bounded, a specific form of the controller \eqref{pcdef} may be based, for example, on the sliding-mode control method (see e.g. \cite{SHTE15}).


  To present the terminal boundedness of the yaw error $\orientationErrorElement{\psi}$, let us introduce the difference
  \begin{align}
    \label{eq:varepsilondef}
  	\auxiliaryDesiredYawDifference&\triangleq\desiredYawAngle-\auxiliaryBoundedYawAngle = \desiredYawAngle-\atantwo(\sin\auxiliaryYawAngle,\cos\auxiliaryYawAngle),
  \end{align}
  where $\auxiliaryBoundedYawAngle$ corresponds to the angle $\auxiliaryYawAngle$ limited to the $[-\pi,\pi)$ range. Based on the transformations presented in \ref{app:epsi}, we can conservatively assess what follows
  \begin{align}
    \label{eq:varepsilonpsiabs}
    \abs{\auxiliaryDesiredYawDifference}&\leq \auxiliaryDesiredYawDifferenceFunction
  \end{align}
  for
  \begin{align}
    \label{eq:fpsidef}
    \auxiliaryDesiredYawDifferenceFunction&=\atantwo\big(2\vfoLongitudinalGain\module{\positionError}+\longitudinalCompensationCoefficient\module{\longitudinalVelocityErrorVector}+\longitudinalCompensationCoefficient\module{\longitudinalVelocityErrorVectorObservationError}, \nonumber \\
    &\big|\pathVelocityProfile\module{\pathReducedTangentialUnitVector}^2-\vfoLongitudinalGain \levelSurface{1} \pathReducedNormalUnitVector{1}^\top\pathReducedTangentialUnitVector-\vfoLongitudinalGain \levelSurface{2} \pathReducedNormalUnitVector{2}^\top\pathReducedTangentialUnitVector+\longitudinalCompensationCoefficient
    \reducedLongitudinalVelocityErrorVectorEstimate^\top\pathReducedTangentialUnitVector\big|),
  \end{align}
  %
  where $\pathReducedTangentialUnitVector = [\pathTangentialUnitVectorElement{x} \ \pathTangentialUnitVectorElement{y}]^\top$.
  \begin{cremark}
    According to the physical interpretation of subsystem \eqref{eq:velocityTrackingErrorDynamics} and to definition \eqref{epsilondef}, it is impossible to obtain  $\auxiliaryDesiredYawDifferenceFunction=0$ for the underactuated vehicle.
    A similar observation was made for  wheeled vehicles moving under skid-slip conditions in \cite{WANG08}.
  \end{cremark}
  In the view of results \eqref{eq:zinfty}, \eqref{eq:epsilonsup}, and \eqref{eq:PositionErrorUltimateBound} and referring to Remark \ref{rem:5}, the boundary condition presented in \eqref{eq:varepsilonpsiabs} is in the domain  $\auxiliaryDesiredYawDifference\in\left(0,\frac{\pi}{2}\right]$, and $\abs{\auxiliaryDesiredYawDifference}$ decreases with the decreasing values of $\module{\positionError}$, $\module{\longitudinalVelocityErrorVector}$, and $\module{\longitudinalVelocityErrorVectorObservationError}$. Finally,  the ultimate bound of $\abs{\auxiliaryDesiredYawDifference}$ can be written down as
  \begin{align}
    \label{eq:limsupepsilonpsi}
    \lsShort\abs{\auxiliaryDesiredYawDifference(t)}\leq\lsShort\auxiliaryDesiredYawDifferenceFunction(t) =: \auxiliaryDesiredYawDifferenceUltimateBound.
  \end{align}
  %
  According to \eqref{eq:e2pidef}, \eqref{eq:3.20} and \eqref{eq:varepsilondef}, we can express the yaw angle error as $\orientationErrorElement{\yaw} = \desiredYawAngle-\auxiliaryBoundedYawAngle+\auxiliaryYawAngle-\yaw+\auxiliaryBoundedYawAngle-\auxiliaryYawAngle = \auxiliaryDesiredYawDifference+\auxiliaryOrientationErrorYaw + 2k\pi$, thus
  \begin{align}
    \orientationErrorElement{\yaw}(t) \ \modulo \ 2\pi=\auxiliaryDesiredYawDifference+\auxiliaryOrientationErrorElement{\yaw}.
  \end{align}
  The ultimate value of $\orientationErrorElement{\yaw}$ is upper bounded by
  %
  \begin{align}
    \lsShort\abs{\orientationErrorElement{\yaw}(t) \ \modulo \ 2\pi}&\leq \lsShort\abs{\auxiliaryDesiredYawDifference} + \lsShort \abs{\auxiliaryOrientationErrorYaw} \nonumber \\
    &\leq \lsShort\abs{\auxiliaryDesiredYawDifference} + \lsShort \module{\auxiliaryOrientationError} \nonumber \\
    &= \auxiliaryDesiredYawDifferenceUltimateBound+\auxiliaryOrientationUltimateBound,
    \label{eq:yawErrorUpperBound}
  \end{align}
  where $\auxiliaryDesiredYawDifferenceUltimateBound$ and $\auxiliaryOrientationUltimateBound$ were respectively introduced in  \eqref{eq:limsupepsilonpsi} and \eqref{eq:AuxiliaryOrientationUltimateBound}.

  Let us now move to the analysis of the pitch angle error $\orientationErrorElement{\pitch}$, beginning with the introduction of a difference
  \begin{align}
    \auxiliaryDesiredPitchDifference &\triangleq \desiredPitchAngle - \auxiliaryPitchAngle \nonumber \\
    &= \atan\left(\frac{{\modifiedConvergenceVectorFieldElementEstimate{z}}\module{\pathReducedTangentialUnitVector}-\pathTangentialUnitVectorElement{z}\sqrt{{\modifiedConvergenceVectorFieldElementEstimate{x}}^2+{\modifiedConvergenceVectorFieldElementEstimate{y}}^2}}{\module{\pathReducedTangentialUnitVector}\sqrt{{\modifiedConvergenceVectorFieldElementEstimate{x}}^2+{\modifiedConvergenceVectorFieldElementEstimate{y}}^2}+{\modifiedConvergenceVectorFieldElementEstimate{z}}\pathTangentialUnitVectorElement{z}}\right)
    \label{eq:epsilonTheta}
  \end{align}
  derived in \ref{app:etheta}. Recalling \eqref{eq:hpstardef}, after some algebraic calculations, one can write down that
  \begin{align}
    \abs{\auxiliaryDesiredPitchDifference} &\leq \atan\left(\frac{\beta_4}{\beta_5}\right) =: \auxiliaryDesiredPitchDifferenceFunction \in \left[0,\frac{\pi}{2}\right),
    \label{eq:epsilonThetaBound}
  \end{align}
  for
  \begin{align}
    \beta_4 &= 4\longitudinalCompensationCoefficient^2\left(\module{\longitudinalVelocityErrorVector}+\module{\longitudinalVelocityErrorVectorObservationError}\right)^2+16\vfoLongitudinalGain\longitudinalCompensationCoefficient\left(\module{\longitudinalVelocityErrorVector}+\module{\longitudinalVelocityErrorVectorObservationError}\right)\module{\positionError} \nonumber \\
    &+16\vfoLongitudinalGain^2\module{\positionError}^2 + 8\pathVelocityProfile\longitudinalCompensationCoefficient\left(\module{\longitudinalVelocityErrorVector}+\module{\longitudinalVelocityErrorVectorObservationError}\right)+16\pathVelocityProfile\vfoLongitudinalGain\module{\positionError} \nonumber \\
    \beta_5 &= \Big\| \left(\module{\pathReducedTangentialUnitVector}\sqrt{{\modifiedConvergenceVectorFieldElementEstimate{x}}^2+{\modifiedConvergenceVectorFieldElementEstimate{y}}^2}+{\modifiedConvergenceVectorFieldElementEstimate{z}}\pathTangentialUnitVectorElement{z}\right) \nonumber \\
    &\cdot \left( {\modifiedConvergenceVectorFieldElementEstimate{z}}\module{\pathReducedTangentialUnitVector}+\pathTangentialUnitVectorElement{z}\sqrt{{\modifiedConvergenceVectorFieldElementEstimate{x}}^2+{\modifiedConvergenceVectorFieldElementEstimate{y}}^2}\right)\Big\|. \nonumber
  \end{align}

  The value of $\auxiliaryDesiredPitchDifference$ decreases with the decrease of $\module{\positionError}$, $\module{\longitudinalVelocityErrorVector}$, and $\module{\longitudinalVelocityErrorVectorObservationError}$ and according to \eqref{eq:zinfty}, \eqref{eq:epsilonsup}, and \eqref{eq:PositionErrorUltimateBound}, its ultimate upper bound can be designated as
  %
  \begin{align}
    \label{eq:auxiliaryDesiredPitchDifferenceUltimateBound}
    \lsShort\abs{\auxiliaryDesiredPitchDifference(t)}\leq \lsShort \auxiliaryDesiredPitchDifferenceFunction(t) =: \auxiliaryDesiredPitchDifferenceUltimateBound.
  \end{align}
  Upon result \eqref{eq:auxiliaryDesiredPitchDifferenceUltimateBound}, we can state that the pitch angle error
  \begin{align}
    \orientationErrorElement{\pitch} = \desiredPitchAngle - \auxiliaryPitchAngle + \auxiliaryPitchAngle - \pitch =  \auxiliaryDesiredPitchDifference+\auxiliaryOrientationErrorElement{\pitch}
  \end{align}
  is ultimately bounded by
  \begin{align}
    \lsShort\abs{\orientationErrorElement{\pitch}(t)}&\leq \lsShort\abs{\auxiliaryDesiredPitchDifference(t)} + \lsShort \abs{\auxiliaryOrientationErrorElement{\pitch}(t)} \nonumber \\
    &\leq \lsShort\abs{\auxiliaryDesiredPitchDifference(t)} + \lsShort \module{\auxiliaryOrientationError(t)} \nonumber \\
    &\leq \auxiliaryDesiredPitchDifferenceUltimateBound+\auxiliaryOrientationUltimateBound,
    \label{eq:pitchAngleUpperBound}
  \end{align}
  where $\auxiliaryDesiredPitchDifferenceUltimateBound$ and $\auxiliaryOrientationUltimateBound$ were introduced in \eqref{eq:auxiliaryDesiredPitchDifferenceUltimateBound} and \eqref{eq:AuxiliaryOrientationUltimateBound}, respectively.

  Recalling the control objective from \eqref{eq:controlobjective}, according to results \eqref{eq:PositionErrorUltimateBound}, \eqref{eq:yawErrorUpperBound}, \eqref{eq:pitchAngleUpperBound}, and upon Assumption \ref{ass:roll}, error $\configurationModuloErrorVector$ defined in \eqref{eq:e2pidef} satisfies
  \begin{align}
    \lsShort&\module{\configurationModuloErrorVector(t)}\leq \lsShort\oneNorm{\positionError(t)}+\lsShort\oneNorm{\orientationError(t) \ \modulo \ {2\pi}} \nonumber \\
    &\leq \sqrt{2}\lsShort\module{\positionError(t)}+\lsShort\abs{\orientationErrorElement{\roll}(t)} \nonumber \\
    &+\lsShort\abs{\orientationErrorElement{\pitch}(t)}+\lsShort\abs{\orientationErrorElement{\yaw}(t) \ \modulo \ {2\pi}} \nonumber \\
    &\leq \sqrt{2}\positionErrorUltimateBound+\underbrace{\rollErrorUltimateBound+\auxiliaryDesiredYawDifferenceUltimateBound+\auxiliaryDesiredPitchDifferenceUltimateBound}_{r_o^\infty}+2\auxiliaryOrientationUltimateBound =: \controlError,
    \label{eq:mainresult}
  \end{align}
  where $\oneNorm{\cdot}$ is the 1-norm of a vector, and the ultimate upper bound $\rollErrorUltimateBound=0$ upon Assumption \ref{ass:roll}. Result \eqref{eq:mainresult} is consistent with \eqref{eq:lem5}, thus the conclusions stated in Lemma~\ref{lem:5} follow.
\end{proof}


\section{Simulation results}
\label{sec:simulations}

\begin{table}
	\caption{Values of the design parameters selected for simulations of the VFO-ADR controller (note that $\rho/m=2.4172$).}
	\label{tab:1}
	\vspace{0.2cm}
	\centering
	\begin{tabular}{cc}
		\hline
		Parameter & Value \\ \hline
		$\adrcGainMatrixLocal$ & \textrm{diag}\{\color{black}2.4172, \ 2.4172, \ 2.4172\color{black}, \ 0.5, \ 0.5, \ 0.5\} \color{black} \\
		$\localInputMatrixEstimate$ & ${\begin{array}{c}\diag\{0.3, \ 0.3, \ 0.3, \ 2.5, \ 0.75, \ 0.75\}\end{array}}$ \\
		$\particularObserverBandwidth{i}$ & 200 \\
		$\rollControllerGain$ & 5 \\
		$\vfoLongitudinalGain$ & 2 \\
		$\vfoOrientationGainPitch$ & 4 \\
		$\vfoOrientationGainYaw$ & 4 \\
		$\pathVelocityDirection$ & 1 \\
		$\longitudinalCompensationCoefficient$ & 0.75 \\
		$\angularCompensationCoefficient$ & 1.0 \\ \hline
	\end{tabular}
\end{table}
A simulation study of the proposed control structure was conducted in Matlab/Simulink environment with a mathematical model of an underactuated ellipsoidally-shaped rigid-body vehicle, with the dynamics modeled according to \cite{FOSS99}, and represented by equations \eqref{DynDefG}-\eqref{DynDefB}. Actuation structure of the vehicle was determined by selecting the input matrix $\actuationMatrixLocal = \diag\{1, 0, 0,  1, 1, 1\}$, implying underactuation along $\yLocalAxis$ and $\zLocalAxis$ axes. The mass and linear damping coefficient matrices were, respectively, set to
\begin{align}
	\massMatrixLocal &= \begin{bmatrix} 4.137 & 0 & 0 & 0 & 0 & 0 \\ 0 & 4.137 & 0 & 0 & 0 & 0 \\ 0 & 0 & 4.137 & 0 & 0 & 0 \\ 0 & 0 & 0 & 0.535 & 0 & -0.390 \\ 0 & 0 & 0 & 0 & 1.653 & 0 \\ 0 & 0 & 0 & -0.390 & 0 & 1.577
	\end{bmatrix}, \nonumber
\end{align}
and $\environmentalDampingMatrixLocal = \diag\{2, \ 10, \ 10, \ 10, \ 10, \ 10\}$, with all of the values expressed in SI units.
We present a detailed results for two simulation scenarios concerning different reference paths, i.e., (A) a helix-like path determined by
\begin{align}
	\begin{cases}
		\levelSurface{1}(\positionVector)=-x+\textrm{sin}(4z) \\
		\levelSurface{2}(\positionVector)=-y+\textrm{cos}(4z)
	\end{cases},
\end{align}
with the velocity profile $\pathVelocityProfile=0.1$m/s, and (B) an elliptically-shaped path described by the equations
\begin{align}
	\begin{cases}
		\levelSurface{1}(\positionVector) = x^2 + \left(\frac{y}{2}\right)^2-1 \\
		\levelSurface{2}(\positionVector) = x+2y+3z-1
	\end{cases},
\end{align}
where the reference velocity along the path was set to $\pathVelocityProfile = 0.2$m/s. In both simulations, a direction along the path was determined by $\pathMotionDirection=1$, while the simulation time horizon was set to $\simulationTime=100$s. The initial configuration vector was equal to \color{black} $\configurationVector(0) = [0 \ -\pi/3 \ \pi/4 \ 1.0 \ 0.6 \ 0.6]^\top$ for simulation (A) and $\configurationVector(0) = [0 \ -\pi/3 \ \pi/4 \ -0.1 \ 0.0 \ 0.3]^\top$ for simulation (B), while the rest of initial conditions were set to  $\configurationVectorDerivative(0)=\zeroMatrix$, \ \textrm{and} \  $\observerEstimatedStateVector{i}(0) = [-\configurationVectorElement{i}(0) \ 0 \ 0]^\top$ (all in SI units) for both simulations.  In simulation (B), an additional external disturbance $\perturbationsVectorGlobal=[2\textrm{sin}t \ 4\textrm{sin}0.8t \ 1.4\textrm{sin}0.6t \ 0 \ 0 \ 0]^\top$ was introduced.

The set of controller parameters utilized in simulations is presented in Table \ref{tab:1}. Particular values of controller gains were selected to meet the conditions imposed in Proposition \ref{prop:1}. In simulation (B), we show a conservativeness of the tuning conditions from Proposition 1, which can be relaxed in practice (see Remark \ref{rem:adrcTuning}), by setting $\adrcGainMatrixLocal=\diag\{0.5, \ 0.5, \ 0.5, \ 0.5, \ 0.5, \ 0.5\}$.
The feedback function of an auxiliary roll controller introduced in \eqref{pcdef} was chosen as $f_\phi(e_\phi,\cdot)=-k_{\phi}\phi(t)$. Choosing this control function assures a practical stability of the equilibrium point $\rollError=0$, i.e. $\lsShort|\rollError(t)|<\delta_\phi$. Since $\rollError(0)=0$, we expect $\delta_\phi$ to be very small, satisfying in practice that $\forall_{t\geq0}\roll(t)\approx0$.
To avoid extremely large values of control signals at the beginning of simulation, the controller action was turned off during the first second of every simulation (see \cite{HUAN14}), so the initial peaking phenomenon didn't make any impact on the presented results. In simulation (B), an initial configuration is located relatively far from the reference path, so to keep the magnitude of  $\controlSignalLocal$ on a reasonable level (that is, a feasible level for actuators implemented on a vehicle), we scaled the commanded velocities $\reducedCommandedVelocityVectorLocal$ during an initial transient stage with the procedure taken from \cite{LAKO19}. The maximal values of commanded velocities were set to $u_{c,\max}=8.0$~m/s and $p_{c,\max}=q_{c,\max}=r_{c,\max}=8.0$~rad/s, while their rates were bounded by $\dot{u}_{c,\max}=2.0$~m/s and $\dot{p}_{c,\max}=\dot{q}_{c,\max}=\dot{r}_{c,\max}=2.0$~rad/s.

\begin{figure}[htpb]
	\centering
	\includegraphics[width=0.31\textwidth]{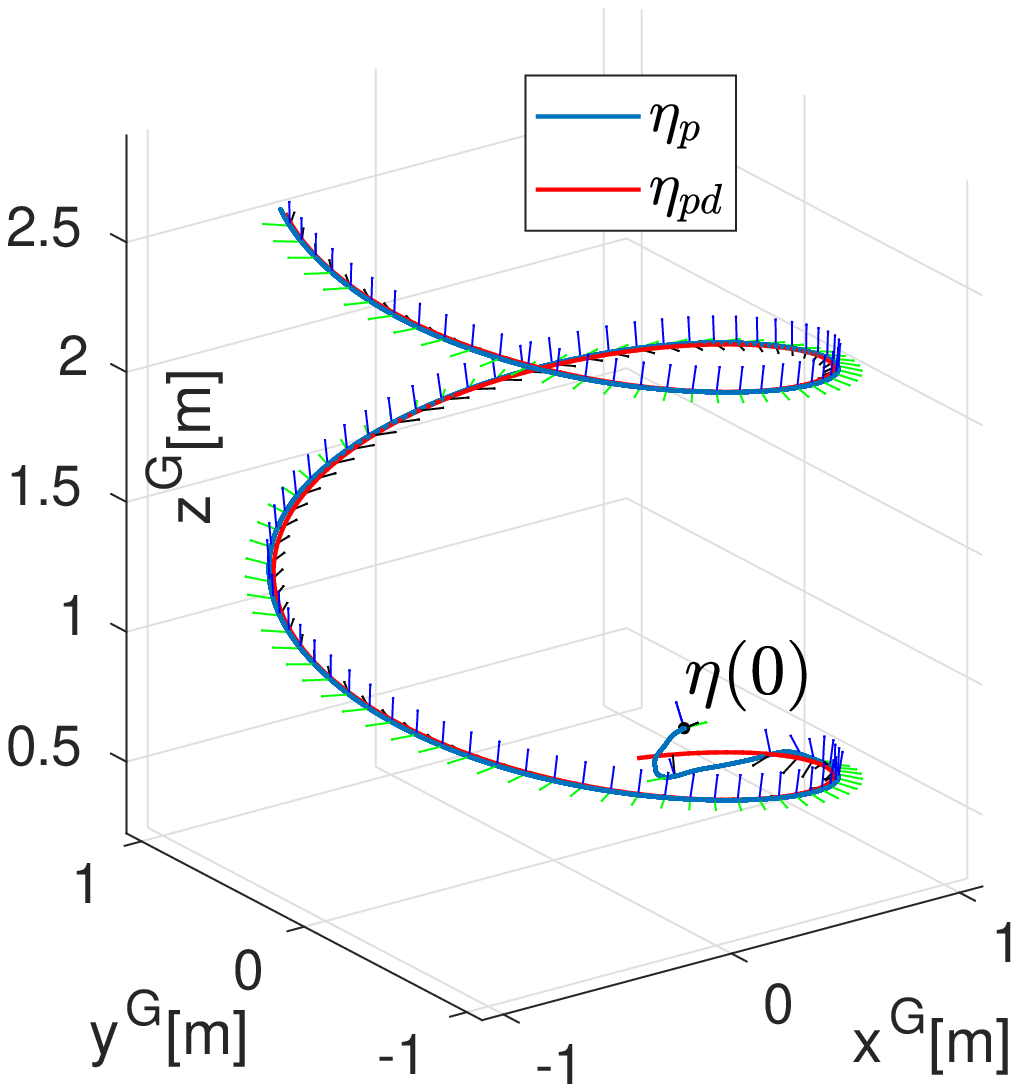} \\
	\includegraphics[width=0.41\textwidth]{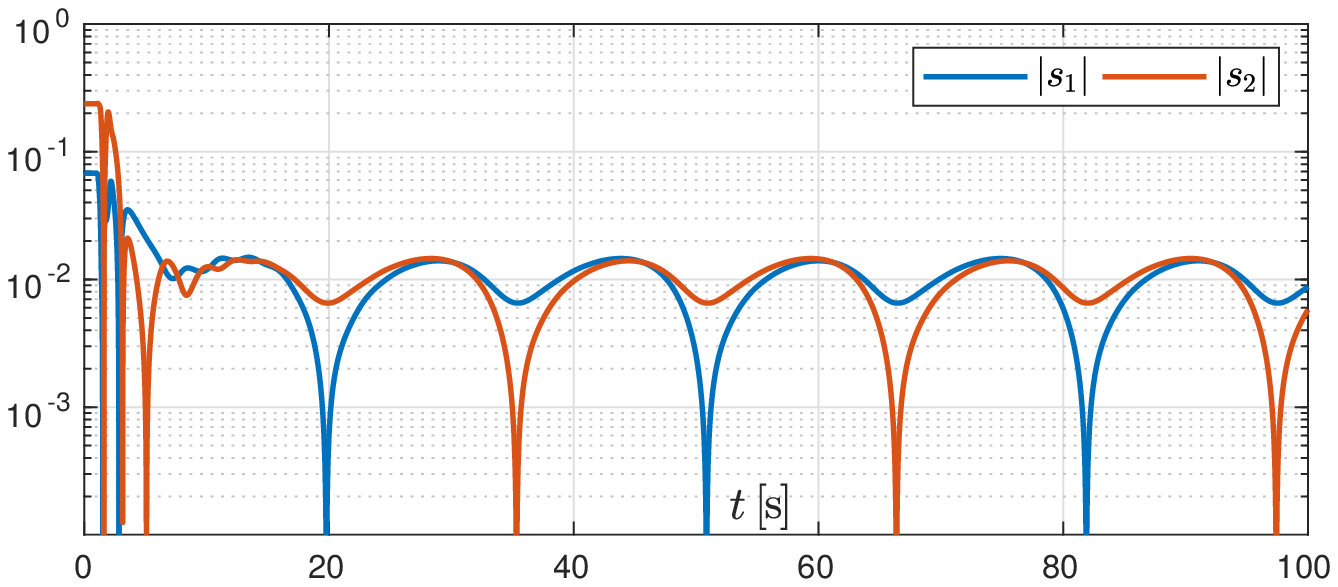}
	\includegraphics[width=0.41\textwidth]{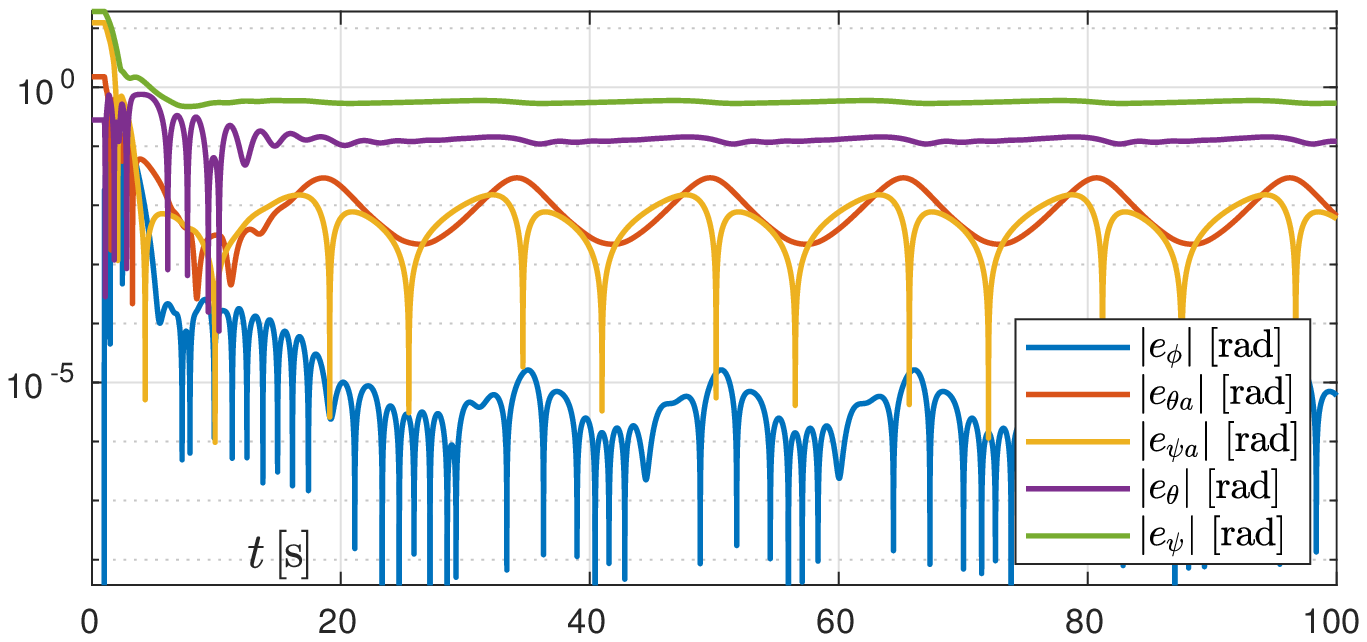}
	\includegraphics[width=0.41\textwidth]{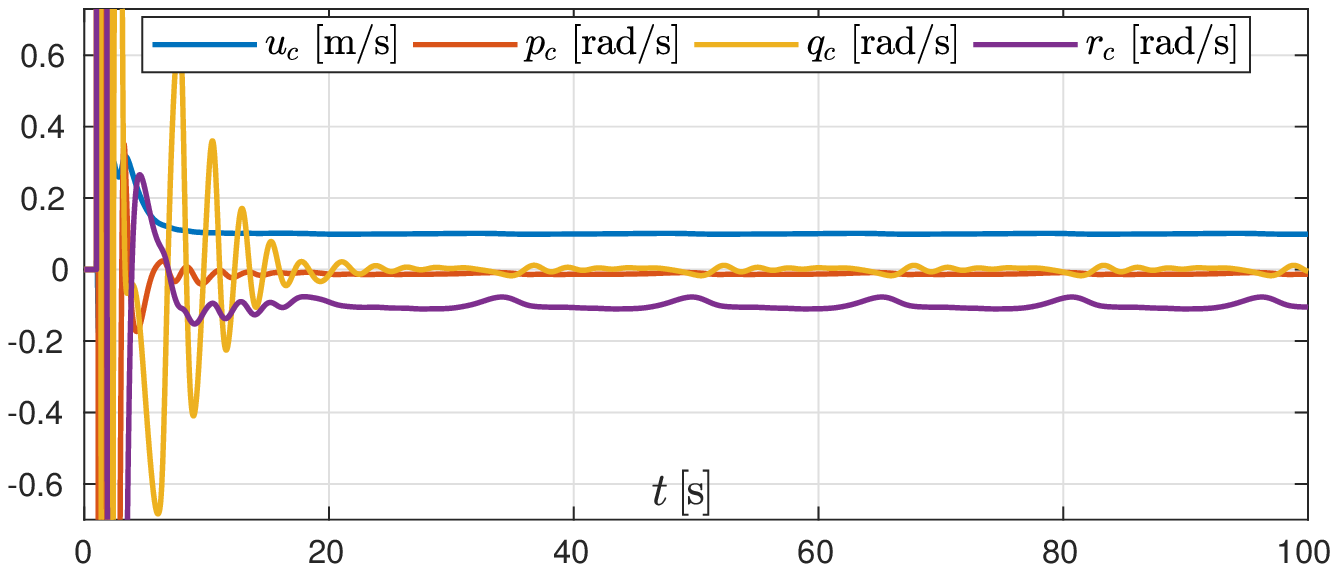}
	\includegraphics[width=0.41\textwidth]{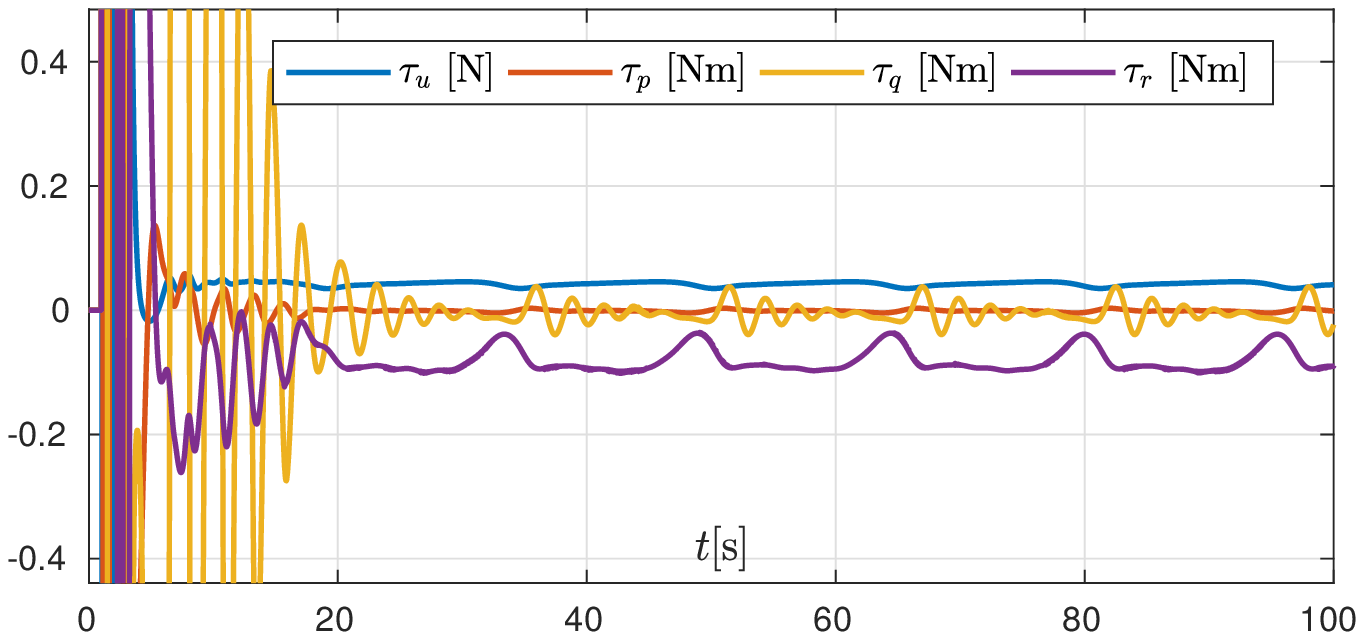}
	\includegraphics[width=0.41\textwidth]{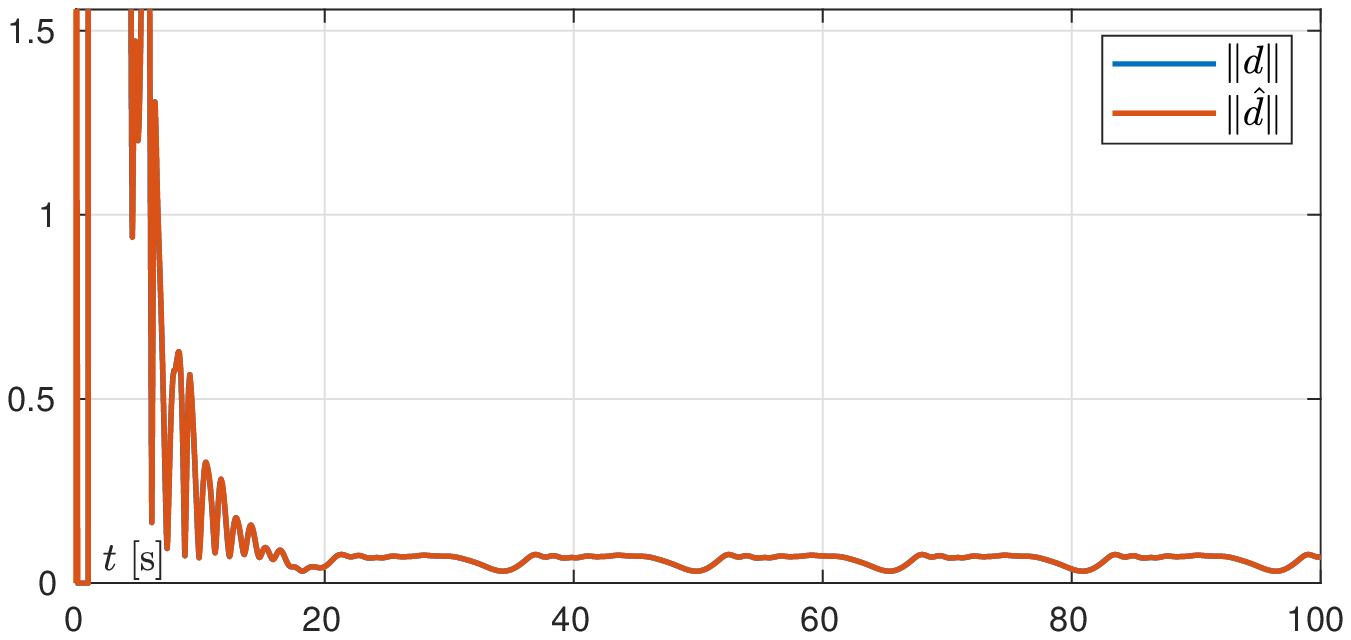}
	\vspace{-0.35cm}
	\caption{(A) Simulation results obtained for the underactuated rigid body following a helix-like path}
	\label{fig:results2}
\end{figure}

\color{black}
The results of simulation (A) are presented in Fig. \ref{fig:results2}. A small coordinate systems has been drawn to represent an attitude of the body-fixed frame $\localCoordinateSystem$ in each second. Signals $\controlSignalLocalAlongY$ and $\controlSignalLocalAlongZ$ are cut out by the selected $\actuationMatrixLocal$ matrix, thus, they are not presented in the figures. Quickly decreasing high values of the commanded velocities, control signals, and total disturbance estimate were not presented in the plots for the sake of a clear presentation of the system steady-state behavior. Particular signals were bounded by $\max_{t\geq0}\{u_c(t),p_c(t),q_c(t),r_c(t)\}<43.41$, $\max_{t\geq0}\{\tau_u(t),\tau_p(t),\tau_q(t),\tau_r(t)\}<3281$, and $\max_{t\geq0}\hat{d}(t)<15\cdot10^4$. The ultimate values of error upper bounds, introduced in objectives  \eqref{eq:controlobjective}-\eqref{eq:controlobjective2}, reached the values $\sup_{t\in[50,100]}\positionalControlError(t)=0.0202, \ \sup_{t\in[50,100]}\controlError(t)=0.6117$. Values of the commanded velocities along $\xLocalAxis$ axis are approximately equal to the velocity profile $\pathVelocityProfile$ and fluctuate around its value when the position error increases, to push the vehicle towards a reference path. Efectiveness of the estimation quality provided by ESO can be assesed upon the plot visualizing $\|\totalDisturbanceVector(t)\|$ and $\|\totalDisturbanceVectorEstimate(t)\|$.

\begin{figure}[ht!]
	\centering
	\includegraphics[width=0.47\textwidth]{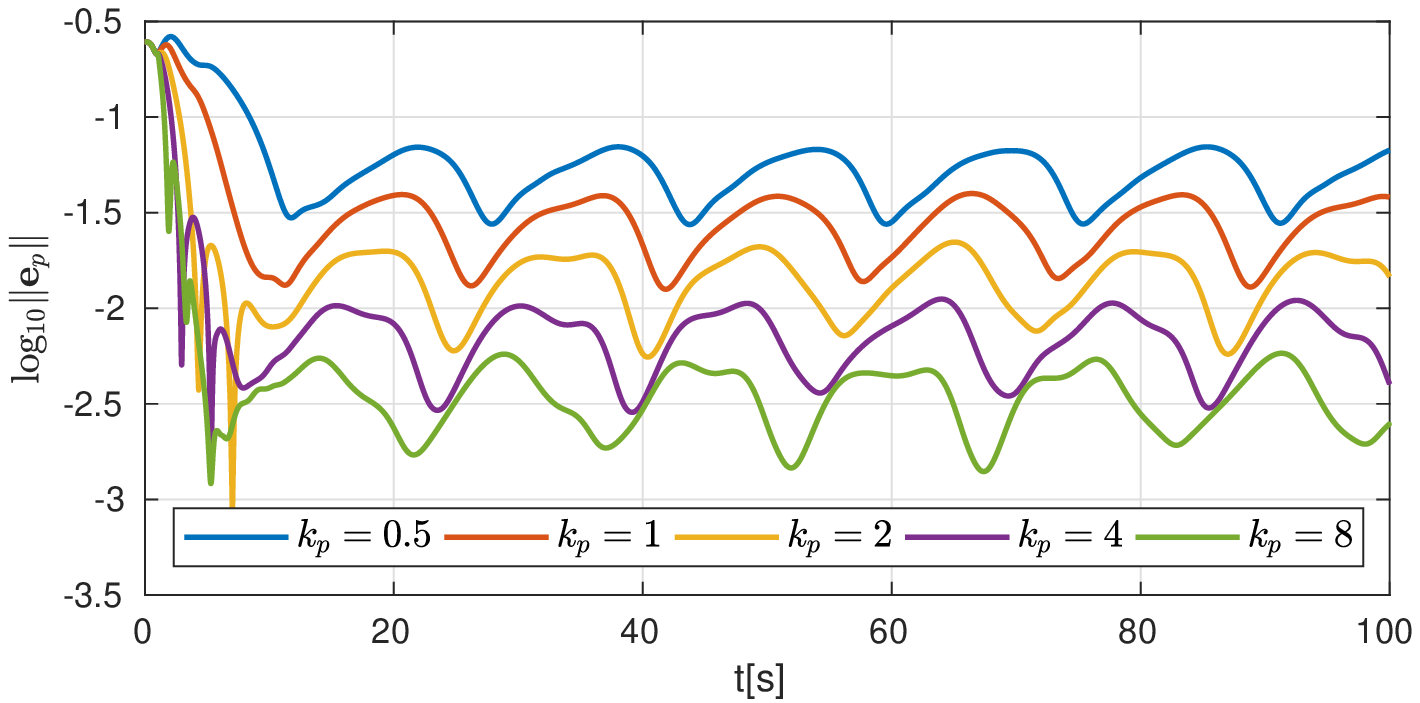}
	\caption{The influence of gain $\vfoLongitudinalGain$ on the positional error $\|\positionError\|$.}
	\label{fig:results3}
\end{figure}

To illustrate the impact of chosen controller parameter values on the achieved control performance, a series of simulations concerning the helix-like path have been conducted with different values of $\vfoLongitudinalGain$. In Fig. \ref{fig:results3}, we can see that increasing a value of $\vfoLongitudinalGain$ results in the decrease of $\|\positionError\|$ values. 

\begin{figure}[ht!]
	\centering
	\includegraphics[width=0.47\textwidth]{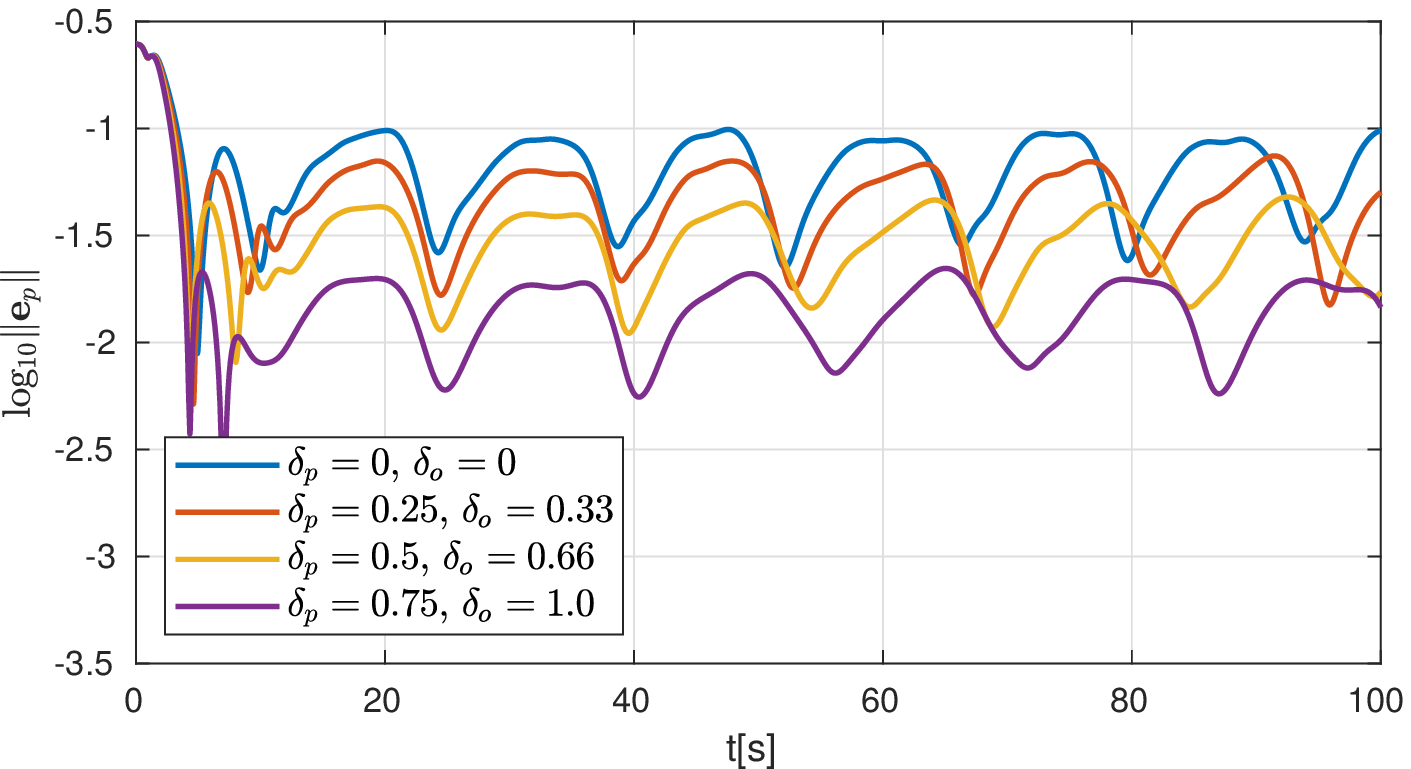}
	\caption{The influence of compensation coefficients $\angularCompensationCoefficient,\longitudinalCompensationCoefficient$ on the positional error $\|\positionError\|$.}
	\label{fig:results4}
\end{figure}

The influence of compensation coefficients has been examined as well, including the case of $\longitudinalCompensationCoefficient=\angularCompensationCoefficient=0$ that corresponds to the VFO-ADR structure with the kinematic-level controller developed for a fully-actuated vehicle, originally proposed in \cite{LAKO17}. The results presented in Fig. \ref{fig:results4} reveal that an introduction of drift compensating components improves the positional control performance. In Table \ref{tab:simresults2}, the average steady-state values of particular signals are provided, which were calculated according to the formula $\alpha_{avg} = \frac{1}{t_2-t_1}\int_{t_1}^{t_2}\alpha(t) dt$ for an exemplary signal $\alpha$ and a time interval determined by $t_1=50s, \ t_2=100s.$ The average steady value of $\|\positionError\|$ decreases with the increasing values of $\longitudinalCompensationCoefficient$ and $\angularCompensationCoefficient$, reaching approximately 4.65 times smaller value for $\longitudinalCompensationCoefficient=0.75$, $\angularCompensationCoefficient=1$ with respect to the VFO-ADR algorithm without transversal-drift compensation (i.e., $\longitudinalCompensationCoefficient=\angularCompensationCoefficient=0$). Although the auxiliary orientation errors $|\auxiliaryOrientationErrorPitch|$ and $|\auxiliaryOrientationErrorYaw|$ are significantly smaller after introducing the drift compensation component, the average value of an orientation error $\module{\orientationError}$ slightly increases. This result is expected - to compensate a transversal drift, the vehicle must be oriented with a certain inclination to the path and goes partially sideways.

\begin{figure}[ht!]
	\centering
	\includegraphics[width=0.43\textwidth]{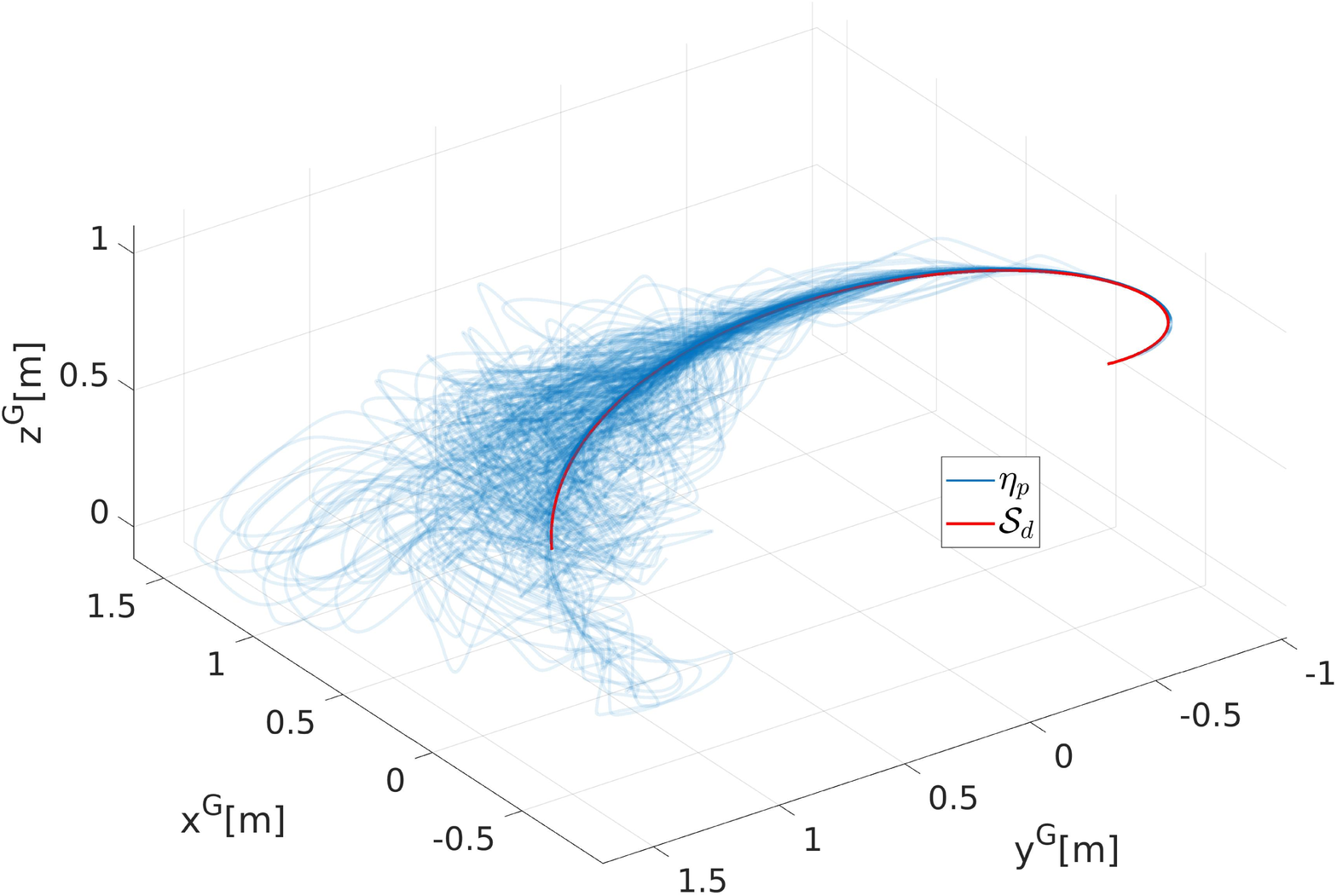}
	\caption{Transient trajectories of a vehicle position obtained for various initial conditions $\configurationVector(0)$.}
	\label{fig:results5}
\end{figure}

\color{black}
Figure \ref{fig:results5} presents a positional transient performance of the vehicle obtained from a set of initial conditions, that was randomly generated in a close neighbourhood of the reference path. A practical convergence of the vehicle's position towards the reference path is illustrated here as a blue funnel-like set that confirms a practical stability of the designed control system derived in Section \ref{sec:stability}.

\begin{figure}[ht!]
	\centering
	\includegraphics[width=0.49\textwidth]{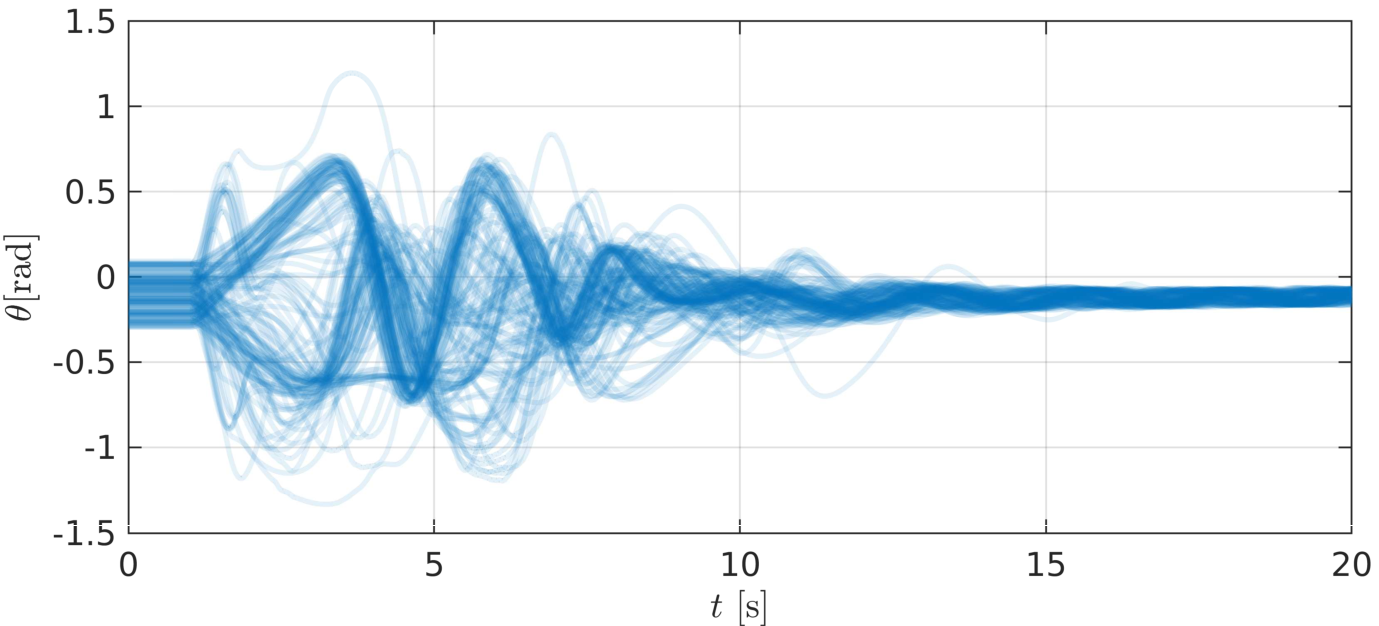}
	\caption{Transient trajectories of the pitch angle obtained for various initial conditions $\configurationVector(0)$.}
	\label{fig:resultsTheta}
\end{figure}

In Fig. \ref{fig:resultsTheta}, we can see the transient stage of pitch angle $\pitch(t)$ obtained from a random set of initial configurations. As far as the reference path is sufficiently close to the initial vehicle position, and neither $\pitch(0)$ nor reference pitch angle \eqref{eq:2.11} are close to their domain boundaries, i.e., $\pm\frac{\pi}{2}$, the admissible configuration domain defined in \eqref{eq:etadef} is not violated, implying that the vehicle configuration  does not reach the singularity of matrix $\velocityTransformationMatrix(\orientationVector)$.

\begin{figure}[ht!]
	\centering
	\includegraphics[width=0.31\textwidth]{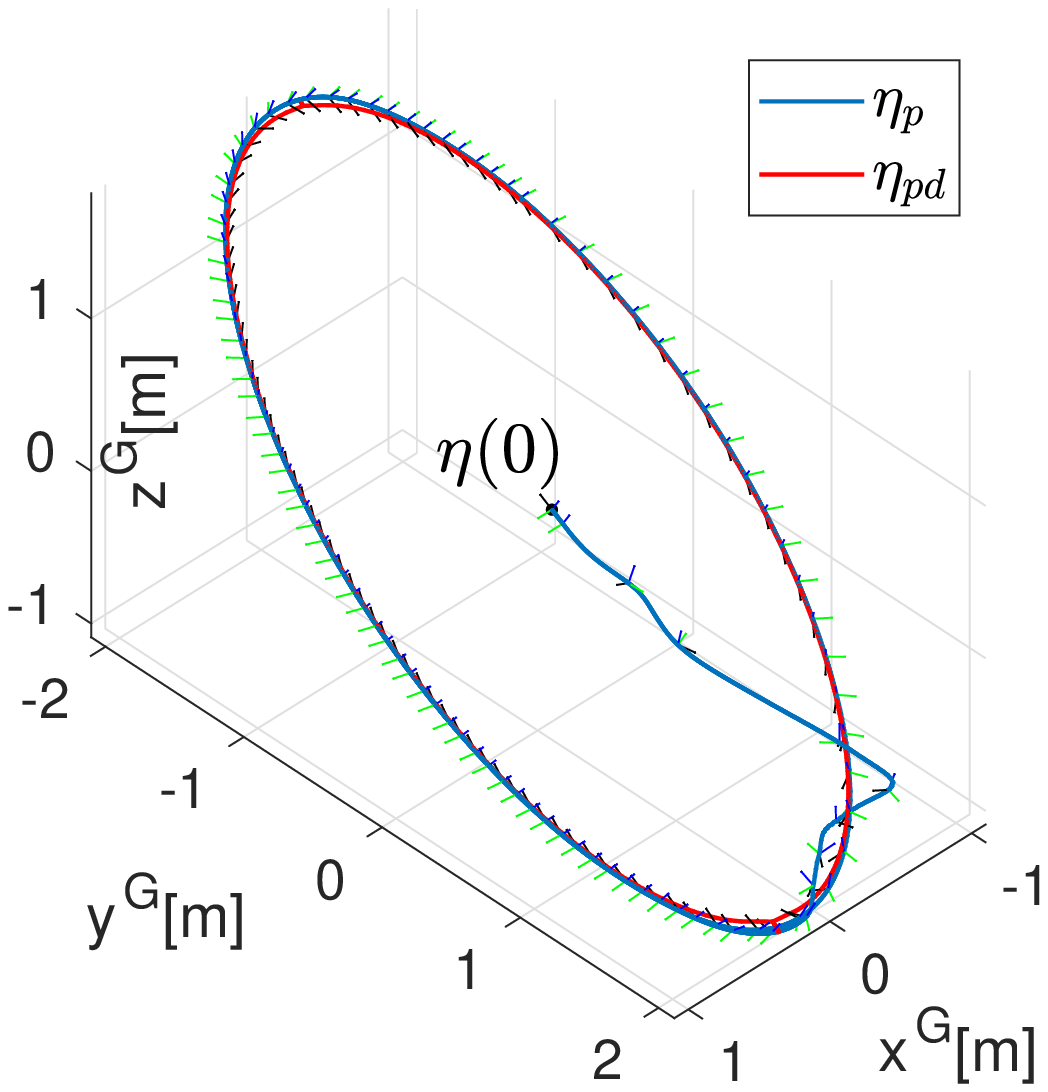} \\
	\includegraphics[width=0.41\textwidth]{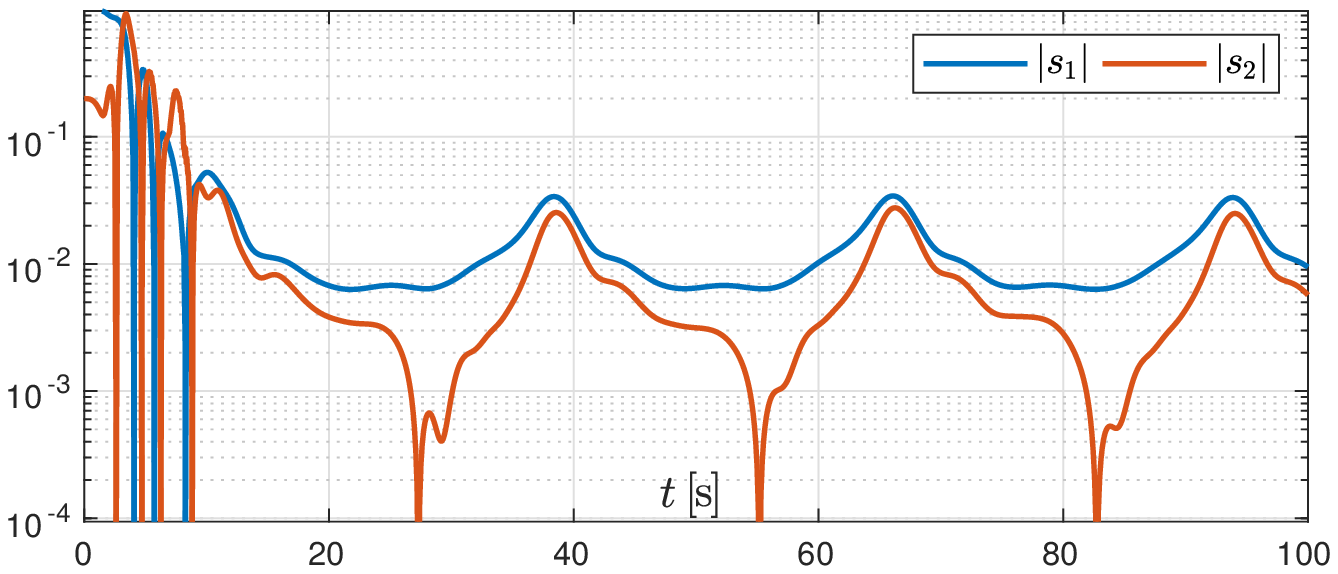}
	\includegraphics[width=0.41\textwidth]{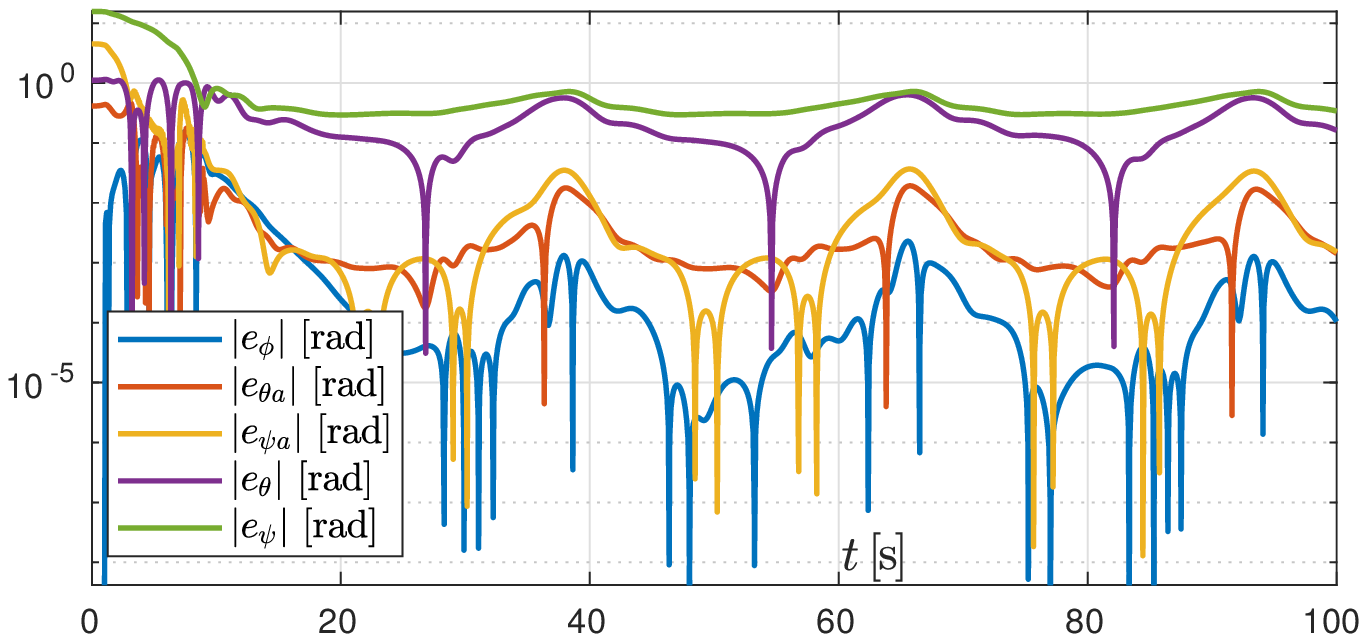}
	\includegraphics[width=0.41\textwidth]{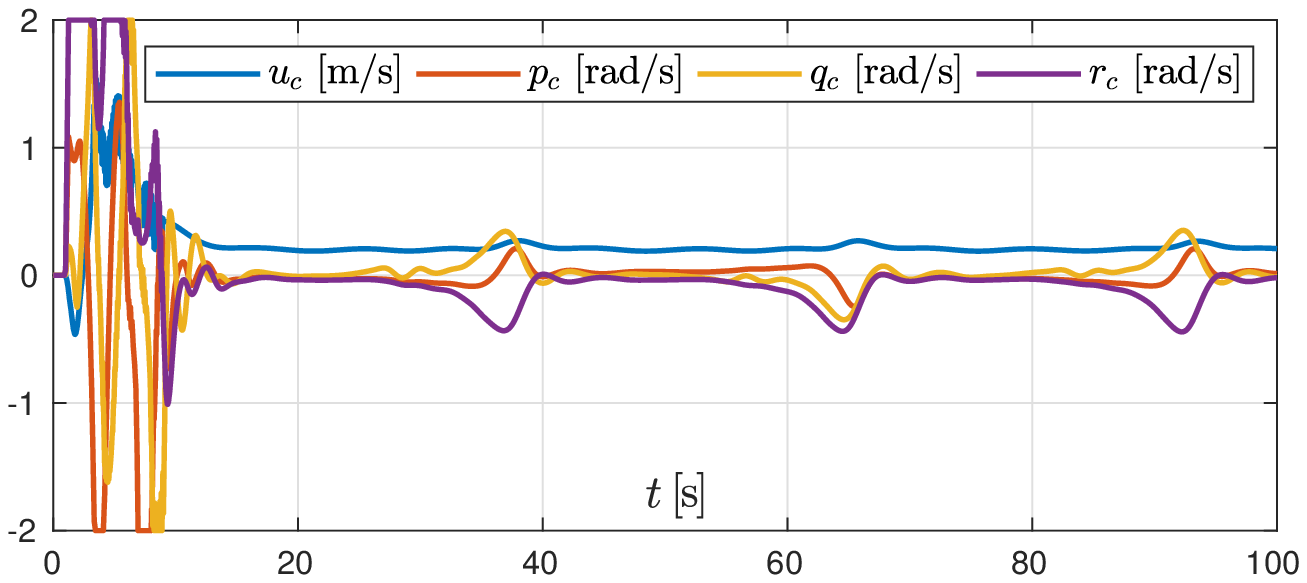}
	\includegraphics[width=0.41\textwidth]{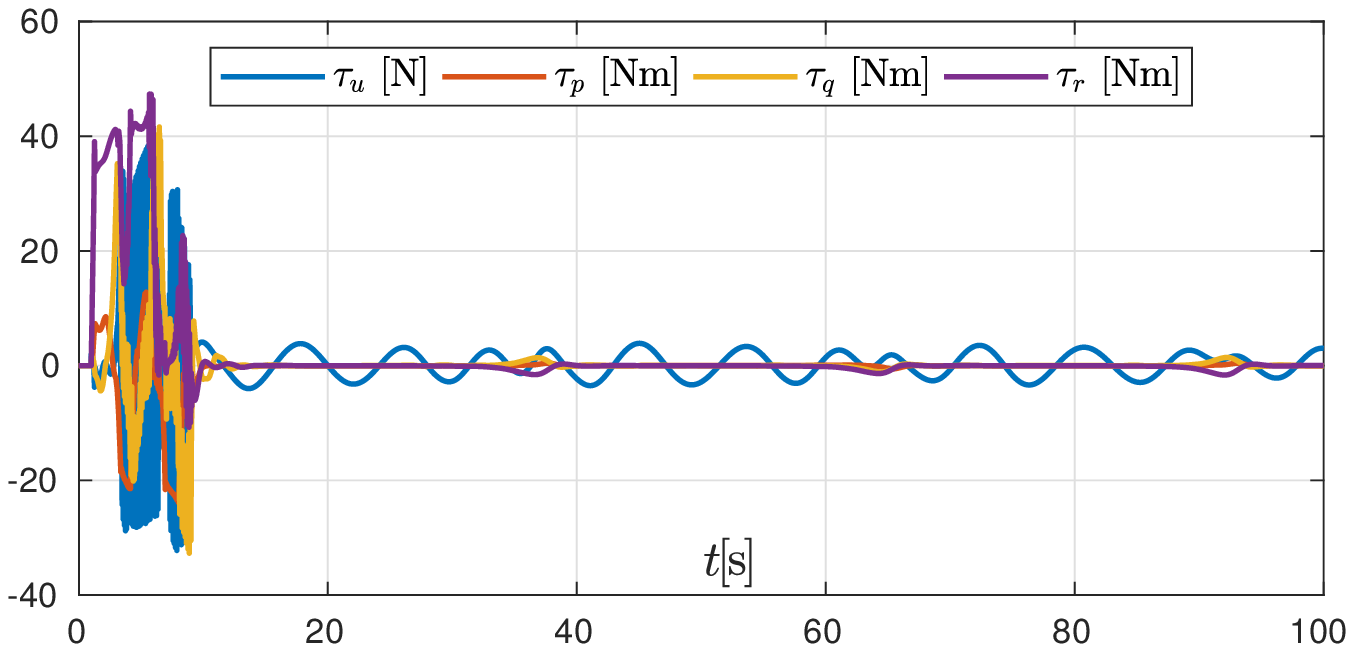}
	\includegraphics[width=0.41\textwidth]{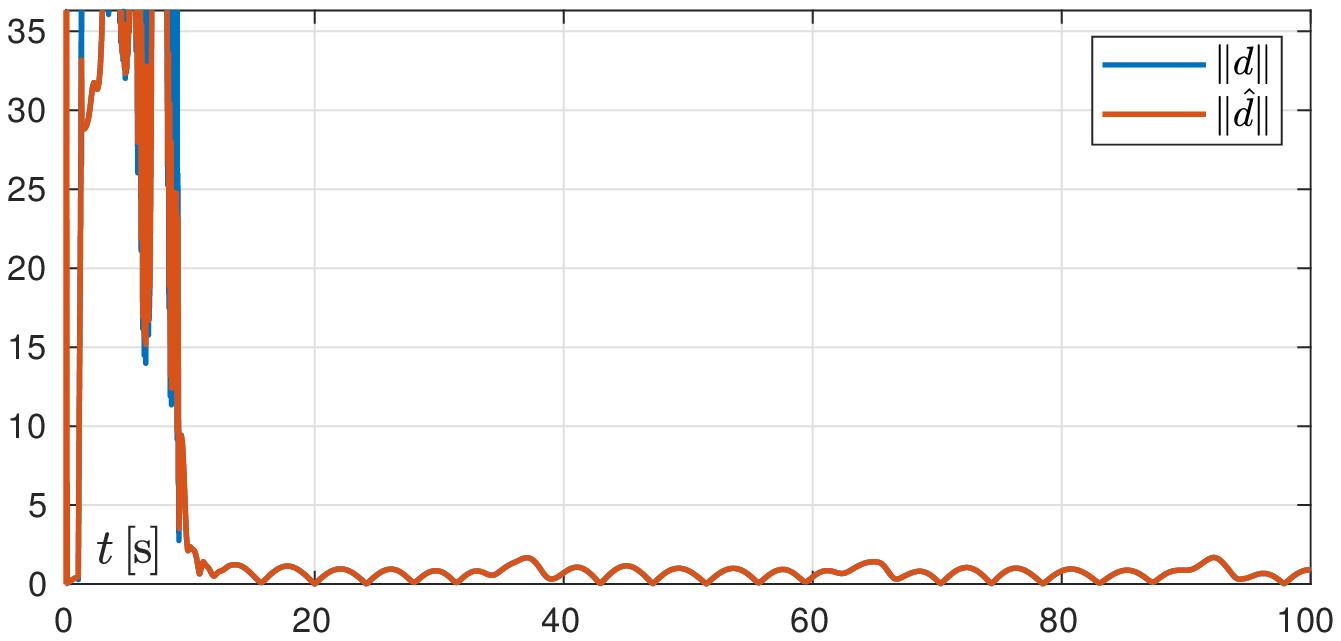}
	\vspace{-0.35cm}
	\caption{(B) Simulation results obtained for the underactuated rigid body following an ellipticaly-shaped path}
	\label{fig:results}
\end{figure}

\color{black}
The results obtained for simulation (B) are presented in Fig. \ref{fig:results}. The ultimate error upper bounds reached in this case the values of $\sup_{t\in[50,100]}\positionalControlError(t)=0.0440$, and $\sup_{t\in[50,100]}\controlError(t)=0.9605$. This result confirms the claim made in Remark \ref{rem:adrcTuning}, and reveals that the conservatively selected gain $\adrcGainMatrixLocal=\blkdiag\{\frac{\rho}{m}\identityMatrix,k_o\identityMatrix\}$ is not the only one that allows the VFO-ADR controller to solve the path-following task. An introduction of the external disturbance $\perturbationsVectorGlobal$ leads to the more dynamic changes of a total disturbance, and higher values of $\controlSignalLocalAlongX$ that has to compensate the introduced perturbation.
 \color{black}

\begin{table}
	\label{tab:simresults2}
	\centering
	\caption{Obtained average steady-state values of particular signals (expressed in the SI units) for different $\longitudinalCompensationCoefficient, \ \angularCompensationCoefficient$ values.}
	\vspace{0.2cm}
	\begin{tabular}{ccccc}
		\hline
		  & $\longitudinalCompensationCoefficient=0$ & $\longitudinalCompensationCoefficient=0.25$ &   $\longitudinalCompensationCoefficient=0.5$ & $\longitudinalCompensationCoefficient=0.75$ \\
		& $\angularCompensationCoefficient=0$ & $\angularCompensationCoefficient=0.33$ & $\angularCompensationCoefficient=0.66$ & $\angularCompensationCoefficient=1$  \\\hline
		$\module{\positionError}_{avg}$ & 0.0670& 0.0475 & 0.0299 & 0.0144 \\
		$\module{\orientationError}_{avg}$ & 0.5493 & 0.5580 & 0.5688 & 0.5782  \\
		$\abs{\rollError}_{avg}$ & $<0.0001$ & $ <0.0001$ & <0.0001 & <0.0001 \\
		$\abs{\auxiliaryOrientationErrorPitch}_{avg}$ & 0.0434 & 0.0319 & 0.0215 & 0.0105 \\
		$\abs{\auxiliaryOrientationErrorYaw}_{avg}$ & 0.0273 & 0.0210 & 0.0139 & 0.0067  \\
		$\module{\commandedVelocityVectorLocal}_{avg}$ & 0.1833 & 0.1706 & 0.1534 & 0.1427 \\
		$\module{\actuationMatrixLocal\controlSignalLocal}_{avg}$ & 0.2062 & 0.1770 & 0.1243 & 0.0934 \\
		 \hline
	\end{tabular}
\end{table}

\section{Conclusions}

The proposed VFO-ADR control system solves the path-following task for underactuated vehicles maintaining a non-banked motion. The derived controller can be applied to vehicles with highly uncertain model with both parametric and structural uncertainties.

The use of an error-based ESO, utilized in the ADR controller, allowed to apply the output-feedback solely from a vehicle configuration which is relatively easily measurable in practice.
With a non-parametrized path definition, the need of calculating a distance between a vehicle and a path has been avoided.

Presented simulation results showed that an application of a drift compensation component in the VFO controller substantially improves the positional control performance. Besides the fact that it is theoretically possible to obtain arbitrarily small positional errors for sufficiently large controller parameters, this result have practical limitations associated with the measurement noises, amplitude and rate limitations of actuators, and a limited frequency of computations that were not considered as a part of this work. 


\appendix
\color{black}
\section{Derivation of the term $\angularModifiedConvergenceVectorFieldEstimate$}
\label{app:angularModifiedConvergenceVectorFieldEstimate}
Since the vehicle configuration $\configurationVector$ is the only measurable signal (see Assumption \ref{ass:measurableConfiguration}), we propose to define
%
\begin{align}
	\label{eq:angularConvergenceVectorField}
	\angularConvergenceVectorField(\configurationVector,\longitudinalVelocityErrorVector,\longitudinalVelocityErrorVectorEstimate,\longitudinalVelocityErrorVectorEstimateDerivative)\triangleq\reducedAuxiliaryOrientationDerivative(\configurationVector,\longitudinalVelocityErrorVector,\longitudinalVelocityErrorVectorEstimate, \longitudinalVelocityErrorVectorEstimateDerivative)+\vfoOrientationGainMatrix\auxiliaryOrientationError(\positionVector,\longitudinalVelocityErrorVectorEstimate),
\end{align}
where $\vfoOrientationGainMatrix = \diag\{\vfoOrientationGainPitch, \ \vfoOrientationGainYaw\}: \vfoOrientationGainPitch,\vfoOrientationGainYaw>0$ is a gain matrix, and  $\reducedAuxiliaryOrientationDerivative(\cdot) = [\auxiliaryPitchAngleDerivative(\cdot) \ \auxiliaryYawAngleDerivative(\cdot)]^\top$. Values of particular $\reducedAuxiliaryOrientationDerivative$ elements result from differentiation of \eqref{eq:3.20} with respect to time, i.e.,
\begin{align}
 \label{eq:auxiliaryYawDerivative}
 \auxiliaryYawAngleDerivative &= \frac{\modifiedConvergenceVectorFieldEstimateDerivativeElement{y}(\configurationVector,\longitudinalVelocityErrorVector,\longitudinalVelocityErrorVectorEstimateDerivative)\modifiedConvergenceVectorFieldElementEstimate{x}(\positionVector,\longitudinalVelocityErrorVectorEstimate)-\modifiedConvergenceVectorFieldEstimateDerivativeElement{x}(\configurationVector,\longitudinalVelocityErrorVector,\longitudinalVelocityErrorVectorEstimateDerivative)\modifiedConvergenceVectorFieldElementEstimate{y}(\positionVector,\longitudinalVelocityErrorVectorEstimate)}{[\modifiedConvergenceVectorFieldElementEstimate{x}(\positionVector,\longitudinalVelocityErrorVectorEstimate)]^2+[\modifiedConvergenceVectorFieldElementEstimate{y}(\positionVector,\longitudinalVelocityErrorVectorEstimate)]^2}, \\
 \label{eq:auxiliaryPitchDerivative}
 \auxiliaryPitchAngleDerivative &= -\frac{{\beta}_2(\configurationVector,\longitudinalVelocityErrorVector,\longitudinalVelocityErrorVectorEstimate, \longitudinalVelocityErrorVectorEstimateDerivative)}{\beta_3(\positionVector,\longitudinalVelocityErrorVectorEstimate)},
\end{align}
where
\begin{align}
 {\beta}_2&(\configurationVector,\longitudinalVelocityErrorVector,\longitudinalVelocityErrorVectorEstimate,\longitudinalVelocityErrorVectorEstimateDerivative) = \nonumber \\ &[\modifiedConvergenceVectorFieldEstimateDerivativeElement{z}(\positionVector,\longitudinalVelocityErrorVector,\longitudinalVelocityErrorVectorEstimateDerivative)(\modifiedConvergenceVectorFieldElementEstimate{x}(\positionVector,\longitudinalVelocityErrorVectorEstimate)\cos\auxiliaryYawAngle+\modifiedConvergenceVectorFieldElementEstimate{y}(\positionVector,\longitudinalVelocityErrorVectorEstimate)\sin\auxiliaryYawAngle) \nonumber \\
 &-\modifiedConvergenceVectorFieldEstimateElement{z}(\positionVector,\longitudinalVelocityErrorVectorEstimate)(\modifiedConvergenceVectorFieldEstimateDerivativeElement{x}(\positionVector,\longitudinalVelocityErrorVector,\longitudinalVelocityErrorVectorEstimateDerivative)\cos\auxiliaryYawAngle \nonumber \\
 &+\modifiedConvergenceVectorFieldEstimateDerivativeElement{y}(\positionVector,\longitudinalVelocityErrorVector,\longitudinalVelocityErrorVectorEstimateDerivative)\sin\auxiliaryYawAngle -\modifiedConvergenceVectorFieldElementEstimate{x}(\positionVector,\longitudinalVelocityErrorVectorEstimate)\auxiliaryYawAngleDerivative(\configurationVector,\longitudinalVelocityErrorVector,\longitudinalVelocityErrorVectorEstimateDerivative)\sin\auxiliaryYawAngle \nonumber \\
 &+\modifiedConvergenceVectorFieldElementEstimate{y}(\positionVector,\longitudinalVelocityErrorVectorEstimate)\auxiliaryYawAngleDerivative(\configurationVector,\longitudinalVelocityErrorVector,\longitudinalVelocityErrorVectorEstimateDerivative)\cos\auxiliaryYawAngle)],
 \label{eq:3.26} \\
 \beta_3(&\positionVector,\longitudinalVelocityErrorVectorEstimate) = \nonumber \\
 &\left[\modifiedConvergenceVectorFieldElementEstimate{z}(\positionVector,\longitudinalVelocityErrorVectorEstimate)\right]^2+\left[\modifiedConvergenceVectorFieldElementEstimate{x}(\positionVector,\longitudinalVelocityErrorVectorEstimate)\cos\auxiliaryYawAngle+\modifiedConvergenceVectorFieldElementEstimate{y}(\positionVector,\longitudinalVelocityErrorVectorEstimate)\sin\auxiliaryYawAngle\right]^2.
 \label{eq:3.27}
\end{align}
A time derivative of a longitudinal part of modified convergence vector field \eqref{eq:hpstardef}, required to calculate \eqref{eq:auxiliaryYawDerivative} and \eqref{eq:auxiliaryPitchDerivative}, takes the form
\begin{align}
 \longitudinalModifiedConvergenceVectorFieldEstimateDerivative(\configurationVector,\longitudinalVelocityErrorVector,\longitudinalVelocityErrorVectorEstimateDerivative) &= \longitudinalConvergenceVectorFieldDerivative(\positionVector,\longitudinalVelocityErrorVector)+\longitudinalCompensationCoefficient\longitudinalVelocityErrorVectorEstimateDerivative,
\end{align}
while
%
\begin{align}
 \longitudinalConvergenceVectorFieldDerivative(\positionVector,\longitudinalVelocityErrorVector) \eq{eq:hpdef} \pathVelocityProfile\pathTangentialUnitVectorDerivative+[&\levelSurfaceDerivative{1}\pathNormalUnitVector{1}+\levelSurface{1}\pathNormalUnitVectorDerivative{1} +\levelSurfaceDerivative{2}\pathNormalUnitVector{2}+\levelSurface{2}\pathNormalUnitVectorDerivative{2}].
 \label{eq:longitudinalModifiedConvergenceVectorFieldEstimateDerivative}
\end{align}
Derivatives $\pathTangentialUnitVectorDerivative$, $\pathNormalUnitVectorDerivative{j}$, $\levelSurfaceDerivative{j}$ for $j\in\{1,2\}$, that are present in \eqref{eq:longitudinalModifiedConvergenceVectorFieldEstimateDerivative}, are described in detail in \ref{app:level-surface}.
%
%
%
Again, to assure the output-feedback characteristics of the control structure, we propose to use the estimate of the longitudinal velocity tracking error $\longitudinalVelocityErrorVectorEstimate$ instead of the original vector $\longitudinalVelocityErrorVector$ utilized in formula \eqref{eq:angularConvergenceVectorField}. Using this substitution, we obtain the modified version of the angular part of the convergence vector field in a form
\begin{align}
 \label{eq:angularConvergenceVectorFieldEstimate}
 \angularConvergenceVectorFieldEstimate(\configurationVector,\longitudinalVelocityErrorVectorEstimate,\longitudinalVelocityErrorVectorEstimateDerivative)=\reducedAuxiliaryOrientationDerivativeEstimate(\configurationVector,\longitudinalVelocityErrorVectorEstimate,\longitudinalVelocityErrorVectorEstimateDerivative)+\vfoOrientationGainMatrix\auxiliaryOrientationError(\positionVector,\longitudinalVelocityErrorVectorEstimate),
\end{align}
where the components of $\reducedAuxiliaryOrientationDerivativeEstimate(\cdot) \triangleq [\auxiliaryPitchAngleDerivativeEstimate(\cdot) \ \auxiliaryYawAngleDerivativeEstimate(\cdot)]^\top$, calculated according to formulas \eqref{eq:auxiliaryPitchDerivative} and \eqref{eq:auxiliaryYawDerivative}, are utilizing
the estimate of the derivative of estimated longitudinal part of convergence vector field
\begin{align}
 \longitudinalModifiedConvergenceVectorFieldEstimateDerivativeEstimate(\positionVector,\longitudinalVelocityErrorVectorEstimate,\longitudinalVelocityErrorVectorEstimateDerivative) &= \longitudinalConvergenceVectorFieldDerivativeEstimate(\positionVector,\longitudinalVelocityErrorVectorEstimate) + \longitudinalCompensationCoefficient\longitudinalVelocityErrorVectorEstimateDerivative,
 \label{eq:longitudinalModifiedConvergenceVectorFieldEstimateDerivativeEstimate}
\end{align}
instead of the elements of vector $\longitudinalModifiedConvergenceVectorFieldEstimateDerivative$
and
\begin{align}
 \longitudinalConvergenceVectorFieldDerivativeEstimate(\positionVector,\longitudinalVelocityErrorVector) = \pathVelocityProfile\pathTangentialUnitVectorDerivativeEstimate+[&\levelSurfaceDerivativeEstimate{1}\pathNormalUnitVector{1}+\levelSurface{1}\pathNormalUnitVectorDerivativeEstimate{1}+\levelSurfaceDerivativeEstimate{2}\pathNormalUnitVector{2}+\levelSurface{2}\pathNormalUnitVectorDerivativeEstimate{2}].
\end{align}
%
Estimates $\pathTangentialUnitVectorDerivativeEstimate$, $\pathNormalUnitVectorDerivativeEstimate{j}$, $\levelSurfaceDerivativeEstimate{j}$ for $j\in\{1,2\}$ are precisely described in \ref{app:level-surface}, while $\longitudinalVelocityErrorVectorEstimateDerivative$ consists of the second elements of vectors $\observerEstimatedStateVectorDerivative{i}$ for $i\in\{1,..,6\}$ calculated in the ESO equations \eqref{LESOdef}.
The final form of the modified convergence vector field is described as
\begin{align}
 \label{eq:angularModifiedConvergenceVectorFieldEstimate2}
 \angularModifiedConvergenceVectorFieldEstimate(\configurationVector, \reducedAngularVelocityErrorVectorEstimate, \longitudinalVelocityErrorVectorEstimateDerivative) = \angularConvergenceVectorFieldEstimate(\configurationVector,\longitudinalVelocityErrorVectorEstimate,\longitudinalVelocityErrorVectorEstimateDerivative) + \angularCompensationCoefficient\reducedAngularVelocityErrorVectorEstimate,
\end{align}
where $\reducedAngularVelocityErrorVectorEstimate = [\velocityErrorVectorElementEstimate{5} \ \velocityErrorVectorElementEstimate{6}]^{\top}$, while $\angularCompensationCoefficient\in[0,1]$ is a design parameter, introduced to deal with a possible overcompensation of the angular velocity tracking errors.

\color{black}
\section{Derivation of some path-related quantities}
\label{app:level-surface}

The nominal values of the level surface derivatives can be expressed, upon \eqref{etacfromnuc} and \eqref{epsilondef}, as
\begin{align}
    \label{eq:3.28}
    \levelSurfaceDerivative{j} &= \levelSurfaceGradientTranspose{j}\positionVectorDerivative=\levelSurfaceGradientTranspose{j}(\commandedPositionVectorDerivative-\longitudinalVelocityErrorVector).
\end{align}
%
According to the desired output-feedback characteristics of the VFO-ADR control structure, we propose to use the modified versions of level surface derivatives that are dependent only on the measurable signals, i.e.,
\begin{align}
  \label{eq:hatdots}
  \levelSurfaceDerivativeEstimate{j} &=\levelSurfaceGradientTranspose{j}(\commandedPositionVectorDerivative-\longitudinalVelocityErrorVectorEstimate).
\end{align}
The nominal forms of $\pathNormalUnitVectorDerivative{j}$ and $\pathTangentialUnitVectorDerivative$ are described by
\begin{align}
    \label{eq:3.29_1}
    \pathTangentialUnitVectorDerivative &= \pathMotionDirection\frac{\levelSurfaceGradientCrossProductDerivative\levelSurfaceGradientCrossProduct^\top  \levelSurfaceGradientCrossProduct-\levelSurfaceGradientCrossProduct\levelSurfaceGradientCrossProduct^\top  \levelSurfaceGradientCrossProductDerivative}{\module{\levelSurfaceGradientCrossProduct}^3}, \\
    \label{eq:3.30_1}
    \pathNormalUnitVectorDerivative{j} &= \frac{-\timeDerivative\levelSurfaceGradient{j}\module{\levelSurfaceGradient{j}} + \levelSurfaceGradient{j}\timeDerivative\module{\levelSurfaceGradient{j}}}{\module{\levelSurfaceGradient{j}}^2},
\end{align}
where $\levelSurfaceGradientCrossProduct \triangleq \levelSurfaceGradient{1}\times \levelSurfaceGradient{2}$, and
\begin{align}
    \levelSurfaceGradientCrossProductDerivative &= \timeDerivative\levelSurfaceGradient{1}\times \levelSurfaceGradient{2} + \levelSurfaceGradient{1}\times\timeDerivative\levelSurfaceGradient{2}, \\
    \label{eq:3.31}
    \timeDerivative\levelSurfaceGradient{j} &= \levelSurfaceHessian{j}\positionVectorDerivative \eq{epsilondef} \levelSurfaceHessian{j}(\commandedPositionVectorDerivative-\longitudinalVelocityErrorVector), \\
    \label{eq:3.32}
    \timeDerivative\module{\levelSurfaceGradient{j}} &=\frac{\levelSurfaceGradientTranspose{j}\levelSurfaceHessian{j}}{\module{\levelSurfaceGradient{j}}}\positionVectorDerivative \eq{epsilondef} \frac{\levelSurfaceGradientTranspose{j}\levelSurfaceHessian{j}}{\module{\levelSurfaceGradient{j}}}(\commandedPositionVectorDerivative-\longitudinalVelocityErrorVector),
\end{align}
for $j\in\{1,2\}$.
Analogously to the procedure utilized to transform \eqref{eq:3.28} into \eqref{eq:hatdots}, the modified versions of \eqref{eq:3.29_1} and \eqref{eq:3.30_1} have the folowing form:
\begin{align}
    \label{eq:3.29_2}
    \pathTangentialUnitVectorDerivativeEstimate &= \pathMotionDirection\frac{\levelSurfaceGradientCrossProductDerivativeEstimate\levelSurfaceGradientCrossProduct^\top  \levelSurfaceGradientCrossProduct-\levelSurfaceGradientCrossProduct\levelSurfaceGradientCrossProduct^\top  \levelSurfaceGradientCrossProductDerivativeEstimate}{\module{\levelSurfaceGradientCrossProduct}^3}, \\
    \label{eq:3.30_2}
    \pathNormalUnitVectorDerivativeEstimate{j} &= \frac{-\widehat{\timeDerivative\levelSurfaceGradient{j}}\module{\levelSurfaceGradient{j}} + \levelSurfaceGradient{j}\widehat{\timeDerivative\module{\levelSurfaceGradient{j}}}}{\module{\levelSurfaceGradient{j}}^2},
\end{align}
where
\begin{align}
  \label{eq:a10}
  \levelSurfaceGradientCrossProductDerivativeEstimate &= \widehat{\timeDerivative\levelSurfaceGradient{1}}\times \levelSurfaceGradient{2} + \levelSurfaceGradient{1}\times\widehat{\timeDerivative\levelSurfaceGradient{2}}, \\
  \widehat{\timeDerivative\levelSurfaceGradient{j}} &= \levelSurfaceHessian{j}(\commandedPositionVectorDerivative-\longitudinalVelocityErrorVectorEstimate), \\
  \widehat{\timeDerivative\module{\levelSurfaceGradient{j}}} &=\frac{\levelSurfaceGradientTranspose{j}\levelSurfaceHessian{j}}{\module{\levelSurfaceGradientTranspose{j}}}(\commandedPositionVectorDerivative-\longitudinalVelocityErrorVectorEstimate),
\end{align}
for $j\in\{1,2\}$.

\section{Velocity error dynamics}
\label{app:velocityErrorDynamics}

Transformations between the velocity dynamics introduced firstly in \eqref{eq:dotepsilon}, and the form represented by  \eqref{eq:velocityTrackingErrorDynamics} are based on the mutual relations between particular matrices and vectors utilized in the rigid body dynamics equations in frames $\localCoordinateSystem$ and $\globalCoordinateSystem$ (see \eqref{DynDefB}, \eqref{DynDefG}), the form of a modified disturbance vector introduced in \eqref{eq:modifiedTotalDisturbance}, and are explained as follows:
\begin{align}
    \velocityErrorVectorDerivative &\eq{eq:dotepsilon} \totalDisturbanceVector-\velocityTransformationMatrix\localInputMatrixEstimate\actuationMatrixLocal\velocityTransformationMatrix^\top\controlSignalGlobal \nonumber \\  
    &= \totalDisturbanceVector - \velocityTransformationMatrix\localInputMatrixEstimate\actuationMatrixLocal\localInputMatrixEstimate^{-1}\velocityTransformationMatrix^{-1}\left[\totalDisturbanceVectorEstimate+\adrcGainMatrixGlobal\velocityErrorVectorEstimate\right] \nonumber \\
    &= \velocityTransformationMatrix\velocityTransformationMatrix^{-1}\totalDisturbanceVector - \velocityTransformationMatrix\actuationMatrixLocal\velocityTransformationMatrix^{-1}\left[\totalDisturbanceVectorEstimate+\adrcGainMatrixGlobal\velocityErrorVectorEstimate\right] \nonumber \\
    &= -\velocityTransformationMatrix\actuationMatrixLocal\adrcGainMatrixLocal\velocityTransformationMatrix^{-1}\velocityErrorVector+\velocityTransformationMatrix\actuationMatrixLocal\adrcGainMatrixLocal\velocityTransformationMatrix^{-1}\velocityErrorVectorObservationError+\velocityTransformationMatrix\actuationMatrixLocal\velocityTransformationMatrix^{-1}\totalDisturbanceVectorObservationError \nonumber \\
    &+\velocityTransformationMatrix(\identityMatrix-\actuationMatrixLocal)\velocityTransformationMatrix^{-1}\totalDisturbanceVector \nonumber \\
    &\eq{eq:modifiedTotalDisturbance} -\velocityTransformationMatrix\actuationMatrixLocal\adrcGainMatrixLocal\velocityTransformationMatrix^{-1}\velocityErrorVector+\velocityTransformationMatrix\actuationMatrixLocal\adrcGainMatrixLocal\velocityTransformationMatrix^{-1}\velocityErrorVectorObservationError+\velocityTransformationMatrix\actuationMatrixLocal\velocityTransformationMatrix^{-1}\totalDisturbanceVectorObservationError \nonumber \\
    &+\velocityTransformationMatrix(\identityMatrix-\actuationMatrixLocal)\velocityTransformationMatrix^{-1}\left[\modifiedTotalDisturbanceVector-\massMatrixGlobal^{-1}\environmentalDampingMatrixGlobal\velocityErrorVector\right] \nonumber \\
    &= -\velocityTransformationMatrix\actuationMatrixLocal\adrcGainMatrixLocal\velocityTransformationMatrix^{-1}\velocityErrorVector+\velocityTransformationMatrix\actuationMatrixLocal\adrcGainMatrixLocal\velocityTransformationMatrix^{-1}\velocityErrorVectorObservationError+\velocityTransformationMatrix\actuationMatrixLocal\velocityTransformationMatrix^{-1}\totalDisturbanceVectorObservationError \nonumber \\
    &+\velocityTransformationMatrix(\identityMatrix-\actuationMatrixLocal)\velocityTransformationMatrix^{-1}\modifiedTotalDisturbanceVector-\velocityTransformationMatrix(\identityMatrix-\actuationMatrixLocal)\massMatrixLocal^{-1}\environmentalDampingMatrixLocal\velocityTransformationMatrix^{-1}\velocityErrorVector \nonumber \\
    &= -\velocityTransformationMatrix\left(\actuationMatrixLocal\adrcGainMatrixLocal+(\identityMatrix-\actuationMatrixLocal)\massMatrixLocal^{-1}\environmentalDampingMatrixLocal\right)\velocityTransformationMatrix^{-1}\velocityErrorVector+\velocityTransformationMatrix\actuationMatrixLocal\adrcGainMatrixLocal\velocityTransformationMatrix^{-1}\velocityErrorVectorObservationError \nonumber \\
    &+\velocityTransformationMatrix\actuationMatrixLocal\velocityTransformationMatrix^{-1}\totalDisturbanceVectorObservationError+\velocityTransformationMatrix(\identityMatrix-\actuationMatrixLocal)\velocityTransformationMatrix^{-1}\modifiedTotalDisturbanceVector,
\end{align}
where $\adrcGainMatrixLocal=\velocityTransformationMatrix^{-1}\adrcGainMatrixGlobal\velocityTransformationMatrix\triangleq\diag\{\adrcGainMatrixElementLocal{1},...,\adrcGainMatrixElementLocal{6}\}$ is the ADR controller gain matrix in the local reference frame.

\section{Derivation of $\reducedAuxiliaryOrientationDerivativeObservationError$}
\label{app:reducedAuxiliaryOrientationDerivativeObservationError}

To calculate the upper-bound of $\reducedAuxiliaryOrientationDerivativeObservationError$, introduced in \eqref{eq:auxiliaryOrientationErrorDynamics}, the error between the nominal and approximated longitudinal convergence vector field derivatives
\begin{align}
    \longitudinalModifiedConvergenceVectorFieldEstimateDerivativeError = [\modifiedConvergenceVectorFieldEstimateDerivativeErrorElement{x} \ \modifiedConvergenceVectorFieldEstimateDerivativeErrorElement{x} \ \modifiedConvergenceVectorFieldEstimateDerivativeErrorElement{z}]^\top\triangleq \longitudinalModifiedConvergenceVectorFieldEstimateDerivative - \longitudinalModifiedConvergenceVectorFieldEstimateDerivativeEstimate
\end{align}
have to be determined and evaluated.
Based on equations \eqref{eq:longitudinalModifiedConvergenceVectorFieldEstimateDerivative} and \eqref{eq:longitudinalModifiedConvergenceVectorFieldEstimateDerivativeEstimate}, we can write down that
\begin{align}
  \longitudinalModifiedConvergenceVectorFieldEstimateDerivativeError = \pathVelocityProfile\pathTangentialUnitVectorDerivativeError+\levelSurfaceDerivativeError{1}\pathNormalUnitVector{1}+\levelSurface{1}\pathNormalUnitVectorDerivativeError{1}+\levelSurfaceDerivativeError{2}\pathNormalUnitVector{2}+\levelSurface{2}\pathNormalUnitVectorDerivativeError{2}
\end{align}
where $\pathTangentialUnitVectorDerivativeError \triangleq \pathTangentialUnitVectorDerivative-\pathTangentialUnitVectorDerivativeEstimate$, $\pathNormalUnitVectorDerivativeError{j} \triangleq \pathNormalUnitVectorDerivative{j}-\pathNormalUnitVectorDerivativeEstimate{j}$, and $\levelSurfaceDerivativeError{j} \triangleq \levelSurfaceDerivative{j} -\levelSurfaceDerivativeEstimate{j}$ for $j\in\{1,2\}$. Then, according to the derivations presented in the \ref{app:level-surface}, we know that
\begin{align}
  \levelSurfaceDerivativeError{j} &= \levelSurfaceDerivative{j}-\levelSurfaceDerivativeEstimate{j} = \levelSurfaceGradientTranspose{j}(\longitudinalVelocityErrorVectorEstimate-\longitudinalVelocityErrorVector) = \levelSurfaceGradientTranspose{j}\longitudinalVelocityErrorVectorObservationError, \\
  \pathTangentialUnitVectorDerivativeError &= \pathMotionDirection\frac{\levelSurfaceGradientCrossProductDerivativeError\levelSurfaceGradientCrossProduct^\top  \levelSurfaceGradientCrossProduct-\levelSurfaceGradientCrossProduct\levelSurfaceGradientCrossProduct^\top  \levelSurfaceGradientCrossProductDerivativeError}{\module{\levelSurfaceGradientCrossProduct}^3}, \\
  \pathNormalUnitVectorDerivative{j} &= \frac{-\widetilde{\timeDerivative\levelSurfaceGradient{j}}\module{\levelSurfaceGradient{j}} + \levelSurfaceGradient{j}\widetilde{\timeDerivative\module{\levelSurfaceGradient{j}}}}{\module{\levelSurfaceGradient{j}}^2},
\end{align}
where
\begin{align}
  \levelSurfaceGradientCrossProductDerivativeError &\triangleq \levelSurfaceGradientCrossProductDerivative - \levelSurfaceGradientCrossProductDerivativeEstimate \eq{eq:a10} \widetilde{\timeDerivative\levelSurfaceGradient{1}}\times\levelSurfaceGradient{2}+\levelSurfaceGradient{1}\times\widetilde{\timeDerivative\levelSurfaceGradient{2}}, \\
  \widetilde{\timeDerivative\levelSurfaceGradient{j}} &\triangleq\timeDerivative\levelSurfaceGradient{2}-\widehat{\timeDerivative\levelSurfaceGradient{2}} =  \levelSurfaceHessian{j}\longitudinalVelocityErrorVectorObservationError, \\
  \widetilde{\timeDerivative\module{\levelSurfaceGradient{j}}} &\triangleq \timeDerivative\module{\levelSurfaceGradient{j}} - \widehat{\timeDerivative\module{\levelSurfaceGradient{j}}} = \frac{\levelSurfaceGradientTranspose{j}\levelSurfaceHessian{j}}{\module{\levelSurfaceGradient{j}}}\longitudinalVelocityErrorVectorObservationError.
\end{align}
Upon the above forms and Assumptions \ref{ass:3} and \ref{ass:4}, we can obtain the following set of conservative bounds
\begin{align}
  \module{\levelSurfaceDerivativeError{j}}&\leq\module{\levelSurfaceGradient{j}}\module{\longitudinalVelocityErrorVectorObservationError}\leq\levelSurfaceGradientUpperBound{j}\module{\observationErrorAggregatedVector}, \\
  \module{\widetilde{\timeDerivative\levelSurfaceGradient{j}}}&\leq \module{\levelSurfaceHessian{j}}\module{\longitudinalVelocityErrorVectorObservationError} \leq \levelSurfaceHessianUpperBound{j}\module{\observationErrorAggregatedVector}, \\
  \module{\widetilde{\timeDerivative\module{\levelSurfaceGradient{j}}}}&\leq \frac{\module{\levelSurfaceGradient{j}}\module{\levelSurfaceHessian{j}}}{\module{\levelSurfaceGradient{j}}}\module{\longitudinalVelocityErrorVectorObservationError}\leq\levelSurfaceHessianUpperBound{j}\module{\observationErrorAggregatedVector}, \\
  \module{\levelSurfaceGradientCrossProductDerivativeError} &\leq (\levelSurfaceHessianUpperBound{1}\levelSurfaceGradientUpperBound{2}+\levelSurfaceGradientUpperBound{1}\levelSurfaceHessianUpperBound{2})\module{\observationErrorAggregatedVector}, \\
  \module{\levelSurfaceGradientCrossProduct}&\geq \levelSurfaceGradientLowerBound{1}\levelSurfaceGradientLowerBound{2}, \\
  \module{\pathTangentialUnitVectorDerivativeError} &\leq \frac{2(\levelSurfaceHessianUpperBound{1}\levelSurfaceGradientUpperBound{2}+\levelSurfaceGradientUpperBound{1}\levelSurfaceHessianUpperBound{2})}{\levelSurfaceGradientLowerBound{1}\levelSurfaceGradientLowerBound{2}}\module{\observationErrorAggregatedVector}, \\
  \module{\pathNormalUnitVectorDerivativeError{j}}&\leq\frac{2\levelSurfaceHessianUpperBound{j}}{\levelSurfaceGradientLowerBound{j}}\module{\observationErrorAggregatedVector},
\end{align}
and determine the upper bound function of the estimation error of the convergence vector field derivative
\begin{align}
  \module{\longitudinalModifiedConvergenceVectorFieldEstimateDerivativeError}&\leq\abs{\longitudinalConvergenceVectorFieldEstimateDerivativeErrorAuxiliaryFunction}\module{\observationErrorAggregatedVector},
  \label{eq:tildedothath}
\end{align}
with
\begin{align}
  \abs{\longitudinalConvergenceVectorFieldEstimateDerivativeErrorAuxiliaryFunction} &= \abs{\pathVelocityProfile}\frac{2(\levelSurfaceHessianUpperBound{1}\levelSurfaceGradientUpperBound{2}+\levelSurfaceGradientUpperBound{1}\levelSurfaceHessianUpperBound{2})}{\levelSurfaceGradientLowerBound{1}\levelSurfaceGradientLowerBound{2}}+(\levelSurfaceGradientUpperBound{1}+\levelSurfaceGradientUpperBound{2}) \nonumber \\
  &+\frac{2\levelSurfaceHessianUpperBound{1}}{\levelSurfaceGradientLowerBound{1}}\abs{\levelSurface{1}}+\frac{2\levelSurfaceHessianUpperBound{2}}{\levelSurfaceGradientLowerBound{2}}\abs{\levelSurface{2}}.
\end{align}

Now, having in mind the result from \eqref{eq:tildedothath}, an analysis of the particular auxiliary angle derivative errors can be written down as
\begin{align}
    \auxiliaryYawAngleDerivativeError &= \frac{\modifiedConvergenceVectorFieldEstimateDerivativeErrorElement{y}\modifiedConvergenceVectorFieldElementEstimate{x}-\modifiedConvergenceVectorFieldEstimateDerivativeErrorElement{x}\modifiedConvergenceVectorFieldElementEstimate{y}}{{\modifiedConvergenceVectorFieldElementEstimate{x}}^2+{\modifiedConvergenceVectorFieldElementEstimate{y}}^2}, \\
    \auxiliaryPitchAngleDerivativeError &=\frac{ -{\beta}_6}{{\modifiedConvergenceVectorFieldElementEstimate{z}}^2+(\modifiedConvergenceVectorFieldElementEstimate{x}\cos\auxiliaryYawAngle+\modifiedConvergenceVectorFieldElementEstimate{y}\sin\auxiliaryYawAngle)^2},
\end{align}
where
\begin{align}
	{\beta}_6&=[\modifiedConvergenceVectorFieldEstimateDerivativeErrorElement{z}(\modifiedConvergenceVectorFieldElementEstimate{x}\cos\auxiliaryYawAngle+\modifiedConvergenceVectorFieldElementEstimate{y}\sin\auxiliaryYawAngle)\nonumber \\
	&-\modifiedConvergenceVectorFieldEstimateElement{z}(\modifiedConvergenceVectorFieldEstimateDerivativeErrorElement{x}\cos\auxiliaryYawAngle+\modifiedConvergenceVectorFieldEstimateDerivativeErrorElement{y}\sin\auxiliaryYawAngle -\modifiedConvergenceVectorFieldElementEstimate{x}\auxiliaryYawAngleDerivativeError\sin\auxiliaryYawAngle+\modifiedConvergenceVectorFieldElementEstimate{y}\auxiliaryYawAngleDerivativeError].
\end{align}
The absolute values of $\reducedAuxiliaryOrientationDerivativeObservationError$ elements can be then estimated as
\begin{align}
  \abs{\auxiliaryYawAngleDerivativeError}&\leq\underbrace{\frac{\abs{\modifiedConvergenceVectorFieldEstimateElement{x}}+\abs{\modifiedConvergenceVectorFieldEstimateElement{y}}}{{\modifiedConvergenceVectorFieldEstimateElement{x}}^2+{\modifiedConvergenceVectorFieldEstimateElement{x}}^2}}_{\abs{\auxiliaryYawDerivativeErrorFunction}}\module{\longitudinalConvergenceVectorFieldEstimateDerivativeError} \leq \abs{\auxiliaryYawDerivativeErrorFunction}\abs{\longitudinalConvergenceVectorFieldEstimateDerivativeErrorAuxiliaryFunction}\module{\observationErrorAggregatedVector}, \\
  \abs{\auxiliaryPitchAngleDerivativeError} &\leq \abs{\auxiliaryPitchDerivativeErrorFunction}\module{\longitudinalConvergenceVectorFieldEstimateDerivativeError} \leq \abs{\auxiliaryPitchDerivativeErrorFunction}\abs{\longitudinalConvergenceVectorFieldEstimateDerivativeErrorAuxiliaryFunction}\module{\observationErrorAggregatedVector},
\end{align}
where
\begin{align}
  \abs{\auxiliaryPitchDerivativeErrorFunction} = \frac{\abs{\modifiedConvergenceVectorFieldEstimateElement{x}}+\abs{\modifiedConvergenceVectorFieldEstimateElement{y}}+2\abs{\modifiedConvergenceVectorFieldEstimateElement{z}}+\abs{\modifiedConvergenceVectorFieldEstimateElement{z}}(\abs{\modifiedConvergenceVectorFieldEstimateElement{x}}+\abs{\modifiedConvergenceVectorFieldEstimateElement{y}})\abs{\auxiliaryYawDerivativeErrorFunction}}{{\modifiedConvergenceVectorFieldElementEstimate{z}}^2+(\modifiedConvergenceVectorFieldElementEstimate{x}\cos\auxiliaryYawAngle+\modifiedConvergenceVectorFieldElementEstimate{y}\sin\auxiliaryYawAngle)^2}.
\end{align}
Now, the upper bound of the auxiliary angle derivative error
\begin{align}
  \module{\reducedAuxiliaryOrientationDerivativeObservationError(t)} &\leq \sup_{t\geq0}(\abs{\auxiliaryYawDerivativeErrorFunction(t)}\abs{\longitudinalConvergenceVectorFieldEstimateDerivativeErrorAuxiliaryFunction(t)}+\abs{\auxiliaryPitchDerivativeErrorFunction(t)}\abs{\longitudinalConvergenceVectorFieldEstimateDerivativeErrorAuxiliaryFunction(t)})\module{\observationErrorAggregatedVector(t)}.
\end{align}

Since the analysis provided in this article is assumed to be valid locally and we assume that the initial conditions of any vector are in some compact set (see Remark \ref{rem:dbound}), the expression $\sup_{t\geq0}(\abs{\auxiliaryYawDerivativeErrorFunction(t)}\abs{\longitudinalConvergenceVectorFieldEstimateDerivativeErrorAuxiliaryFunction(t)}+\abs{\auxiliaryPitchDerivativeErrorFunction(t)}\abs{\longitudinalConvergenceVectorFieldEstimateDerivativeErrorAuxiliaryFunction(t)})$ is finite.

\section{Estimation of the  $\positionErrorLyapunovFunctionDerivative$ }
\label{app:A}
According to the definition of the Lyapunov function for the position error dynamics \eqref{eq:Vp}, and upon dynamics  \eqref{eq:s1dot} and \eqref{eq:s2dot}, a derivative of the Lyapunov function can be written down as
\begin{align}
  \positionErrorLyapunovFunctionDerivative&=\frac{1}{\module{\levelSurfaceGradient{1}}}\levelSurface{1}\levelSurfaceDerivative{1}+\frac{1}{\module{\levelSurfaceGradient{2}}}\levelSurface{2}\levelSurfaceDerivative{2}\nonumber \\
  &=\left(\levelSurface{1}\pathNormalUnitVector{1}^\top+\levelSurface{2}\pathNormalUnitVector{2}^\top\right)(\commandedPositionVectorDerivative-\longitudinalVelocityErrorVector-\longitudinalModifiedConvergenceVectorFieldEstimate+\longitudinalCompensationCoefficient\longitudinalVelocityErrorVectorEstimate) \nonumber \\
  &-\vfoLongitudinalGain\positionError^\top\underbrace{\begin{bmatrix} 1 & \cos\angleBetweenlevelSurfaceGradients \\ \cos\angleBetweenlevelSurfaceGradients & 1\end{bmatrix}}_{\positionErrorStateMatrix}\positionError \nonumber \\
  &= -\vfoLongitudinalGain\positionError^\top\positionErrorStateMatrix\positionError-\positionError^\top \begin{bmatrix} \pathNormalUnitVector{1}^\top \\ \pathNormalUnitVector{2}^\top \end{bmatrix}\positionErrorResidueVector - \positionError^\top \begin{bmatrix} \pathNormalUnitVector{1}^\top \\ \pathNormalUnitVector{2}^\top \end{bmatrix}\left((1-\longitudinalCompensationCoefficient)\longitudinalVelocityErrorVector+\longitudinalCompensationCoefficient\longitudinalVelocityErrorVectorObservationError\right),
  \label{eq:VpDotDerivation}
\end{align}
where $\angleBetweenlevelSurfaceGradients\angle(\levelSurfaceGradient{1},\levelSurfaceGradient{2})\neq k\pi, \ k\in\integerNumbers$ correspond to the angle between the level-surface gradients with the values constrained according to Assumption \ref{ass:nonCollinearity}, and the residual vector
\begin{align}
  \label{eq:residualVector}
  \positionErrorResidueVector \triangleq \longitudinalModifiedConvergenceVectorFieldEstimate - \commandedPositionVectorDerivative = \longitudinalModifiedConvergenceVectorFieldEstimate - \rotationMatrix\longitudinalCommandedVelocityVectorLocal =  \longitudinalModifiedConvergenceVectorFieldEstimate-\commandedXLocalVelocity\positionErrorDirectionVector,
\end{align}
for
\begin{align}
    \positionErrorDirectionVector \triangleq [\cos\pitch\cos\yaw \ \cos\pitch\sin\yaw \ -\sin\pitch]^\top, \  \module{\positionErrorDirectionVector}=1.
\end{align}
Referring to equation \eqref{nucbardef}, we can express the residual vector from \eqref{eq:residualVector} as
\begin{align}
    \positionErrorResidueVector &= \module{\longitudinalModifiedConvergenceVectorFieldEstimate}\left(\begin{bmatrix}\frac{\modifiedConvergenceVectorFieldEstimateElement{x}}{\module{\longitudinalModifiedConvergenceVectorFieldEstimate}} \\ \frac{\modifiedConvergenceVectorFieldEstimateElement{y}}{\module{\longitudinalModifiedConvergenceVectorFieldEstimate}} \\ \frac{\modifiedConvergenceVectorFieldEstimateElement{z}}{\module{\longitudinalModifiedConvergenceVectorFieldEstimate}}\end{bmatrix}-\begin{bmatrix}\cos\positionErrorResidueAngle \cos\pitch \cos\yaw \\ \cos\positionErrorResidueAngle \cos\pitch \sin\yaw \\ -\cos\positionErrorResidueAngle \sin\pitch \end{bmatrix}\right),
\end{align}
where $\positionErrorResidueAngle = \angle(\longitudinalModifiedConvergenceVectorFieldEstimate,\positionErrorDirectionVector)$,
and write down that
\begin{align}
  \module{\positionErrorResidueVector}^2 &= \module{\longitudinalModifiedConvergenceVectorFieldEstimate}^2\Big[\frac{{\modifiedConvergenceVectorFieldEstimateElement{x}}^2}{\module{\longitudinalModifiedConvergenceVectorFieldEstimate}^2}-\frac{2\modifiedConvergenceVectorFieldEstimateElement{x}}{\module{\longitudinalModifiedConvergenceVectorFieldEstimate}}\cos\positionErrorResidueAngle \cos\pitch \cos\yaw+\cos^2\positionErrorResidueAngle \cos^2\pitch \cos^2\yaw \nonumber\\
	&+ \frac{{\modifiedConvergenceVectorFieldEstimateElement{y}}^2}{\module{\longitudinalModifiedConvergenceVectorFieldEstimate}^2}-\frac{2\modifiedConvergenceVectorFieldEstimateElement{y}}{\module{\longitudinalModifiedConvergenceVectorFieldEstimate}}\cos\positionErrorResidueAngle \cos\pitch \sin\yaw + \cos^2\positionErrorResidueAngle \cos^2\pitch \sin^2\yaw \nonumber\\
  &+\frac{{\modifiedConvergenceVectorFieldEstimateElement{z}}^2}{\module{\longitudinalModifiedConvergenceVectorFieldEstimate}^2}+\frac{2\modifiedConvergenceVectorFieldEstimateElement{z}}{\module{\longitudinalModifiedConvergenceVectorFieldEstimate}}\cos\positionErrorResidueAngle \sin\pitch +\cos^2\positionErrorResidueAngle \sin^2\pitch \Big] \nonumber\\
	&= \module{\longitudinalModifiedConvergenceVectorFieldEstimate}^2\Big[1-2\cos\positionErrorResidueAngle \underbrace{\frac{\modifiedConvergenceVectorFieldEstimateElement{x} \cos\pitch \cos\yaw + \modifiedConvergenceVectorFieldEstimateElement{y} \cos\pitch \sin\yaw - \modifiedConvergenceVectorFieldEstimateElement{z} \sin\pitch}{\module{\longitudinalModifiedConvergenceVectorFieldEstimate}}}_{{\cos\positionErrorResidueAngle}}+\cos^2\positionErrorResidueAngle\Big]\nonumber\\
	&=\module{\longitudinalModifiedConvergenceVectorFieldEstimate}^2(1-\cos^2\positionErrorResidueAngle). \label{eq:r2}
\end{align}
Using the interpretation of auxiliary angles $\auxiliaryPitchAngle$  and $\auxiliaryYawAngle$ defined in \eqref{eq:3.20}, the longitudinal part of the modified convergence vector field can be rewritten as
\begin{align}
    \longitudinalModifiedConvergenceVectorFieldEstimate &= \module{\longitudinalModifiedConvergenceVectorFieldEstimate}\begin{bmatrix}\pathVelocityDirection \cos \auxiliaryPitchAngle \cos\auxiliaryYawAngle \\ \pathVelocityDirection \cos\auxiliaryPitchAngle \sin\auxiliaryYawAngle \\ -\pathVelocityDirection \sin\auxiliaryPitchAngle\end{bmatrix},
\end{align}
and
\begin{align}
    \cos\positionErrorResidueAngle &= \frac{{\longitudinalModifiedConvergenceVectorFieldEstimate}^\top\positionErrorDirectionVector}{\module{\longitudinalModifiedConvergenceVectorFieldEstimate}\module{\positionErrorDirectionVector}} = \begin{bmatrix} \pathVelocityDirection \cos\auxiliaryPitchAngle \cos\auxiliaryYawAngle & \pathVelocityDirection \cos\auxiliaryPitchAngle \sin\auxiliaryYawAngle & -\pathVelocityDirection \sin\auxiliaryPitchAngle \end{bmatrix}\begin{bmatrix}\cos\pitch \cos\yaw \\ \cos\pitch \sin\yaw \\ -\sin\pitch \end{bmatrix} \nonumber \\
    &= \pathVelocityDirection\cos\auxiliaryPitchAngle\cos\auxiliaryYawAngle\cos\pitch\cos\yaw+\pathVelocityDirection\cos\auxiliaryPitchAngle\sin\auxiliaryYawAngle\cos\pitch\sin\yaw+\pathVelocityDirection\sin\auxiliaryPitchAngle\sin\pitch \nonumber\\
		&=\pathVelocityDirection\cos\auxiliaryPitchAngle\cos\pitch(\cos\auxiliaryYawAngle\cos\yaw+\sin\auxiliaryYawAngle\sin\yaw)+\pathVelocityDirection\sin\auxiliaryPitchAngle\sin\pitch \nonumber\\
    &= \pathVelocityDirection\cos\auxiliaryPitchAngle\cos\pitch\cos\auxiliaryOrientationErrorElement{\yaw}+\pathVelocityDirection\sin\auxiliaryPitchAngle\sin\pitch.
\end{align}
Using the trygonometric identities
\begin{align}
    &\cos\pitch = \cos(\auxiliaryPitchAngle-\auxiliaryOrientationErrorElement{\pitch}) = \cos\auxiliaryPitchAngle\cos\auxiliaryOrientationErrorElement{\pitch}+\sin\auxiliaryPitchAngle\sin\auxiliaryOrientationErrorElement{\pitch} \\
    &\sin\pitch = \sin(\auxiliaryPitchAngle - \auxiliaryOrientationErrorElement{\pitch}) = \sin\auxiliaryPitchAngle\cos\auxiliaryOrientationErrorElement{\pitch} - \cos\auxiliaryPitchAngle \sin\auxiliaryOrientationErrorElement{\pitch} \\
    &\cos\yaw = \cos(\auxiliaryYawAngle-\auxiliaryOrientationErrorElement{\yaw}) = \cos\auxiliaryYawAngle\cos\auxiliaryOrientationErrorElement{\yaw}+\sin\auxiliaryYawAngle\sin\auxiliaryOrientationErrorElement{\yaw} \\
    &\sin\yaw = \sin(\auxiliaryYawAngle - \auxiliaryOrientationErrorElement{\yaw}) = \sin\auxiliaryYawAngle\cos\auxiliaryOrientationErrorElement{\yaw} - \cos\auxiliaryYawAngle \sin\auxiliaryOrientationErrorElement{\yaw} \\
    &(1-\cos\auxiliaryOrientationErrorElement{\yaw}) = 2\sin^2\frac{\auxiliaryOrientationErrorElement{\yaw}}{2},
\end{align}
together with the equivalence
\begin{align}
    1 &\equiv (\sin^2\auxiliaryPitchAngle+\cos^2\auxiliaryPitchAngle)^2(\cos^2\auxiliaryOrientationErrorElement{\pitch}\cos^2\auxiliaryOrientationErrorElement{\yaw}+\cos^2\auxiliaryOrientationErrorElement{\pitch}\sin^2\auxiliaryOrientationErrorElement{\yaw}+\sin^2\auxiliaryOrientationErrorElement{\pitch}),  \label{eq:jeden}
\end{align}
after some algebraic transformations, we can write down that
{\small
\begin{align}
    1-\cos^2\positionErrorResidueAngle &= 1-\left[\cos^2\auxiliaryPitchAngle \cos^2\pitch\cos^2\auxiliaryOrientationErrorElement{\yaw}+2\cos\auxiliaryPitchAngle \cos\pitch\cos\auxiliaryOrientationErrorElement{\yaw}\sin\auxiliaryPitchAngle \sin\pitch+\sin^2\auxiliaryPitchAngle \sin^2\pitch\right] \nonumber\\
    &= -\sin^4\auxiliaryPitchAngle\cos^2\auxiliaryOrientationErrorElement{\pitch}\sin^2\auxiliaryOrientationErrorElement{\yaw}+\sin^4\auxiliaryPitchAngle\cos^2\auxiliaryOrientationErrorElement{\pitch}\sin^2\auxiliaryOrientationErrorElement{\yaw} \nonumber\\
		&+\sin^2\auxiliaryOrientationErrorElement{\pitch}(\sin^4\auxiliaryPitchAngle+\cos^4\auxiliaryPitchAngle)-4\sin^2\auxiliaryPitchAngle\cos^2\auxiliaryPitchAngle\cos^2\auxiliaryOrientationErrorElement{\pitch}\sin^2\frac{\auxiliaryOrientationErrorElement{\yaw}}{2}\cos\auxiliaryOrientationErrorElement{\yaw} \nonumber\\
		&+2\sin^2\auxiliaryPitchAngle\cos^2\auxiliaryPitchAngle\cos^2\auxiliaryOrientationErrorElement{\pitch}\sin^2\auxiliaryOrientationErrorElement{\yaw} \nonumber\\
		&+2\cos^2\auxiliaryPitchAngle\sin^2\auxiliaryPitchAngle\sin^2\auxiliaryOrientationErrorElement{\pitch}(1+\cos\auxiliaryOrientationErrorElement{\yaw})+\cos^4\auxiliaryPitchAngle\cos^2\auxiliaryOrientationErrorElement{\pitch}\sin^2\auxiliaryOrientationErrorElement{\yaw} \nonumber\\
    &+4\cos\auxiliaryPitchAngle\sin\auxiliaryPitchAngle\cos\auxiliaryOrientationErrorElement{\pitch}\sin\auxiliaryOrientationErrorElement{\pitch}\sin^2\frac{\auxiliaryOrientationErrorElement{\yaw}}{2}(\cos^2\auxiliaryPitchAngle\cos\auxiliaryOrientationErrorElement{\yaw}+\sin^2\auxiliaryPitchAngle)\nonumber\\
		&-\cos^2\auxiliaryPitchAngle\sin^2\auxiliaryPitchAngle\sin^2\auxiliaryOrientationErrorElement{\pitch}(1+\cos^2\auxiliaryOrientationErrorElement{\yaw}).
    \label{eq:1mcalfa}
\end{align}}
Knowing that  $\abs{\cos\rho}\leq1$, $\abs{\sin\rho}\leq1$, and  $\abs{\sin\rho}\leq\abs{\rho}$, an upper bound of expression \eqref{eq:1mcalfa} can be (conservatively) assesed as follows
\begin{align}
    \abs{1-\cos^2\positionErrorResidueAngle}&\leq7\abs{\auxiliaryOrientationErrorElement{\pitch}}^2+5\abs{\auxiliaryOrientationErrorElement{\yaw}}^2 \nonumber\\
		&= \begin{bmatrix} \auxiliaryOrientationErrorElement{\pitch} & \auxiliaryOrientationErrorElement{\yaw}\end{bmatrix}\begin{bmatrix} 7 & 0 \\ 0 & 5 \end{bmatrix}\begin{bmatrix} \auxiliaryOrientationErrorElement{\pitch} \\ \auxiliaryOrientationErrorElement{\yaw}\end{bmatrix} \nonumber\\
		&= \auxiliaryOrientationError\positionErrorResidueAngleStateMatrix\auxiliaryOrientationError\leq\maxEigenvalue{\positionErrorResidueAngleStateMatrix}\module{\auxiliaryOrientationError}^2 =7\module{\auxiliaryOrientationError}^2,
    \label{eq:cosAlpha}
\end{align}
implying that
\begin{align}
    \sqrt{1-\cos^2\positionErrorResidueAngle}=\sqrt{\abs{1-\cos^2\positionErrorResidueAngle}}\leq\sqrt{7\module{\auxiliaryOrientationError}^2}=\sqrt{7}\module{\auxiliaryOrientationError}. \label{eq:calfaass}
\end{align}
The upper bound function of the longitudinal modified convergence vector field estimate, according to \eqref{eq:hpstardef}, \eqref{eq:hpdef}, and \eqref{eq:ObsErrorVect}, may be written down in the form
\begin{align}
  \label{eq:longitudinalModifiedConvergenceVectorFieldUpperBound}
  \module{\longitudinalModifiedConvergenceVectorFieldEstimate} &\leq \abs{\pathVelocityProfile}+\vfoLongitudinalGain\module{\positionError}+\longitudinalCompensationCoefficient(\module{\longitudinalVelocityErrorVector}+\module{\longitudinalVelocityErrorVectorObservationError}).
\end{align}
According to results \eqref{eq:r2}, \eqref{eq:cosAlpha}, and \eqref{eq:longitudinalModifiedConvergenceVectorFieldUpperBound},
we can say that the derivative of $\positionErrorLyapunovFunction$ is bounded by
\begin{align}
    \positionErrorLyapunovFunctionDerivative&= -\vfoLongitudinalGain\positionError^\top\positionErrorStateMatrix\positionError-\positionError^\top \begin{bmatrix} \pathNormalUnitVector{1}^\top \\ \pathNormalUnitVector{2}^\top \end{bmatrix}\positionErrorResidueVector - \positionError^\top \begin{bmatrix} \pathNormalUnitVector{1}^\top \\ \pathNormalUnitVector{2}^\top \end{bmatrix}\left((1-\longitudinalCompensationCoefficient)\longitudinalVelocityErrorVector+\longitudinalCompensationCoefficient\longitudinalVelocityErrorVectorObservationError\right) \nonumber \\
    &\leq  -\vfoLongitudinalGain\positionError^\top\positionErrorStateMatrix\positionError+2\module{\positionError}\left(\module{\positionErrorResidueVector}+(1-\longitudinalCompensationCoefficient)\module{\longitudinalVelocityErrorVector}+\longitudinalCompensationCoefficient\module{\longitudinalVelocityErrorVectorObservationError}\right)  \nonumber \\
    &\leq  -\vfoLongitudinalGain(1-\positionErrorMajorizationCoefficient)\positionError^\top\positionErrorStateMatrix\positionError+\module{\positionError}\Big[2(1-\longitudinalCompensationCoefficient)\module{\longitudinalVelocityErrorVector} \nonumber \\
    &+2\longitudinalCompensationCoefficient\module{\longitudinalVelocityErrorVectorObservationError}+2\sqrt{7}\abs{\pathVelocityProfile}\module{\auxiliaryOrientationError}+2\sqrt{7}\longitudinalCompensationCoefficient(\module{\longitudinalVelocityErrorVector}+\module{\longitudinalVelocityErrorVectorObservationError})\module{\auxiliaryOrientationError} \nonumber \\
    &-\vfoLongitudinalGain\module{\positionError}(\minEigenvalue{\positionErrorStateMatrix}\positionErrorMajorizationCoefficient-2\sqrt{7}\module{\auxiliaryOrientationError})\Big].
\end{align}


\section{Difference between the desired and auxiliary yaw angles}
\label{app:epsi}

The transformation from \eqref{eq:varepsilondef} into \eqref{eq:varepsilonpsiabs}, was done using the relation
\begin{align}
\atantwo(Y_1, X_1) - \atantwo(Y_2, X_2) = \atantwo(\rho_6, \rho_7),
\end{align}
where $\rho_6:=Y_1X_2-Y_2X_1$, and $\rho_7:= X_1X_2+Y_1Y_2$. Subsequent transformation can be explained as follows:
\begin{align}
	\auxiliaryDesiredYawDifference&\triangleq\desiredYawAngle-\auxiliaryBoundedYawAngle = \desiredYawAngle-\atantwo(\sin\auxiliaryYawAngle,\cos\auxiliaryYawAngle) \nonumber \\
  &=\atantwo(\pathVelocityDirection\pathTangentialUnitVectorElement{y},\pathVelocityDirection\pathTangentialUnitVectorElement{x})-\atantwo(\pathVelocityDirection\modifiedConvergenceVectorFieldElementEstimate{y},\pathVelocityDirection\modifiedConvergenceVectorFieldElementEstimate{x}) \nonumber \\
  &= \atantwo(\modifiedConvergenceVectorFieldElementEstimate{x}\pathTangentialUnitVectorElement{y} - \modifiedConvergenceVectorFieldElementEstimate{y}\pathTangentialUnitVectorElement{x}, \modifiedConvergenceVectorFieldElementEstimate{x}\pathTangentialUnitVectorElement{x}+\modifiedConvergenceVectorFieldElementEstimate{y} \pathTangentialUnitVectorElement{y}) \nonumber \\
  &\eq{eq:hpstardef}\atantwo(-\vfoLongitudinalGain \levelSurface{1}[\pathNormalUnitVectorElement{1}{x}\pathTangentialUnitVectorElement{y}-\pathNormalUnitVectorElement{1}{y}\pathTangentialUnitVectorElement{x}] \nonumber \\
  &-\vfoLongitudinalGain \levelSurface{2}[\pathNormalUnitVectorElement{2}{x}\pathTangentialUnitVectorElement{y}-\pathNormalUnitVectorElement{2}{y}\pathTangentialUnitVectorElement{x}]+\longitudinalCompensationCoefficient(\velocityErrorVectorElementEstimate{1}\pathTangentialUnitVectorElement{x} - \velocityErrorVectorElementEstimate{2}\pathTangentialUnitVectorElement{y}), \nonumber \\
  &\pathVelocityProfile[\pathTangentialUnitVectorElement{x}^2+\pathTangentialUnitVectorElement{y}^2]-\vfoLongitudinalGain \levelSurface{1}[\pathTangentialUnitVectorElement{y}\pathNormalUnitVectorElement{1}{y}+\pathTangentialUnitVectorElement{x}\pathNormalUnitVectorElement{1}{x}] \nonumber  \\
  &-\vfoLongitudinalGain \levelSurface{2}[\pathTangentialUnitVectorElement{x}\pathNormalUnitVectorElement{2}{x}+\pathTangentialUnitVectorElement{y}\pathNormalUnitVectorElement{2}{y}]+\longitudinalCompensationCoefficient(\velocityErrorVectorElementEstimate{1}\pathTangentialUnitVectorElement{x}+\velocityErrorVectorElementEstimate{2}\pathNormalUnitVectorElement{y})) \nonumber  \\
  &= \atantwo(-\vfoLongitudinalGain\levelSurface{1}\pathReducedNormalUnitVector{1}\otimes\pathReducedTangentialUnitVector-\vfoLongitudinalGain\levelSurface{2}\pathReducedNormalUnitVector{2}\otimes\pathReducedTangentialUnitVector \nonumber  \\
  &+\longitudinalCompensationCoefficient(\velocityErrorVectorElementEstimate{1}\pathTangentialUnitVectorElement{x} - \velocityErrorVectorElementEstimate{2}\pathTangentialUnitVectorElement{y}), \nonumber  \\
  &\pathVelocityProfile\pathReducedTangentialUnitVector^\top\pathReducedTangentialUnitVector - \vfoLongitudinalGain\levelSurface{1}\pathReducedNormalUnitVector{1}^\top\pathReducedTangentialUnitVector-\vfoLongitudinalGain\levelSurface{2}\pathReducedNormalUnitVector{2}^\top\pathReducedTangentialUnitVector+\longitudinalCompensationCoefficient\reducedLongitudinalVelocityErrorVectorEstimate^\top\pathReducedTangentialUnitVector),
  \label{eq:yawDesiredAuxiliaryDifference}
\end{align}
where $\pathReducedNormalUnitVector{i}\triangleq[\pathNormalUnitVectorElement{i}{x} \ \pathNormalUnitVectorElement{i}{y}]^\top$ for $i\in\{1,2\}$, and $\pathReducedTangentialUnitVector \triangleq [\pathTangentialUnitVectorElement{x} \ \pathTangentialUnitVectorElement{y}]^\top$, while the operation $\bfa\otimes\bfb \triangleq a_1b_2 - a_2b_1$ for some vectors $\bfa = [a_1 \ a_2]^\top\in\realNumbers^2$ and $\bfb = [b_1 \ b_2]^\top\in\realNumbers^2$. The upper-bound function of the absolute value of \eqref{eq:yawDesiredAuxiliaryDifference} can be expressed as
\begin{align}
  \abs{\auxiliaryDesiredYawDifference}&\leq\atantwo(\vfoLongitudinalGain\module{\positionError}\module{\pathReducedNormalUnitVector{1}}\module{\pathReducedTangentialUnitVector}\abs{\sin\beta_1}+\vfoLongitudinalGain\module{\positionError}\module{\pathReducedNormalUnitVector{2}}\module{\pathReducedTangentialUnitVector}\abs{\sin\beta_2} \nonumber \\
  &+\longitudinalCompensationCoefficient\module{\reducedLongitudinalVelocityErrorVectorEstimate}\module{\pathReducedTangentialUnitVector}, \nonumber  \\
  &\big|\pathVelocityProfile\module{\pathReducedTangentialUnitVector}^2-\vfoLongitudinalGain \levelSurface{1} \pathReducedNormalUnitVector{1}^\top\pathReducedTangentialUnitVector-\vfoLongitudinalGain \levelSurface{2} \pathReducedNormalUnitVector{2}^\top\pathReducedTangentialUnitVector+\longitudinalCompensationCoefficient
  \reducedLongitudinalVelocityErrorVectorEstimate^\top\pathReducedTangentialUnitVector\big|) \nonumber  \\
  &\leq\atantwo\big(2\vfoLongitudinalGain\module{\positionError}+\longitudinalCompensationCoefficient\module{\longitudinalVelocityErrorVector}+\longitudinalCompensationCoefficient\module{\longitudinalVelocityErrorVectorObservationError}, \nonumber  \\
  &\big|\pathVelocityProfile\module{\pathReducedTangentialUnitVector}^2-\vfoLongitudinalGain \levelSurface{1} \pathReducedNormalUnitVector{1}^\top\pathReducedTangentialUnitVector-\vfoLongitudinalGain \levelSurface{2} \pathReducedNormalUnitVector{2}^\top\pathReducedTangentialUnitVector+\longitudinalCompensationCoefficient
  \reducedLongitudinalVelocityErrorVectorEstimate^\top\pathReducedTangentialUnitVector\big|),
\end{align}
where $\beta_i = \angle(\pathReducedNormalUnitVector{i}, \pathReducedTangentialUnitVector)$.

\section{Difference between the desired and auxiliary pitch angles}
\label{app:etheta}

In order to obtain $\auxiliaryDesiredPitchDifference$ (see \eqref{eq:epsilonTheta}), we have used the relationship
\begin{align}
    \atan(X_1)-\atan(X_2) = \atan\left(\frac{X_1-X_2}{1+X_1X_2}\right),
\end{align}
leading to
\begin{align}
  \auxiliaryDesiredPitchDifference &\triangleq \desiredPitchAngle - \auxiliaryPitchAngle \nonumber \\
  &\eqtext{\eqref{eq:2.11}\eqref{eq:3.20}} \atan\left(\frac{-\pathTangentialUnitVectorElement{z}}{\pathTangentialUnitVectorElement{x}\cos\desiredYawAngle+\pathTangentialUnitVectorElement{y}\sin\desiredYawAngle}\right)-\atan\left(\frac{-\modifiedConvergenceVectorFieldEstimateElement{z}}{\modifiedConvergenceVectorFieldElementEstimate{x}\cos\auxiliaryYawAngle+\modifiedConvergenceVectorFieldElementEstimate{y}\sin\auxiliaryYawAngle}\right) \nonumber \\
  &\eqtext{\eqref{eq:2.10}\eqref{eq:3.20}} \atan\left(\frac{-\pathVelocityDirection\pathTangentialUnitVectorElement{z}}{\sqrt{\pathTangentialUnitVectorElement{x}^2+\pathTangentialUnitVectorElement{y}^2}}\right) - \atan\left(\frac{-\pathVelocityDirection\modifiedConvergenceVectorFieldElementEstimate{z}}{\sqrt{{\modifiedConvergenceVectorFieldElementEstimate{x}}^2+{\modifiedConvergenceVectorFieldElementEstimate{y}}^2}}\right) \nonumber \\
  &= \atan\left(\frac{\pathVelocityDirection\modifiedConvergenceVectorFieldElementEstimate{z}\sqrt{\pathTangentialUnitVectorElement{x}^2+\pathTangentialUnitVectorElement{y}^2}-\pathVelocityDirection\pathTangentialUnitVectorElement{z}\sqrt{{\modifiedConvergenceVectorFieldElementEstimate{x}}^2+{\modifiedConvergenceVectorFieldElementEstimate{y}}^2}}{\sqrt{{\pathTangentialUnitVectorElement{x}}^2+{\pathTangentialUnitVectorElement{y}}^2}\sqrt{{\modifiedConvergenceVectorFieldElementEstimate{x}}^2+{\modifiedConvergenceVectorFieldElementEstimate{y}}^2}+{\modifiedConvergenceVectorFieldElementEstimate{z}}\pathTangentialUnitVectorElement{z}}\right) \nonumber \\
  &= \atan\left(\frac{{\pathVelocityDirection\modifiedConvergenceVectorFieldElementEstimate{z}}\module{\pathReducedTangentialUnitVector}-\pathVelocityDirection\pathTangentialUnitVectorElement{z}\sqrt{{\modifiedConvergenceVectorFieldElementEstimate{x}}^2+{\modifiedConvergenceVectorFieldElementEstimate{y}}^2}}{\module{\pathReducedTangentialUnitVector}\sqrt{{\modifiedConvergenceVectorFieldElementEstimate{x}}^2+{\modifiedConvergenceVectorFieldElementEstimate{y}}^2}+{\modifiedConvergenceVectorFieldElementEstimate{z}}\pathTangentialUnitVectorElement{z}}\right).
\end{align}

{\bf References} \\
\bibliographystyle{plain}
\bibliography{bibliography,MyIEEEBib}
\end{document}